\documentclass[twocolumn,pra,aps,longbibliography,nofootinbib]{revtex4-1}

\usepackage[caption=false]{subfig}

\usepackage{amsmath}
\usepackage{amssymb}
\usepackage[dvips]{graphicx}
\usepackage{pstricks}
\usepackage{pst-node}
\usepackage{pst-plot}
\usepackage[normalem]{ulem}

\usepackage{bbm}
\usepackage{algcompatible}
\usepackage{newfloat}
\DeclareFloatingEnvironment[
    fileext=loa,
    listname=List of Algorithms,
    name=ALGORITHM,
    placement=tbhp,
]{algorithm}

\usepackage{amsthm}
\usepackage{hyperref}
\hypersetup{
  colorlinks   = true, 
  urlcolor     = blue, 
  linkcolor    = blue, 
  citecolor   = blue
}
\usepackage{tikz}
\usepackage{multirow}
\usepackage{enumitem}

\makeatletter
\newtheorem*{rep@theorem}{\rep@title}
\newcommand{\newreptheorem}[2]{%
\newenvironment{rep#1}[1]{%
 \def\rep@title{#2 \ref{##1}}%
 \begin{rep@theorem}}%
 {\end{rep@theorem}}}
\makeatother

\newtheorem{theorem}{Theorem}
\newreptheorem{theorem}{Theorem}
\newtheorem{corollary}{Corollary}
\newtheorem{lemma}{Lemma}
\newtheorem{definition}{Definition}

 \newcommand{\sset}[1]{ \{#1\} }

 \newcommand{\ket}[1]{|#1\rangle}

\begin{document}

\title[Short Title]{
Quantum Error Correction with the Semion Code}

\author{G. Dauphinais, L. Ortiz, S. Varona, and M.A. Martin-Delgado}
\affiliation{
Departamento de F\'{\i}sica Te\'orica, Universidad Complutense, 28040 Madrid, Spain
}

\begin{abstract}
We present a full quantum error correcting procedure with the semion code: an off-­shell extension of the double ­semion model. We construct open-string operators that recover the quantum memory from arbitrary errors and closed-string operators that implement the basic logical operations for information processing. Physically, the new open-string operators provide a detailed microscopic description of the creation of semions at their end­points. Remarkably, topological properties of the string operators are determined using fundamental properties of the Hamiltonian, namely, the fact that it is composed of commuting local terms squaring to the identity. In all, the semion code is a topological code that, unlike previously studied topological codes, it is of non-­CSS type and fits into the stabilizer formalism. This is in sharp contrast with previous attempts yielding non-­commutative codes.
\end{abstract}

\maketitle
\section{Introduction}

Topological properties of quantum systems have become a resource of paramount importance to construct
quantum memories that are more robust to external noise and decoherence \cite{kitaev97,dennis_etal02,bravyikitaev98,colorcode1,colorcodeII,homologicalBM,comparativeBM} than
standard quantum error correcting codes \cite{shor1,steane,shor2,knill_etal,kitaev1,aharonov,NC00,GM}. The latter  are based on a  special class of codes - concatenated codes - which enable us to perform longer quantum computations reliably, as we increase the block size.

 The Kitaev code is the simplest topological code yielding a quantum memory \cite{kitaev97}. It can be thought of as a simple two-dimensional lattice gauge theory with gauge group $G=\mathbb{Z}_2$. In $D=2$ spatial dimensions, there is another lattice gauge theory with the same gauge group but different topological properties: the Double Semion (DS) model \cite{levinwen,DS2,FH16,PhysRevB.90.195148,PhysRevB.90.035117,1367-2630-16-1-013015,PhysRevB.92.155105}. Although the Kitaev and the DS models are lattice gauge theories sharing the same gauge group, $G=\mathbb{Z}_2$, the braiding properties of their quasiparticle excitations are radically different. For example, whereas braiding two elementary quasiparticle excitations (either an electric or a magnetic charge) gives a $+1$ phase in the Kitaev code, doing so in the DS model yields $\pm i$ phase factors.

The DS model was introduced in the context of the search of new topological orders in strongly correlated systems, gapped, non-chiral and based on string-net mechanisms in $D=2$ dimensions~\cite{levinwen,DS2}. Generalizations of the DS model to $D=3$ and to higher dimensions have appeared recently \cite{FH16}. While the properties of the Kitaev code has been extensively studied in quantum computation and condensed matter, barely nothing is known about the quantum error correcting properties of the DS model despite recent efforts towards realizing such models \cite{FURUSAKI2017,LI2017,SYED2019}. In this work we remedy this situation by introducing a new formulation of the DS model that is suitable for a complete treatment as a quantum memory with topological properties. 

\begin{figure}
\subfloat[\label{fig:vertex_on_lattice}]{%
\raisebox{.5cm}{
\includegraphics[scale=0.2]{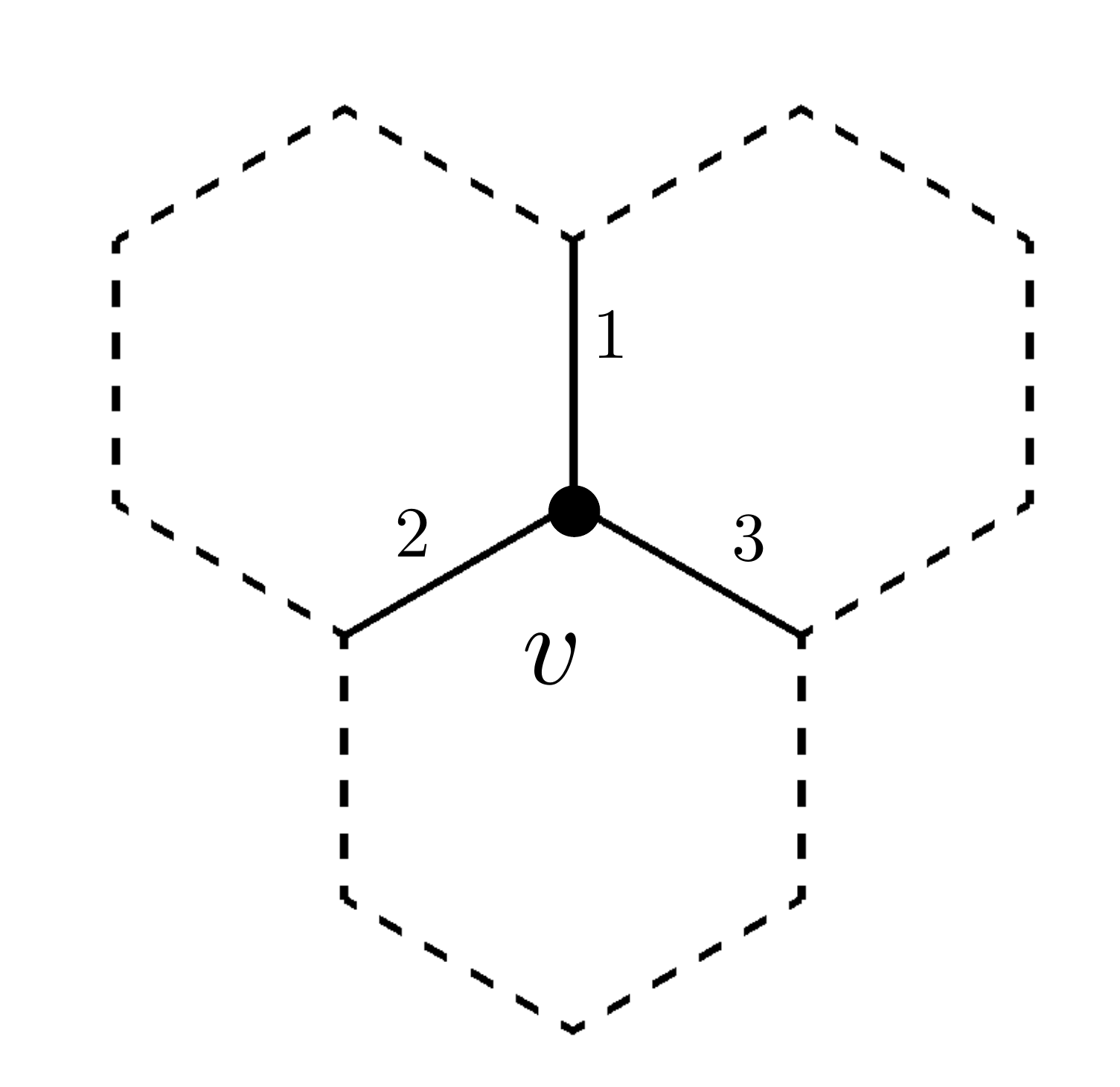}%
}}\hfill
\subfloat[\label{fig:plaquette_on_lattice}]{%
\includegraphics[scale=0.2]{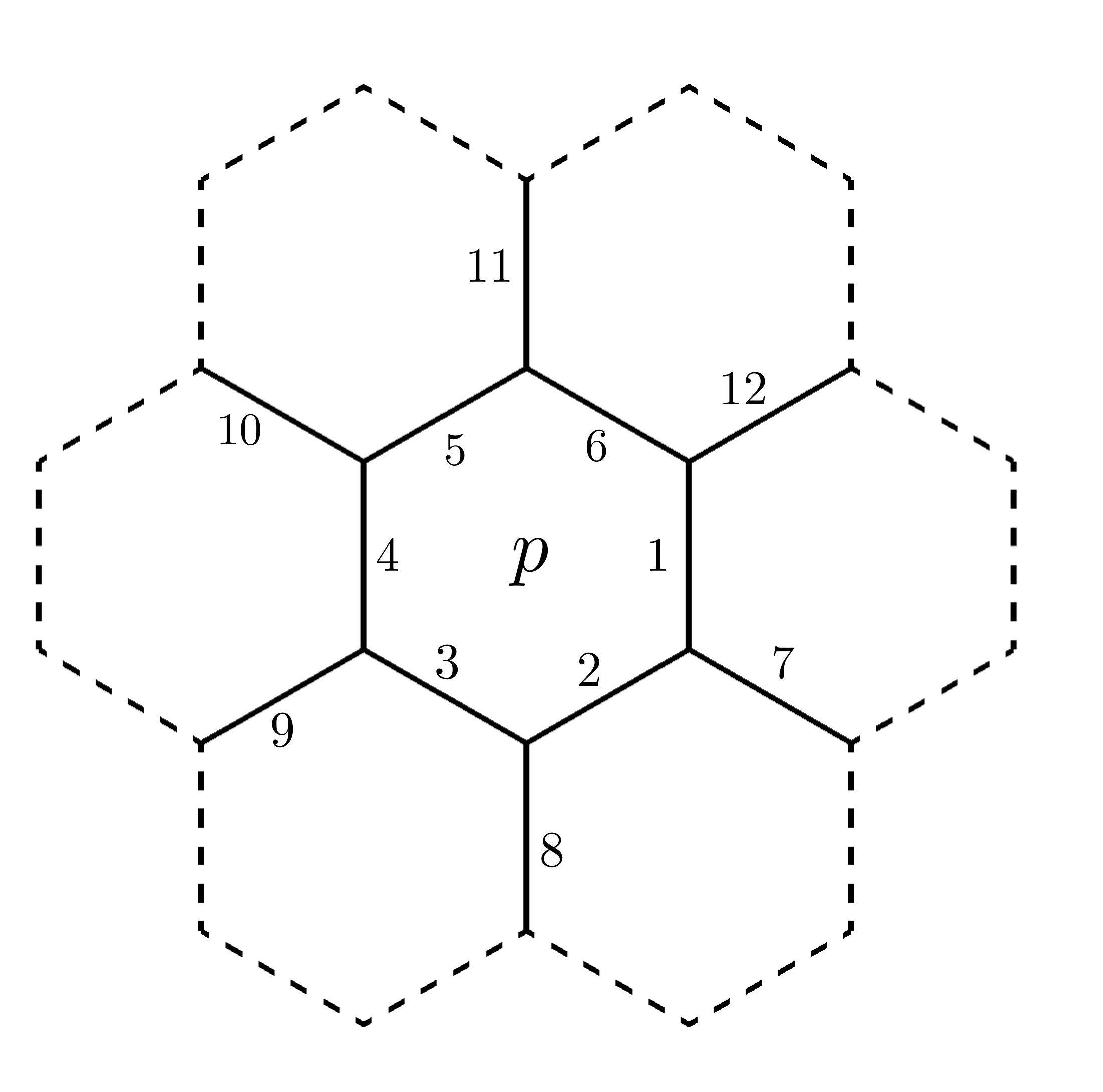}%
}
\caption{The support of the vertex operator $Q_v = \sigma^z_1 \sigma^z_2 \sigma^z_3$ and plaquette operator $\tilde{B}_p$. The qubits are placed on the edges.}
\label{fig:support_plaquette}
\end{figure}

The first obstacle to tackle the DS model as a quantum memory is the original formulation as a string-net model \cite{levinwen}. In this formulation, the Hamiltonian is only Hermitian and exactly solvable in a particular subspace, where plaquette operators are Hermitian and commute. Only linear combinations of closed-string configurations, implying the absence of vertex excitations, are allowed in this subspace \cite{levinwen,MR13,fionna}. The microscopic formulation of the original DS model starts with a hexagonal lattice $\Lambda $ with qubits placed at links $e$. Vertex operators $Q_v$ are attached to the three links meeting at a vertex $v\in \Lambda$. Plaquette operators $\tilde{B}_p$ are attached to hexagons $p\in \Lambda$ with the novel feature that their outer links carry additional phase factors that are missing in the corresponding Kitaev model.
Explicitly, 
\begin{equation}\label{vertex_operators}
Q_{v} := \sigma^z_i \sigma^z_j \sigma^z_k ,
\end{equation}
with $i,\ j ,\ k$, being the three qubits belonging to vertex $v$ (see Fig.~\ref{fig:vertex_on_lattice}) and
\begin{equation}\label{plaquette_operators_old}
\tilde{B}_p := \left( \prod_{i\in \partial p} \sigma^x_i\right) \prod_{j\in o(p)} i^{\frac{1}{2} \left(1-\sigma^z_j\right)},
\end{equation}
where $\partial p$ are the six links of the hexagon and $o(p)$ is the set of six edges outgoing from each plaquette $p$, as it is shown in Fig.~\ref{fig:plaquette_on_lattice}. Unlike the Kitaev code, these plaquette operators $\tilde{B}_p$ are Hermitian and commute among themselves only in a subspace of the whole Hilbert space, defined by the so-called zero-flux rule \cite{levinwen,MR13,fionna}. This is given by a vertex-free condition on states,
\begin{equation}\label{vertex-free}
{\cal H}_v := \sset{ \ket{\psi}: Q_v \ket{\psi} = + \ket{\psi}, \forall v \in \Lambda }.
\end{equation}

The set of  vertex and plaquette operators defines a Hamiltonian
\begin{equation}\label{hamDS}
\tilde{H}_{\mathrm{DS}} := - \sum_{v\in \Lambda} Q_v + \sum_{p\in \Lambda} \tilde{B}_p.
\end{equation}
Due to the involved structure of phases that plaquette operators in Eq.~\eqref{plaquette_operators_old} have, it was aforementioned that these operators do not commute out of the vertex-free subspace. This implies that the model is only well-defined when there are no vertex excitations.

Therefore, in order to treat the DS model as a quantum error correcting code \cite{PhysRevA.54.1098,Steane2551,preskillnotes,knill_etal,RevModPhys.87.307}, it is necessary to have a formulation of the model that is valid in the whole Hilbert space and not just for the vertex-free subspace \eqref{vertex-free}, since generic noise processes will make the system leave the mentioned subspace. To this end, we introduce an off-shell DS model that we call the semion code. This new model is achieved by making a deformation of the original plaquette operators \eqref{plaquette_operators_old} such that they become commuting and Hermitian operators without imposing the vertex-free condition \cite{fionna}. In addition, we are able to develop the complete program of quantum error correction with the semion code.

\subsection{Summary of main results}

In order to summarize the main contributions that we present in this paper, we hereby advance a list of some of our most relevant results.

\begin{enumerate}[label=(\emph{\roman*})]

\item We perform a thorough analysis of a formulation of the plaquette operators which commute in the whole Hilbert space \cite{fionna}. This construction consists in adding extra phases to the plaquette operators, which depend on the configuration of the three edges at each vertex. When the vertex-free condition is imposed, we recover the standard definition of the DS model.
\item We give an explicit construction for string operators along arbitrary paths. They are complete in the sense that any operator acting on the system can be decomposed as a linear superposition of such operators. Additionally, the string operators can be constructed efficiently despite the complex structure of the plaquette operators. Remarkably, a microscopic formulation to create semions was not proven until now.
\item We analytically show that the excitations of the system behave as semions, via the detailed study of the constructed string operators, which allows us to explicitly calculate the topological S-matrix. Interestingly enough, most of the string operator properties rely on very generic arguments about the structure of the local operators making up the Hamiltonian, namely that they commute and square to the identity.
\item  Closed-string operators are constructed, which allow us to perform logical operations on the quantum memory built from the semion code. Logical operators are closed-string operators whose paths are homologically non-trivial and  act non-trivially on the degenerate ground space.
\item  Given the above properties, we define a topological quantum error correcting code based on a non-trivial extension of the DS model. We build a code, which is characterized by the following key properties: is topological, satisfies  the stabiliser formalism, is non-CSS, non-Pauli and additive. 
\end{enumerate}
The topological nature of the code becomes apparent through the fact that global degrees of freedom are used to encode information and only local interactions are considered \cite{kitaev97,colorcode1}. Another remarkable feature of the semion code is that is of non-CSS type since in the plaquette operators both Pauli $X$ and $Z$ operators enter in the definition \cite{PhysRevA.54.1098,steane}. Consequently, errors in an uncorrelated error model such as independent bit-flip and phase-flip errors cannot be treated separately. However, they can be decomposed as a linear superpositions of fundamental \textit{anyonic} errors (string operators creating pairs of anyons) with a known effect on the Hilbert space. Moreover, the semion code is not a subgroup of the Pauli group since the complex phases entering its definition (see Eq.\ \eqref{plaquette_operators_old}) makes impossible to express its generators in terms of tensor products of Pauli matrices \cite{gottesman96,doi:10.1063/1.4920923}. Nevertheless, the semion code is still an additive code: the sum of quantum codewords is also a codeword \cite{calderbank_etal}. This last fact is intimitely related to the abelian nature of the semionic excitations \cite{kitaev05, Pfeifer2012}.

\subsection{Outline}
The article is organized as follows. In Sec.~\ref{sec:off_shell} we introduce the off-shell DS model, which is suitable for quantum error correction. In Sec.~\ref{sec:string_operators} we build string operators creating vertex excitations at their endpoints. Sec.~\ref{sec:semion_code} is devoted to logical operators and quantum error correction. We conclude in Sec.~\ref{sec:conclusions}. Appendices deserve special attention since they contain the detailed explanations of all the constructions used throughout the text. Specifically, App.~\ref{app:example_string} presents an explicit example of a string operator, App.~\ref{app:proof_thm_string_operators} gives the detailed proof of Theorem \ref{thm:string_operators}, which presents a systematic way to construct string operators, as well as several key properties of the string operators and finally App.\ \ref{app:openstrings_commutation} is devoted to the proof of Theorem \ref{thm:crossing_strings}, which gives the commutation relations among string operators.

\section{Off-Shell double semion: microscopic model}\label{sec:off_shell}

We begin by considering a microscopic description of the DS model on the entire Hilbert space of states, since it is much better suited for quantum error correction. We call this an off-shell DS code by borrowing the terminology from quantum field theory and other instances in physics where a shell condition amounts to a constraint on the phase space of a system. For instance, the equation of motion is a shell condition for quantum particles but the phase space is more general. In our case, the shell condition is the vertex-free subspace or zero-flux rule introduced in Eq.\ \eqref{vertex-free}. 

\subsection{ Double semion model in the vertex-free subspace}


Let us start by introducing the DS model in a new presentation that is more suitable for building an off-shell formulation of it. We consider the same hexagonal lattice $\Lambda$ with qubits attached to the edges $e\in \Lambda$. The vertex operators $Q_v$ will remain the same as in Eq.\ \eqref{vertex_operators}, but the plaquette operators in the zero-flux subspace can be rewritten in an equivalent form to that shown in Eq.\ \eqref{plaquette_operators_old}, i.e.,
\begin{equation} \label{eq:bp_simple}
\tilde{B}_p = \left( \prod_{j=1}^6 \sigma^x_{j} \right) \left( \prod_{j=1}^{6} \left( -1 \right)^{n^{-}_{j-1} n^{+}_{j}} \right),
\end{equation}
in which $n^{\pm}_i := \frac{1}{2} (1 \pm \sigma^z_i)$ is the projector on the state $|0\rangle$ ($n^+$) or $|1\rangle$ ($n^-$) of qubit $i$,  qubits are labeled as shown in Fig.~\ref{fig:support_plaquette} and we use the convention that $n^{\pm}_0$ refers to qubit `6' for simplicity. Remarkably, this expression avoids any reference to the outgoing links of hexagonal plaquettes $p$. One readily sees that the vertex operators fulfill the following relations:
\begin{equation} \label{eq:vertex_relations}
Q^2_v = 1, \quad [Q_v,Q_{v'}] = 0, \quad [Q_v, \tilde{B}_p] = 0;
\end{equation}
$\forall v,v',p \in \Lambda$. As for the plaquette operators $\tilde{B}_p$, they also satisfy
\begin{equation} \label{eq:bp_relations}
\tilde{B}^2_p = 1, \quad \tilde{B}^\dag_p = \tilde{B}_p, \quad [\tilde{B}_p, \tilde{B}_{p'}] = 0;
\end{equation}
$\forall p,p' \in \Lambda$, but only in the vertex-free subspace \eqref{vertex-free}.
Furthermore, the product of all the vertex and plaquette operators is the identity. A simple counting argument reveals that the ground space is $4^g$ degenerate\footnote{This is strictly true only if the total number of plaquettes of the system is even. If it is odd, then the ground state must contain a single flux excitation, which can be placed in any of the plaquettes. In that latter case the ground space degeneracy is an extensive quantity. However, any given flux configuration is $4^g$-degenerate. For simplicity, we assume in this work that the system contains an even number of plaquettes.}, which $g$ being the genus of the orientable compact surface onto which the lattice is placed.

An explicit unnormalized wavefunction belonging to the ground space is obtained in the following way:  we start from the vacuum, i.e., $|0\rangle^{\otimes N}$, which has +1 eigenvalue for all vertex operators. Then, plaquette operators are used to build projectors and apply them onto the vacuum,
\begin{align}\label{eq: gs}
 \ket{\Omega}=\prod_{p\in \Lambda}\frac{1-\tilde{B}_p}{2}\ |0\rangle^{\otimes N}.
\end{align} 
It is straightforward to check that this state fulfills the lowest energy condition for the Hamiltonian \eqref{hamDS} within the vertex-free subspace. Expanding the product in Eq.\ \eqref{eq: gs}, one can see that the ground state is a superposition of closed loops configurations. Due to the condition  $\tilde B_p=-1$ for the ground state, the coefficients of this superposition of closed loops alternate sign. Thus, we can write the ground state in a different way:
\begin{align}
\ket{\Omega}=\sum_{\vec{i}\ \in \left\lbrace \text{C-S conf.} \right\rbrace} (-1)^{N_L\left(\vec{i}\right)}|\vec{i}\;\rangle,
\end{align}
where $\vec{i}$ is a bitstring representing a qubit configuration and  $ \left\lbrace \text{C-S conf.} \right\rbrace$ is the set of all possible closed-string configurations. Each configuration in this set has a certain number of closed loops, $N_L(\vec{i})$, whose parity determines the sign of the coefficient in the ground state superposition. 

Of course the above construction only gives rise to one of the ground states. To find the other ones, the starting configuration can simply be replaced by a configuration containing an homologically non-trivial closed loop (which necessarily belongs to the vertex-free subspace), and proceed with the same construction. Every different homological class for the closed loop corresponds to a different ground state.

\begin{figure}
\includegraphics[scale=.4]{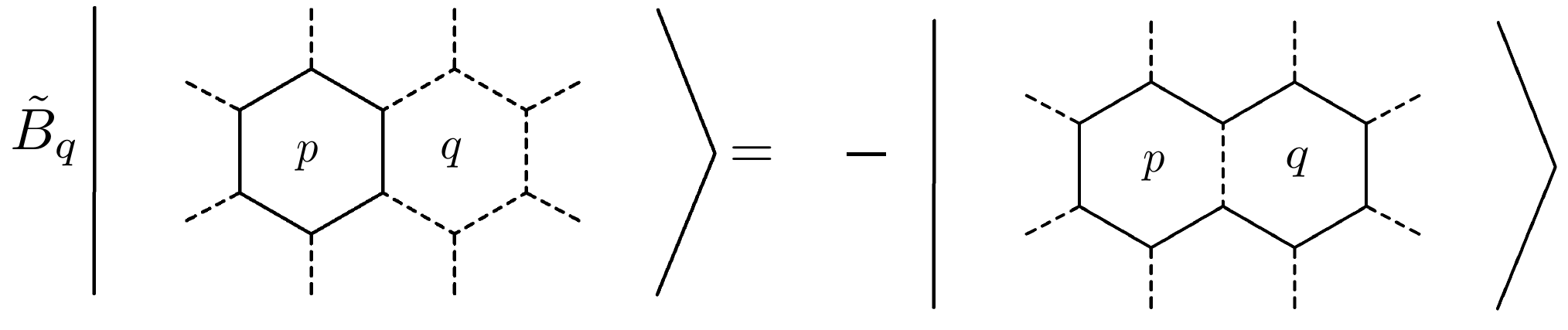}
\caption{Applying plaquette operator $\tilde{B}_q$ adds a closed loop around plaquette $q$ and multiply the wave function by a $-1$ phase. Each qubit in a given configuration is represented by a filled link of the lattice if it is in state $|1\rangle$, whereas links in $|0\rangle$ are left empty (dashed line).}
\label{fig:Bp_tilde}
\end{figure}

Applying the $\tilde{B}_p$ operator on a specific loop configuration flips the string occupancy of the interior edges of plaquette $p$ while acquiring a phase that depends on the specific configuration under consideration. Applying $\tilde{B}_p$ on the vacuum simply adds a closed loop around plaquette $p$, while applying $\tilde{B}_p$ next to a closed loop either enlarges (or shrinks) the existing loop to include (exclude) plaquette $p$, while multiplying the wave function by $-1$ factor (see Fig. \ref{fig:Bp_tilde}).

Due to the lack of commutativity of the plaquette operators and the fact that they are not Hermitian, the original DS model is only well-defined when there are no vertex excitations.  Moreover, the strings creating vertex excitations are not properly defined either. A naive attempt to construct these strings as a chain of $\sigma^x$ operators, following the similarities with the Kitaev code, resoundingly fails. As a consequence of the phases on the external legs of plaquette operators, $\sigma^x$ operators create vertex excitations but also plaquette excitations. In order to get a string that creates only two vertex excitations at the endpoints but commutes with all the plaquette operators along the path $\mathcal P$, it is necessary to add some extra phases to the chain of $\sigma^x$ on the outer legs. An approach to this problem is described in \cite{fionna} but it is not  successfully solved since the strings are only  well-defined in the vertex-free subspace.

 The DS model gives rise to quasiparticle excitations behaving like anyons. They are called semions, due to the fact that their topological charge is  `half' of that of a fermion, i.e., $\pm i$. There exist two types of semions in the model, one corresponding to a vertex excitation, while the other corresponds to both vertex and adjacent  plaquettes excitations. From now on, we name these two different possibilities as {\em chiralities}, positive chirality for the former kind, and negative chirality for the latter one. We warn the reader that this choice has been made arbitrarily, and does not necessarily reflect the topological charge of a given specie.

Taking into account all these caveats, we present in the following a formulation of the DS model which gives a microscopic approach to this interesting topological order, fulfilling all the necessary properties in the whole Hilbert space.

\subsection {Exactly solvable model in the whole Hilbert space}

If we want to consider encoding quantum information in the degenerate ground-state manifold of the standard (on-shell) DS model in Eq.~\eqref{hamDS}, we immediately run into major problems: $X$ Pauli errors make the state of the system leave the vertex-free subspace. The non-commutativity  of the $\tilde{B}_p$ operators poses difficulties when interpreting the DS model as a stabilizer code.

\begin{figure}
\center
\includegraphics[scale=0.3]{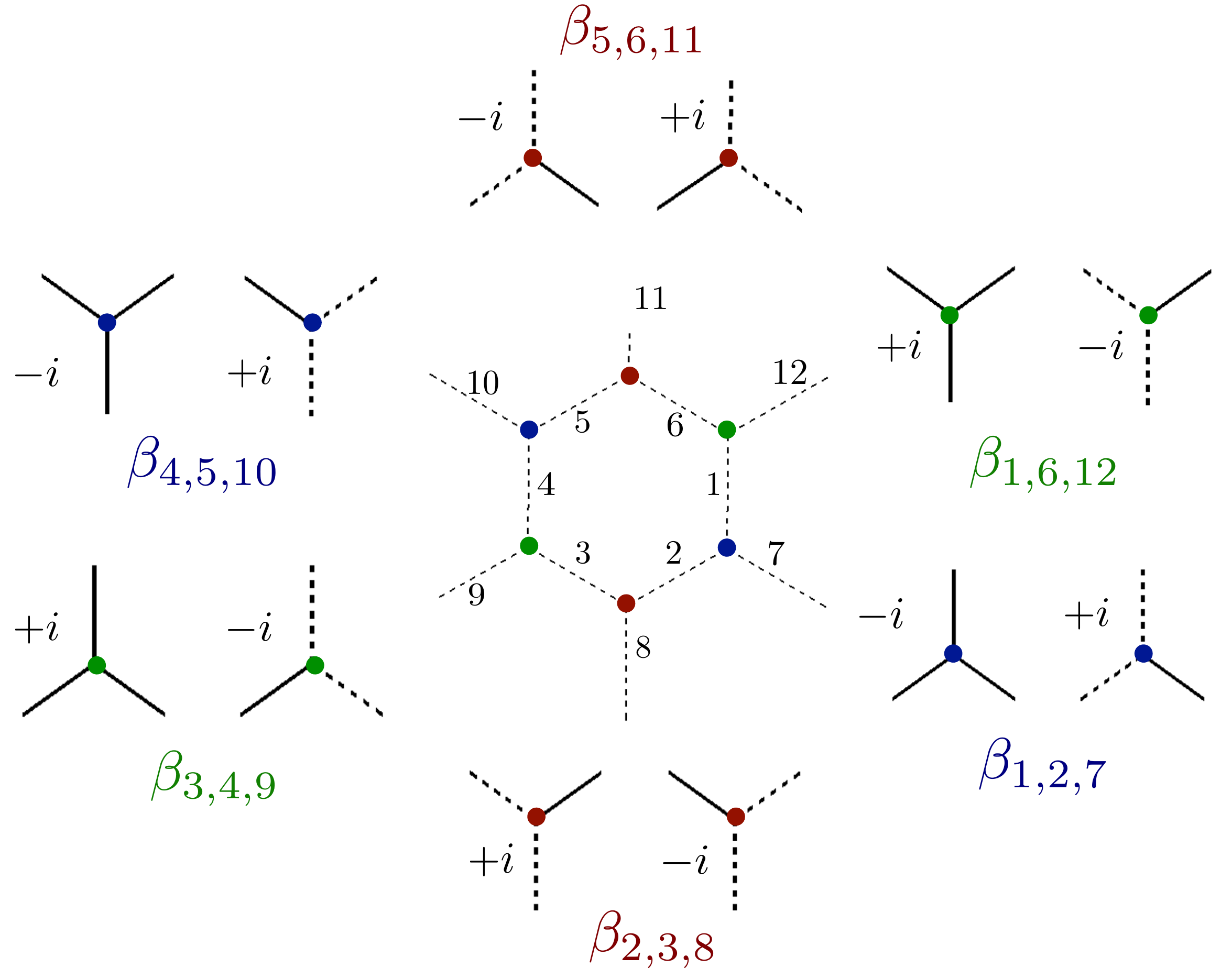}
\caption{A graphical representation of the various non-trivial phase factors of the form $\beta_{(i, j, k)}$. A trivial phase of $+1$ is applied for any configuration not shown on this picture. The labeling of the edges have been omitted for simplicity, and corresponds to the same as the one on Fig. \ref{fig:support_plaquette}.}
\label{fig:phase_factor}
\end{figure}

To avoid such difficulties, we consider a modified version of the plaquette operators in Eq.~\eqref{plaquette_operators_old}, which we call the off-shell DS model or semion code
\begin{equation} \label{eq:hamil_gen}
H_{\rm{DS}} = - \sum_{v} Q_{v} + \sum_p B_p,
\end{equation}
where the \emph{generalized} plaquette operator $B_p$ is a modification of $\tilde{B}_p$ obtained by multiplying it by a phase factor that depends on the configuration on which it is applied. More specifically, we have
\begin{equation} \label{eq:Bp_generalized}
B_p := \tilde{B}_p \times \tilde{\beta}_p,
\end{equation}
with
\begin{equation} \label{eq:theta_op}
\tilde{\beta}_p := \sum_{\vec i} \tilde{b}_p ( \vec i ) \vert \vec i \rangle \langle \vec i \vert, 
\end{equation}
where the sum runs over all possible configurations of edges $1$ through $12$ shown in Fig.~\ref{fig:support_plaquette}. $\tilde{b}_p ( \vec i )$ is the phase factor corresponding to the string configuration $\vec i$. $\vec i$ represents a state in the computational basis. A qubit in the state $\vert 0 \rangle$ is interpreted as the absence of a string on its corresponding edge, while the state $\vert 1 \rangle$ reflects the presence of a string. The phase operator $\beta_p$ can be decomposed as
\begin{equation} \label{eq:decomp_phase_factor}
\tilde{\beta}_p = \beta_{(6,1,12)} \,  \beta_{(1,2,7)} \,  \beta_{(2,3,8)} \, \beta_{(3,4,9)} \, \beta_{(4,5,10)} \, \beta_{(5,6,11)},
\end{equation}
where $\beta_{(i, j, k)}$ is a function of the string configuration of edges $i, j$ and $k$ connected to vertex $v(i,j,k)$. The specific values for each factor $\beta_{(i,j,k)}$ are shown graphically in Fig.~\ref{fig:phase_factor}. Note that their specific form differ depending on their position on the plaquette.

For future reference, notice that the generalised plaquette operator $B_p$ can be written as
\begin{equation} \label{eq:trivalent_plaquette}
B_p = \prod_{i \in \partial p} \sigma^x_i \left( \prod_{j \in \partial p} \left( -1 \right)^{n^{-}_{j-1} n^{+}_{j}} \right) \prod_{v \in p} \beta_{v},
\end{equation}
where we use the notation $j \in \partial p$ to indicate the qubit associated to edge $j$  in plaquette $p$. $ v \in p$  identify the vertices belonging to plaquette $p$. Notice that this last expression clearly shows that the phase factor appearing in $B_p$ is a product of phases, $\beta_v$, depending on the string configuration of the three edges connected to each vertex of plaquette $p$. The complete algebraic expression of the product of all $\beta_v$ in a plaquette $p$ is \cite{fionna}
\begin{align} \label{eq:beta_algebraic}
\prod_{v \in p} \beta_{v} &=\  i^{n_{12}^- \left( n^-_1 n^-_6 - n^+_1 n^+_6 \right)} i^{n^-_7 \left( n^+_1 n^+_2 - n^-_1 n^-_2 \right)}\nonumber \\ 
&\times\ i^{n^+_8 \left( n^-_2 n^+_3 - n^+_2 n^-_3 \right)}  i^{n^-_9 \left( n^-_3 n^-_4 - n^+_3 n^+_4 \right)} \\
 &\times\ i^{n^-_{10} \left( n^+_4 n^+_5 - n^-_4 n^-_5 \right)} i^{n^+_{11} \left( n^-_5 n^+_6 - n^+_5 n^-_6 \right)}.\nonumber
\end{align}
We can easily check that in the zero-flux rule the factors in Eq.\ \eqref{eq:beta_algebraic} reduce to 1, recovering expression \eqref{eq:bp_simple} for the plaquette operators.

The crucial point now is that the new generalized plaquette operators, $B_p$, satisfy the desired properties needed by the stabilizer formalism of quantum error correction. Namely,
\begin{equation} \label{eq:gen_bp_relations}
B^2_p = 1,\ \ B^\dag_p = B_p,\ \  [B_p, B_{p'}] = 0,\ \ [Q_v, B_p]=0;
\end{equation}
$\forall p,\ p',\ v \in \Lambda,$ regardless of the vertex-free condition \eqref{vertex-free}. The study of $H_{\rm{DS}}$ is rendered much simpler than that of $\tilde{H}_{\rm{DS}}$ on the whole Hilbert space of the qubits by the fact that the new plaquette operators commute.

%
\section{String operators} \label{sec:string_operators}

We seek open-string operators creating excitations at their endpoints without affecting the rest of vertex and plaquette operators, as well as closed-string operators that commute with vertex and plaquette operators. In our case, excited states correspond to states in a $-1$ eigenstate for a vertex operator or a $+1$ eigenstate of a plaquette operator. We say that an excitation is present at vertex $v$ (plaquette $p$) if the state of the system is in a $-1$ ($+1$) eigenstate of $Q_{v}$ ($B_p$). Since we have that $\prod_{v \in \Lambda} Q_{v} = \prod_{p \in \Lambda} B_p = 1$, excitations are always created in pairs.

In order to find such string operators, it is convenient to reexpress the generalized plaquette operators as
\begin{equation} \label{eq:Bp_def}
B_p = \prod_{i \in p} \sigma^x_i \sum_{\vec i} b_p \left( \vec i \right) \vert \vec i \rangle \langle \vec i \vert,
\end{equation}
where $p$ denotes the interior edges of a plaquette (edges $1$ through $6$ in Fig.~\ref{fig:support_plaquette}) and the string configurations in the sum are taken on edges $1$ through $12$. $b_p (\vec i)$ denotes the complex phase picked up when applying operator $B_p$ to the configuration $\vec i$. Note that $\sum_{\vec i} b_p ( \vec i ) \vert \vec i \rangle \langle \vec i \vert$ differs from the product of the $\beta_p$'s in the $-1$ factors appearing in Eq.~\eqref{eq:trivalent_plaquette}. $\sum_{\vec i} b_p ( \vec i ) \vert \vec i \rangle \langle \vec i \vert$ includes the $-1$ factors as well as the product of $\beta_v$.

Given two string configurations $\vec i $ and $\vec{\alpha}$ on a set of edges, it is useful to define the string configuration $\vec i \oplus \vec{\alpha}$ to be the configuration $\vec{i}'$ where the edges occupied in configuration $\vec{\alpha}$ has been flipped. It is equivalent to sum (mod 2) the two bitstrings. Additionally, we define the configuration $\vec{\alpha}^p$ of plaquette $p$  as the string configuration which is empty everywhere except for the six edges in the interior of plaquette $p$, corresponding to edges $1$ through $6$ in Fig.~\ref{fig:support_plaquette}. Likewise, $\vec{\alpha}^\mathcal{P}$, will be the configuration which is only occupied for edges of path $\mathcal{P}$.

 Given a path $\mathcal P$, we construct string operators $S^+_{\mathcal P}$ creating vertex excitations at its endpoints and commuting with every other operators in Hamiltonian \eqref{eq:hamil_gen}. Negative chirality strings are defined as $S^-_{\mathcal P}:= S^+_{\mathcal P}S^z_\mathcal{P'_{\textnormal{dual}}}$, where $S^z_\mathcal{P'_{\textnormal{dual}}}$ is a product of $\sigma^z$ operators forming a path $\mathcal{P'_{\textnormal{dual}}}$ in the dual lattice which is contained in the support of $S^+_{\mathcal P}$. If $\mathcal P$ is open, $ S^z_\mathcal{P'_{\textnormal{dual}}}$ creates excitations at plaquettes containing the vertices at the endpoints of $\mathcal P$ and opposite to the first and last edges of $\mathcal P$ ($p_1$ and $p_9$ in Fig.\ \ref{fig:string_structure}), while if $\mathcal P$ is closed, $S^z_\mathcal{P'_{\textnormal{dual}}}$ forms a closed path in the same homological class as $\mathcal P$. Note that for a given path $\mathcal P$, various paths $\mathcal P'_{\textnormal{dual}}$ are possible, and each one gives rise to a different string operator $S^-_{\mathcal P}$.

\subsection{An algorithm to generate string operators}
\label{sec:algorithm}

\begin{figure}
\center
\includegraphics[scale=0.3]{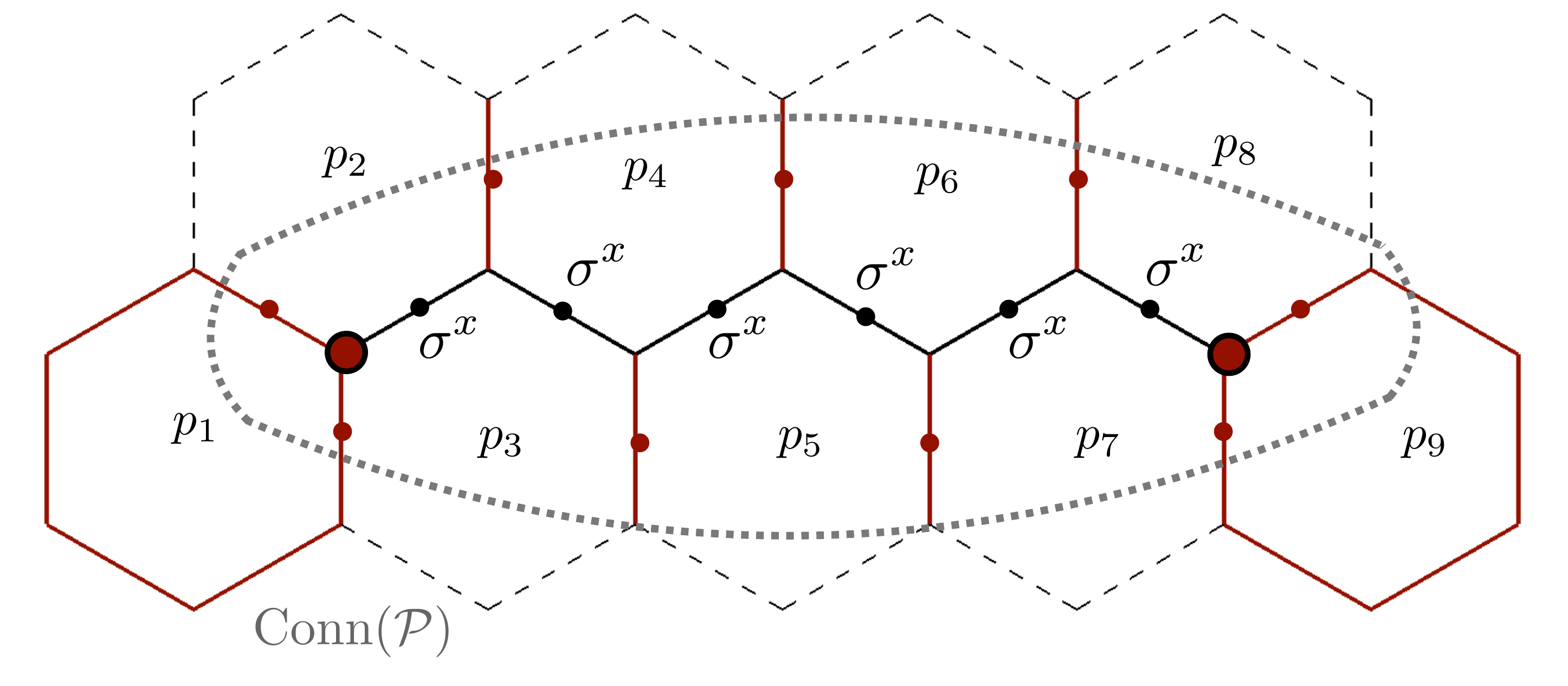}
\caption{The structure of $S^{+}_{\mathcal P}$, where the path $\mathcal P$ is indicated by the full black edges where the $\sigma^x$ operators are applied. The phase factors $F^{+}_{\mathcal P}$ depend on the configuration of all the  edges identified with a dot, which are collectively denoted by $\textnormal{Conn} (\mathcal P)$. The effect of $S^{+}_{\mathcal P}$ on the ground state of the system is to create a pair of vertex excitations at the vertices located at the endpoints of the path $\mathcal P$, which are identified by big red dots. Plaquettes $p_1$ to $p_9$ constitute the set $\mathcal B_{\mathcal P}$.}
\label{fig:string_structure}
\end{figure}

In order to find these string operators, we consider the following ansatz:

\begin{equation} \label{eq:String_ansatz}
S^{+}_{\mathcal P} := \prod_{i \in \mathcal P} \sigma^x_i  \sum_{\vec i \, \in \, \textnormal{Conn} (\mathcal P) } F_{\mathcal P} \left( \vec i \right) \vert \vec i \rangle \langle \vec i \vert,
\end{equation}
where $F_{\mathcal P} ( \vec i )$ is a phase factor acquired when $S^{+}_{\mathcal P}$ is applied on configuration $\vec i$. $F_{\mathcal P} ( \vec i )$ only depends on qubits belonging to $\textnormal{Conn} (\mathcal P)$, defined as:
\begin{equation}
\textnormal{Conn} (\mathcal P):=\{\textnormal{links of} \; \mathcal P \textnormal{ and its external legs} \}.
\end{equation}
 It is also useful to define the set of plaquettes
 \begin{equation}
  \mathcal B_{\mathcal P}:=\{  p: \partial p \; \cap \;\textnormal{Conn} (\mathcal P) \neq  \varnothing\},
  \end{equation}
which is the set of plaquettes that have at least one of their interior edges contained in Conn($\cal{P}$). Equivalently, one can define $\mathcal B_{\mathcal P}$ to be the set of plaquettes such that for at least one string configuration $\vec i$, $b_p (\vec i \oplus \vec{\alpha}^{\mathcal P}) \neq b_p (\vec i)$. The structure of $S^{+}_{\mathcal P}$ is illustrated in Fig.~\ref{fig:string_structure}. Note that depending on the context, a configuration $\vec i$ is either understood to be on the full system, or is the configuration restricted to $\textnormal{Conn} (\mathcal P)$. The specific case considered is explicitly stated in each case.

Ansatz \eqref{eq:String_ansatz} should satisfy the following properties:
\begin{enumerate}[label=(\emph{\roman*})]
\item Anticommutes with vertex operators at the endpoints of $\cal{P}$ if it is open, while it commutes with every other vertex and plaquette operators. \label{property_1}

\item Acts trivially on edges outside Conn($\cal{P}$). \label{property_2}
\end{enumerate}
Operators satisfying (\emph{i}) and (\emph{ii}) are called \emph{string operators}. Additionally, we may be interested in the properties:
\begin{enumerate}[label=(\emph{\roman*})] 
\setcounter{enumi}{2}
\item Squares to the identity. \label{property_3}
\item Hermitian. \label{property_4}
\end{enumerate}
If these are satisfied, they will be called \emph{canonical} string operators.

Properties  \ref{property_3} and \ref{property_4} are satisfied (see Lemmas \ref{lem:square_identity} and \ref{lem:hermiticity} in App. \ref{app:proof_thm_string_operators}) if 
\begin{equation} \label{eq:Hermicity_constraint}
F_{\mathcal P} \left( \vec i \oplus \vec{\alpha}^\mathcal{P} \right) = \left[ F_{\mathcal P} \left( \vec i \right) \right]^*.
\end{equation}

Since we want $S^{+}_{\mathcal P}$ to commute with all plaquettes in $\mathcal B_{\mathcal P}$ (for the rest of plaquettes, it commutes by construction), we impose that the commutator vanishes, $\left[S^{+}_{\mathcal P}, B_p\right]=0$, which yields the equation
\begin{equation} \label{eq:commutation_string}
F_{\mathcal P} \left( \vec i \oplus \vec{\alpha}^p \right) = \frac{ b_p \left( \vec i  \oplus \vec{\alpha}^{\mathcal P} \right) }{ b_p \left( \vec i \right) } F_{\mathcal P} \left( \vec i \right).
\end{equation}
It is useful to define the function $\theta_\mathcal{P} ( \vec i ,p ) := { b_p ( \vec i  \oplus \vec{\alpha}^{\mathcal P} ) }/{ b_p ( \vec i ) }$, which relates $F_{\mathcal P} ( \vec i \oplus \vec{\alpha}^p)$ to $F_{\mathcal P} ( \vec i )$. We can generalize Eq.\ \eqref{eq:commutation_string} and function $\theta_\mathcal{P} ( \vec i ,p )$ for an arbitrary number of plaquettes, namely,
\begin{equation} \label{eq:F_theta_F}
F_{\mathcal P} \left( \vec i \oplus \vec{\alpha}^{p_1}\oplus...\oplus \vec{\alpha}^{p_m} \right) = \theta_\mathcal{P} \left( \vec i ,p_1,...,p_m \right) F_{\mathcal P} \left( \vec i \right),
\end{equation}
where $\theta_\mathcal{P} ( \vec i ,p_1,...,p_m )$ can be expressed as
\begin{equation}\label{eq:def_theta}
\theta_{\mathcal P} \left( \vec i , p_1, \dots, p_m \right) =\prod_{i=1}^m \frac{b_{p_i} \left( \vec i \oplus \vec{\alpha}^{\mathcal P} \bigoplus\limits_{j=1}^{i-1} \vec{\alpha}^{p_j} \right)}{b_{p_i} \left( \vec i \bigoplus\limits_{j=1}^{i-1} \vec{\alpha}^{p_j} \right)}.
\end{equation}
 Note that while we use the same symbol for the configurations $\vec i$ in $\theta_{\mathcal P}$ and in $F_{\mathcal P}$, the one in $\theta_{\mathcal P}$ is over the whole system in order for Eq.~\eqref{eq:def_theta} to be well-defined. The specific way that the configuration $\vec i$ is extended over the whole system (\emph{i.e} which configuration on the rest of the system is appended to it) does not matter, since as a consequence of the structure of the plaquette operators, it does not affect the value of $\theta_{\mathcal P}$. As a consequence of the fact that plaquette operators commute, the order of the plaquettes $p_1,\ ...,\ p_m$ in $\theta_\mathcal{P} ( \vec i ,p_1,...,p_m )$ does not matter (see Lemma \ref{lem:theta_permutation}). The function $\theta_\mathcal{P} ( \vec i ,p_1,...,p_m )$ relates the value of $F_{\mathcal P}$ for configuration $\vec i$ to that of configuration $\vec i \oplus \vec{\alpha}^{p_1}\oplus...\oplus \vec{\alpha}^{p_m}$. These two configurations differ by a sum of plaquettes and may be considered part of the same configuration class $\mathcal{C}_\mathcal{P}\left( \vec i \right)$, defined as
\begin{equation}
\mathcal{C}_\mathcal{P}\left( \vec i \right) := \left\lbrace \vec j : \vec j = \vec i \bigoplus_{p \in \rm{subset}( \mathcal B_{\mathcal P})} \vec\alpha^p \right\rbrace,
\end{equation}
where the configurations are restricted to $\textnormal{Conn} (\mathcal P)$. These configurations can be regarded as the set of configurations related to $\vec i$ by adding loops associated with plaquettes in $\mathcal B_{\mathcal P}$, only in the region where $S^+_{\mathcal P}$ acts non-trivially. Taking all this into account, one can obtain an algorithm to compute the phases $F^{+}_{\mathcal P}$ of the ansatz in Eq.\ \eqref{eq:String_ansatz}. This is given by Algorithm \ref{alg:F_finding}.

\begin{algorithm}[h!]
\begin{algorithmic}
\FOR{every configuration class $\mathcal C_{\mathcal P} $ }
\STATE{Pick a class representative $\vec i \in \mathcal C_{\mathcal P}$}
\STATE{Set $F_{\mathcal P} (\vec i) = e^{i \phi(\vec i)}$ \;}
\FOR{every subset $(p_1, \dots, p_m) \subseteq \mathcal B_{\mathcal P}$}
   \STATE{$F_{\mathcal P} ( \vec i \bigoplus\limits_{i=1}^m \vec{\alpha}^{p_i} ) =  \theta_{\mathcal P} ( \vec i, p_1, \dots, p_m) F_{\mathcal P} ( \vec i ) $ \;}
\ENDFOR
\ENDFOR
\end{algorithmic}
\caption{Determination of the $F_{\mathcal P}$ functions for a path $\mathcal P$. $e^{i \phi(\vec i)}$ are initial phases that can take  any value.}
\label{alg:F_finding}
\end{algorithm}

Algorithm \ref{alg:F_finding} begins by picking up a configuration $\vec i$, which we call its class representative, and by setting its value to an arbitrary phase $e^{i \phi(\vec i)}$. Any phase picked up by the algorithm yields a valid string operator. Once the value for the class representative $\vec i$ is fixed, the algorithm assigns values to the rest of the configurations in the same configuration class by making use of Eq.~\eqref{eq:F_theta_F}. Afterwards, a configuration, belonging to a different configuration class, where the values have not yet been fixed, is chosen and the same procedure is repeated until $F_{\mathcal P}$ has been fixed for all possible configurations. An explicit example of this can be seen in App. \ref{app:example_string}. 

As it is shown in App.~\ref{app:proof_thm_string_operators}, it is always possible to determine $F_{\mathcal P}$ using Algorithm \ref{alg:F_finding} such that the resulting $S^{+}_{\mathcal P}$ is a string operator. Furthermore, it is also possible to enforce the constraint given in Eq.~\eqref{eq:Hermicity_constraint} such that we obtain canonical string operators. Those important results are summarized in the following theorem:

\begin{theorem} \label{thm:string_operators}
Let $\mathcal P$ be a path. Any function $F_{\mathcal P}$ defined by Algorithm \ref{alg:F_finding} is such that $S^{+}_{\mathcal P}$ is a string operator. Furthermore, it is possible to choose the phases $e^{i \phi}$ such that the string operator is canonical.
\end{theorem}
Note that the open-string operators generated by Algorithm\ \ref{alg:F_finding}, $S^+_\mathcal{P}$, have positive chirality, because they anticommute with vertex operators at the endpoints of $\mathcal{P}$ and commute with every other vertex and plaquette operators, satisfying property \ref{property_1}. However, the same does not apply to closed-string operators, since closed strings do not have endpoints. Algorithm \ref{alg:F_finding} produces, in general, closed-string operators without a definite chirality.

\subsubsection{Concatenation of open-string operators}
It is very useful to build strings as a concatenation of smaller strings. This is specially relevant for constructing non-trivial closed strings, since, as it was mentioned before,  Algorithm \ref{alg:F_finding} yields, in general, closed strings which have no definite chirality. By building these closed strings out of a multiplication of open strings, which have definite chirality, we obtain positive- and negative-chirality closed strings. Observe that given two paths, $\mathcal{P}_1$ and $\mathcal{P}_2$, meeting at one endpoint or forming a closed path, the multiplication of both $S^{+}_{\mathcal{P}_1 \# \mathcal{P}_2} = S^{+}_{\mathcal{P}_2}S^{+}_{\mathcal{P}_1}$ is a string operator satisfying properties \ref{property_1} and \ref{property_2}.  In this way we can build long string operators by concatenating short ones.
\subsection{Crossing string operators}

In order to understand the algebra of the string  operators of the semion code that we build in Sec.\ \ref{subsec:logical_operators}, it is essential to know the commutation relations between them acting on different paths.

\begin{figure}
\center
    \subfloat[\label{fig:self_crossing}]{
    \includegraphics[scale=0.195]{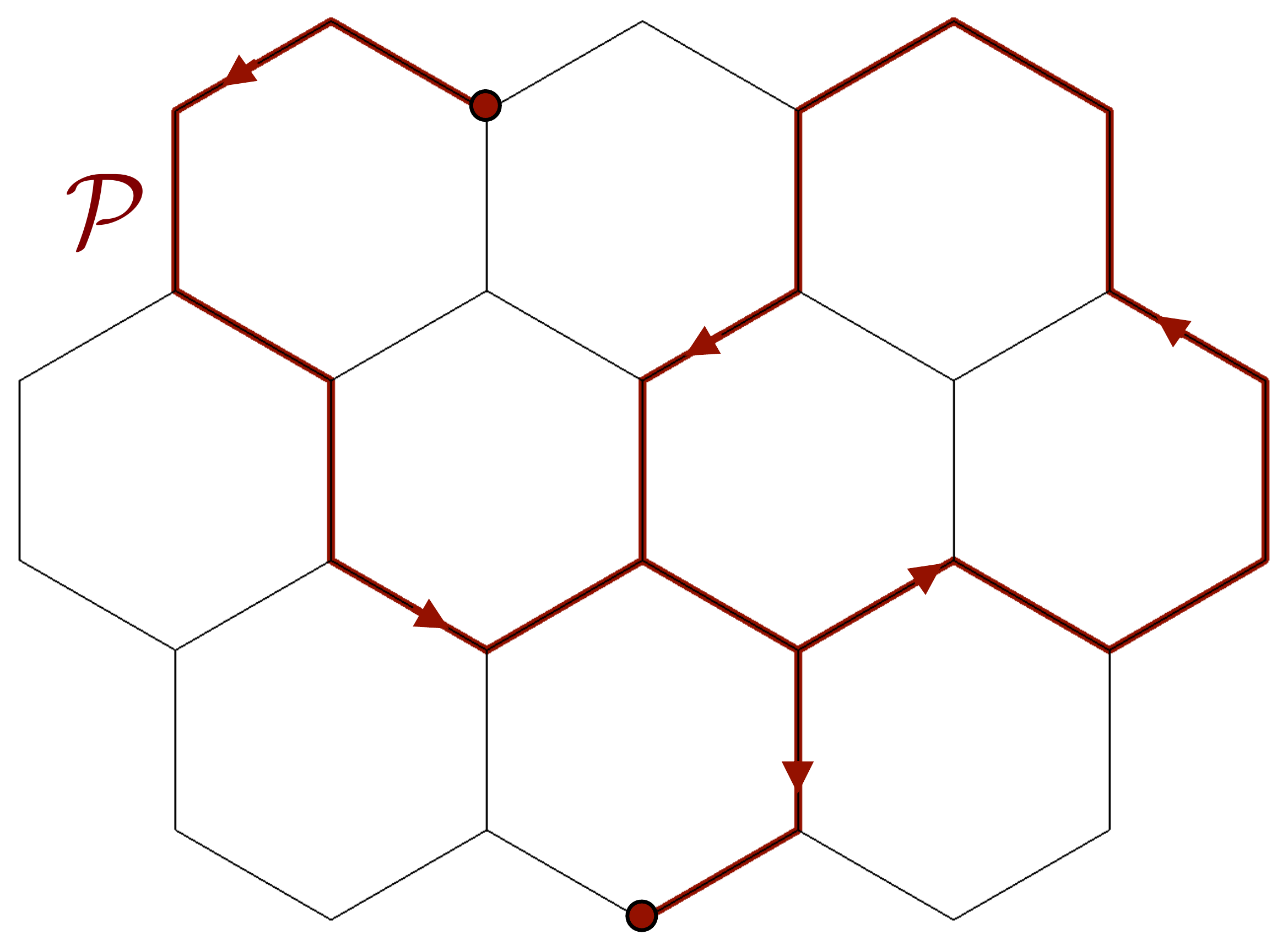}
    }    
    \subfloat[\label{fig:self_overlapping}]{
    \includegraphics[scale=0.195]{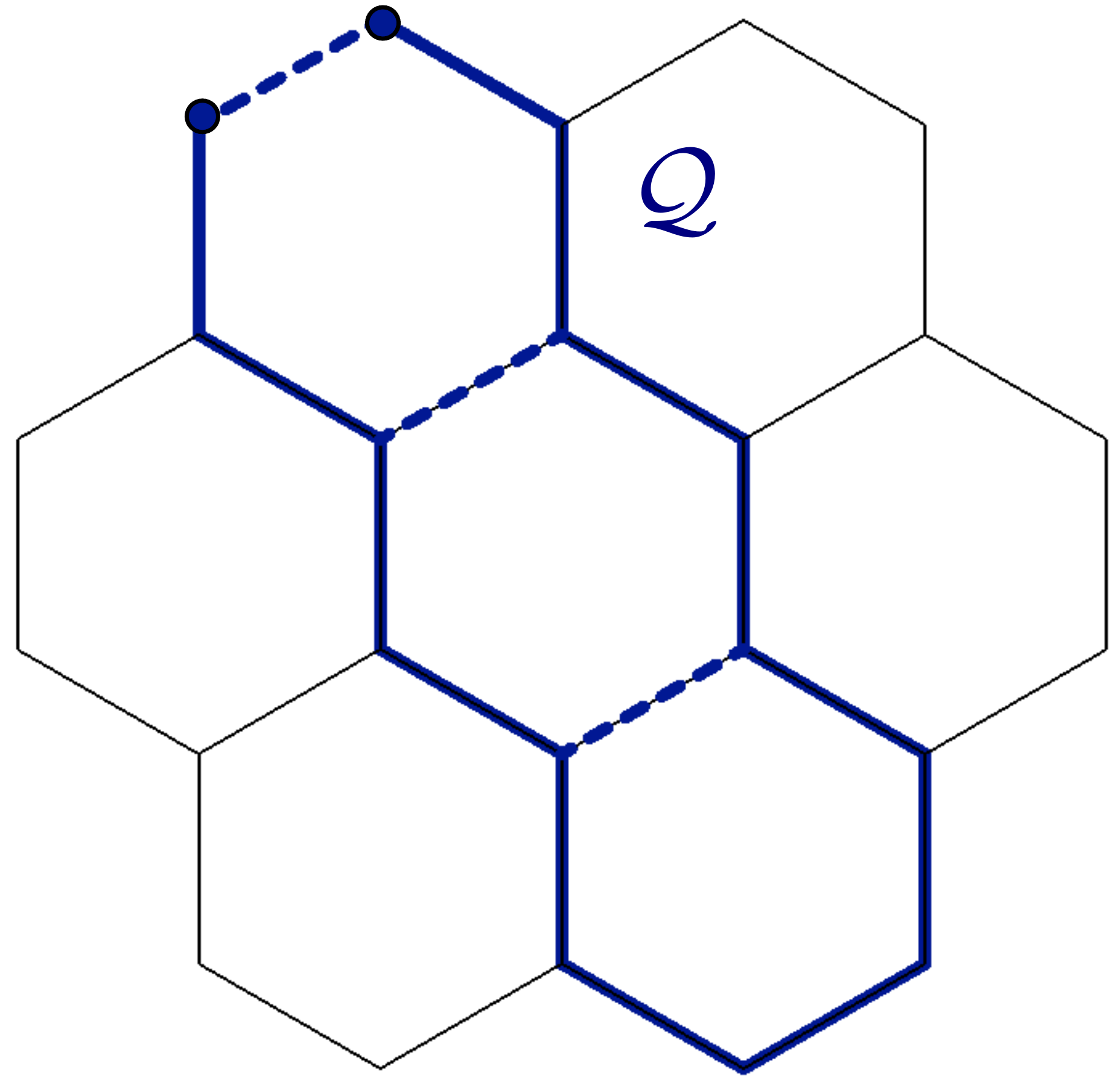}
    }
    \caption{Examples of (a) a self-crossing path $\mathcal P$ and of (b) a self-overlapping path $\mathcal Q$. The dashed blue edges indicate the links where the string \cal Q self-overlaps, since some distant parts of the path are connected.}
    \label{fig:self_cross_overlap}
\end{figure}

The notion of crossing paths need to be precisely defined since the region on which the string operators act non-trivially, $\textnormal{Conn} ( \mathcal P )$, has a finite thickness. Heuristically, in order to consider that two paths are crossing, the commun edges to both paths must not contain the first nor last vertex of neither of the paths. Note that two paths can cross more than once.

We further need to define the notions of \emph{self-crossing} and \emph{self-overlapping} paths. Essentially, a path is self-crossing if an observer moving on the path passes more than once on any given edge. A path is said to be \emph{self-overlapping} if some regions of the support of the string operator but not the path itself overlap and connect some distant parts of the paths.
Fig.~\ref{fig:self_cross_overlap} illustrates the previous concepts. We refer the reader to App.\  \ref{app:openstrings_commutation} for rigorous definitions, as well as the proof of Theorem~\ref{thm:crossing_strings}. This theorem summarizes the commutation relations among the string operators.

\begin{theorem} \label{thm:crossing_strings}
Let $\mathcal P$ and $\mathcal Q$ be two paths crossing $n$ times, composed of non self-overlapping nor self-crossing individual open paths, i.e., $\mathcal P = \mathcal P_1 \# \dots \# \mathcal P_m$ and $\mathcal Q = \mathcal Q_1 \# \dots \# \mathcal Q_{m'}$ for some integers $m$ and $m'$. We have that
\begin{eqnarray}
\begin{aligned}
\lbrack S^+_{\mathcal P}, S^+_{\mathcal Q} \rbrack &= 0 \textnormal{ if } n \textnormal{ is even}, \\
\{ S^+_{\mathcal P}, S^+_{\mathcal Q} \} &= 0 \textnormal{ if } n \textnormal{ is odd}.
\end{aligned}
\end{eqnarray}
\end{theorem}

Given the definition of negative chirality string operators, and using the above results, we find that $\lbrack S^-_{\mathcal P}, S^-_{\mathcal Q} \rbrack = 0$ if $n$ is even, while $\{ S^-_{\mathcal P}, S^-_{\mathcal Q} \} =  0$ if $n$ is odd. We also find $\lbrack S^{\pm}_{\mathcal P}, S^{\mp}_{\mathcal Q} \rbrack = 0$ for any $n$.

Interpreting a vertex (in the case of $S^+_{\mathcal P}$) or the combination of a vertex and plaquette excitations (in the case of $S^-_{\mathcal P}$) as the presence of a quasiparticle labeled by $s^+$ and $s^-$ respectively, the topological $S$ matrix \cite{kitaev05}, written in the basis $(1, s^+, s^-, s^+s^-)$ where $1$ denotes the vacuum, i.e. the absence of excitation, and $s^+s^-$ the composite object excitation, is found to be
\begin{equation}
S = \frac{1}{2} \left( \begin{matrix}
1 & 1 & 1 & 1 \\
1 & -1 & 1 & -1 \\
1 & 1 & -1 & -1 \\
1 & -1 & -1 & 1
\end{matrix} \right).
\end{equation}
We can thus interpret the string operators $S^+_{\mathcal P}$ and $S^-_{\mathcal P}$ as creating pairs of semions of different chirality at their endpoints.

\subsubsection{The need for path concatenation}
Notice that Theorem~\ref{thm:crossing_strings} does not state anything about closed paths (homologically trivial or not) which are composed of a single path. One can check that when such a path crosses another one, in general they do not commute nor anti-commute. Such paths thus cannot be considered as `fundamental' string operators in the sense that they do not possess a definite chirality.

Algorithm~\ref{alg:F_finding} enforces that the $\sum_{\vec i} F_{\mathcal P} (\vec i ) \vert \vec i \rangle \langle \vec i \vert$ operator does not contain any open $S^z_{\mathcal P'_{\textnormal{dual}}}$ operator for an open path, since by construction, Algortihm~\ref{alg:F_finding} builds a string operator which commutes with every plaquette operator. For a closed path however, one can add $S^z_{\mathcal P'_{\textnormal{dual}}}$ to $\sum_{\vec i} F_{\mathcal P} (\vec i ) \vert \vec i \rangle \langle \vec i \vert \rightarrow \left( \sum_{\vec i} F_{\mathcal P} (\vec i ) \vert \vec i \rangle \langle \vec i \vert \right) S^z_{\mathcal P'_{\textnormal{dual}}}$, which is also a valid output of Algorithm~\ref{alg:F_finding}. In fact, such a $S^z_{\mathcal P'_{\textnormal{dual}}}$ operator can be added selectively to only a subset of configuration classes, causing a `mixing' of the chiralities. This is explained in detail in App. \ref{Sec:app_initial_phases}.

By producing closed paths starting from smaller open paths as their basic constituents, one can enforce the production of strings of a definite chirality. This is caused by the fact that the individual components cannot carry flux excitations by construction, and so neither can their concatenation. The physical intuition is that each small open string operator creates a pair of semions of positive chirality at their endpoints. Since the created semions are their own anti-particles, they subsequently all fuse to the vacuum, returning the system to the ground space.

\subsection{Completeness of the string operators}\label{sec:completeness_strings}
In this section, we seek to decompose a string of $\sigma^x$ operators into the strings operators defined in our model, $S^{+}$ and $S^z$. 

We first note that any matrix $\rho$ of size $2^n \times 2^n$ can be written as a linear combination of Pauli operators, i.e.,
\begin{equation}
\rho = \sum_{P_x, P_z} c (P_x, P_z) P_x P_z,
\end{equation}
where $P_x (P_z)$ are Pauli operators acting on $n$ qubits and formed of products of identities and $\sigma^x$ $(\sigma^z)$ operators only and where $c (P_x, P_z)$ are non-zero complex numbers. Given the matrix $\rho$, one can recover the coefficients $c(P_x, P_z)$ using the formula
\begin{equation} \label{eq:Pauli_decomp}
c(P_x, P_z) = \frac{1}{2^n} \textnormal{Tr} \left( P_x P_z \rho \right).
\end{equation}

In our case, a chain of $\sigma^x$ operators on path $\mathcal P$, denoted by $X_{\mathcal P}$, can be written as
\begin{equation} \label{eqn:Pauli_form1}
X_{\mathcal P} = S^+_{\mathcal P} \times \sum_{\vec i} \lbrack F_{\mathcal P} ( \vec i)  \rbrack^{\ast} \vert \vec i \rangle \langle \vec i \vert.
\end{equation}
Given the form of $ \sum_{\vec i} \lbrack F_{\mathcal P} ( \vec i)  \rbrack^{\ast} \vert \vec i \rangle \langle \vec i \vert$, we can write
\begin{equation} \label{eqn:Pauli_form2}
X_{\mathcal P} = S^+_{\mathcal P} \times \sum_{P_z \in {\textnormal{Conn} (\mathcal P)}} c ( P_z ) P_z,
\end{equation}
where $P_z$ are Pauli operators containing only identities and $\sigma^z$ and where we write $P_z \in {\textnormal{Conn} (\mathcal P)}$ in an abuse of notation to signify that $P_z$ acts non-trivially only on the qubits in $\textnormal{Conn} (\mathcal P)$.
The coefficients $c(P_z)$ are given by
\begin{equation} \label{eq:string_decomp}
c (P_z) = \frac{1}{2^{\vert \textnormal{Conn} (\mathcal P) \vert} } \sum_{\vec i} \langle \vec i \vert P_z \vert \vec i \rangle \lbrack F_{\mathcal P} (\vec i) \rbrack^{\ast},
\end{equation}
where $\vert \textnormal{Conn} ( \mathcal P ) \vert$ is the number of edges in $ \textnormal{Conn} ( \mathcal P )$. Chains of $\sigma^z$ operators form valid string operators $S^z$, which create flux excitations at their endpoints.  This clearly shows that any Pauli operator acting on the system can be written in terms of string operators. Since any operator acting on the system can be decomposed as a linear combination of Pauli operators, we find that the string operators are complete in the sense that any operator can be expressed in terms of them.

\section{The semion code}\label{sec:semion_code}
The tools we have developed in previous sections can be used to build a quantum error correction code using  as code space the ground space of the off-shell DS model, given by Hamiltonian \eqref{hamDS}. The information encoded is topologically protected since we are using global degrees of freedom that cannot be affected by local errors. Additionally, we perform quantum error correction using the DS model as a stabilizer code, where plaquette and vertex operators are our stabilizers.
\subsection{Logical operators} \label{subsec:logical_operators}
\begin{figure}
\begin{center}
\includegraphics[width=0.7\linewidth]{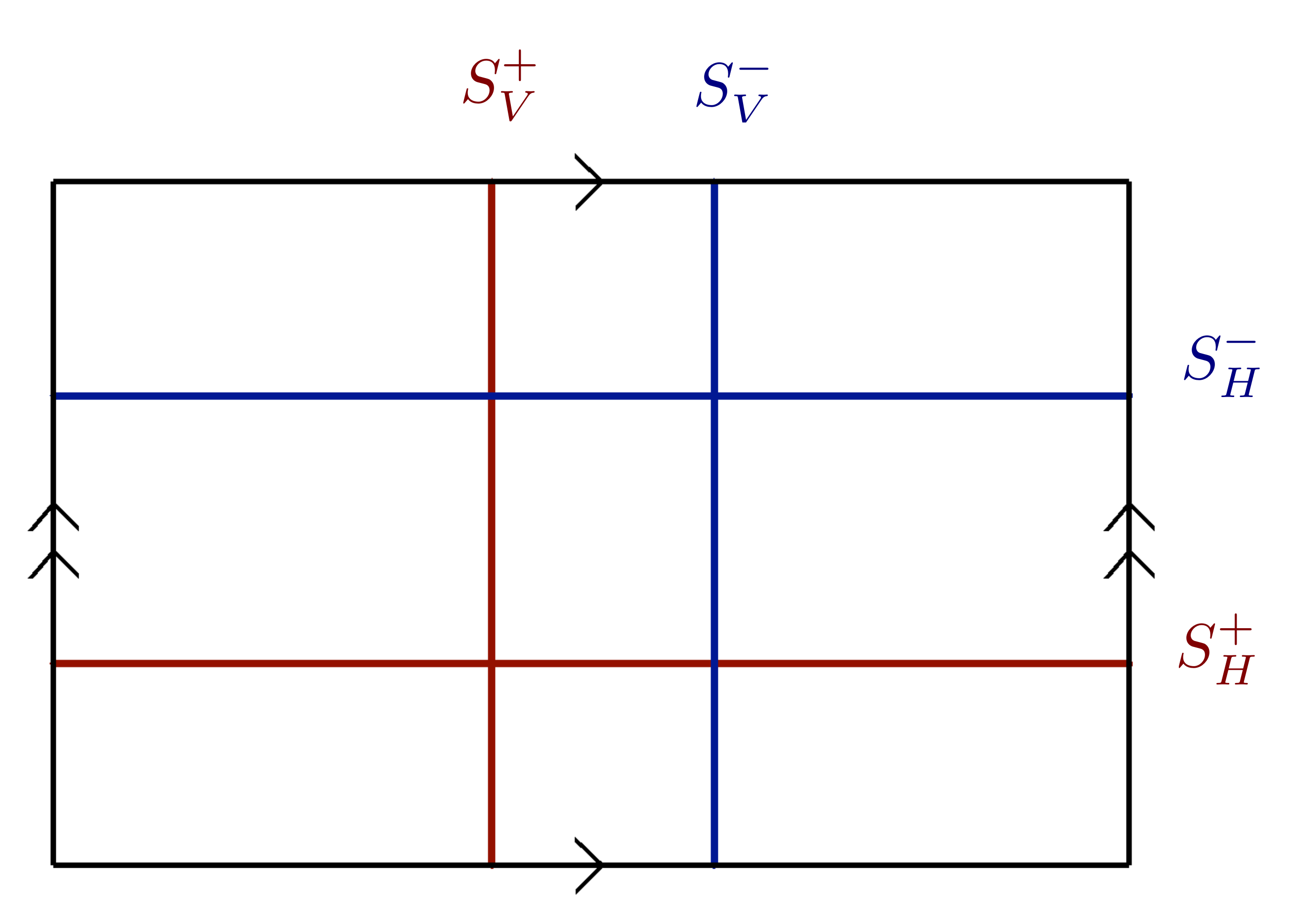}
\caption{\label{fig:example_logicals} A cartoon example of $2$ sets of logical operators on a $1$-torus. $\{S^+_V ,S^+_H\}=\{S^-_V, S^-_H\}=0$, while $[S^+_V, S^-_H]=[S^-_V, S^+_H]=0 $. Arrows indicate identified boundaries. }
\label{fig:logicals}
\end{center}
\end{figure}

 Recalling that a surface of genus $g$ can be seen as the connected sum of $g$ tori~\cite{Nakahara}, we can define two pairs of anti-commutating logical operators for every torus in the connected sum \cite{doi:10.1063/1.2731356}. One pair consists of string operators $S^+_{V}$ and $S^+_{H}$, with $V$($H$) any homologically non-trivial path along the vertical (horizontal) direction,while the other pair consists of $S^-_{V}$ and $S^-_{H}$, Both pairs are made up of open non self-crossing nor self-overlapping individual paths, as prescribed by Theorem~\ref{thm:crossing_strings}. Fig.~\ref{fig:logicals} illustrates two such pairs for a genus $1$ torus.

\subsection{Quantum error correction}

 The stabilizer operators, vertices and plaquette defined in Eqs.\ \eqref{vertex_operators} and \eqref{eq:Bp_generalized}, can be periodically measured to detect any errors occurring in the system. Once the syndrome pattern is obtained, assuming a given noise model, it is fed into a decoder. It outputs a recovery operation using the string operators developed in this work in order to bring the system back to the encoded subspace, where the probability of applying a non-trivial logical operation is minimized. While we leave the development of decoders specifically designed for the semion code for future work, one could imagine adapting some of the various existing decoders developed for topological codes  \cite{dennis_etal02,Wang:2010:TER:2011362.2011368,PhysRevLett.104.050504,PhysRevLett.111.200501,PhysRevLett.108.180501,1367-2630-16-6-063038,delfosse,delfosse2,maskara,kubica,herold,Wootton_2015,PhysRevA.85.022317,harrington,2058-9565-3-4-044002,1367-2630-19-6-063012,Dauphinais2017,Sweke2018,Breuckmann2017,Breuckmann2018scalableneural,PhysRevLett.109.160503,PhysRevA.89.022326,PhysRevA.90.032326,PhysRevE.97.051302}.
 
\begin{figure}
    \subfloat[\label{fig:orientationa}]{
    \includegraphics[scale=0.175]{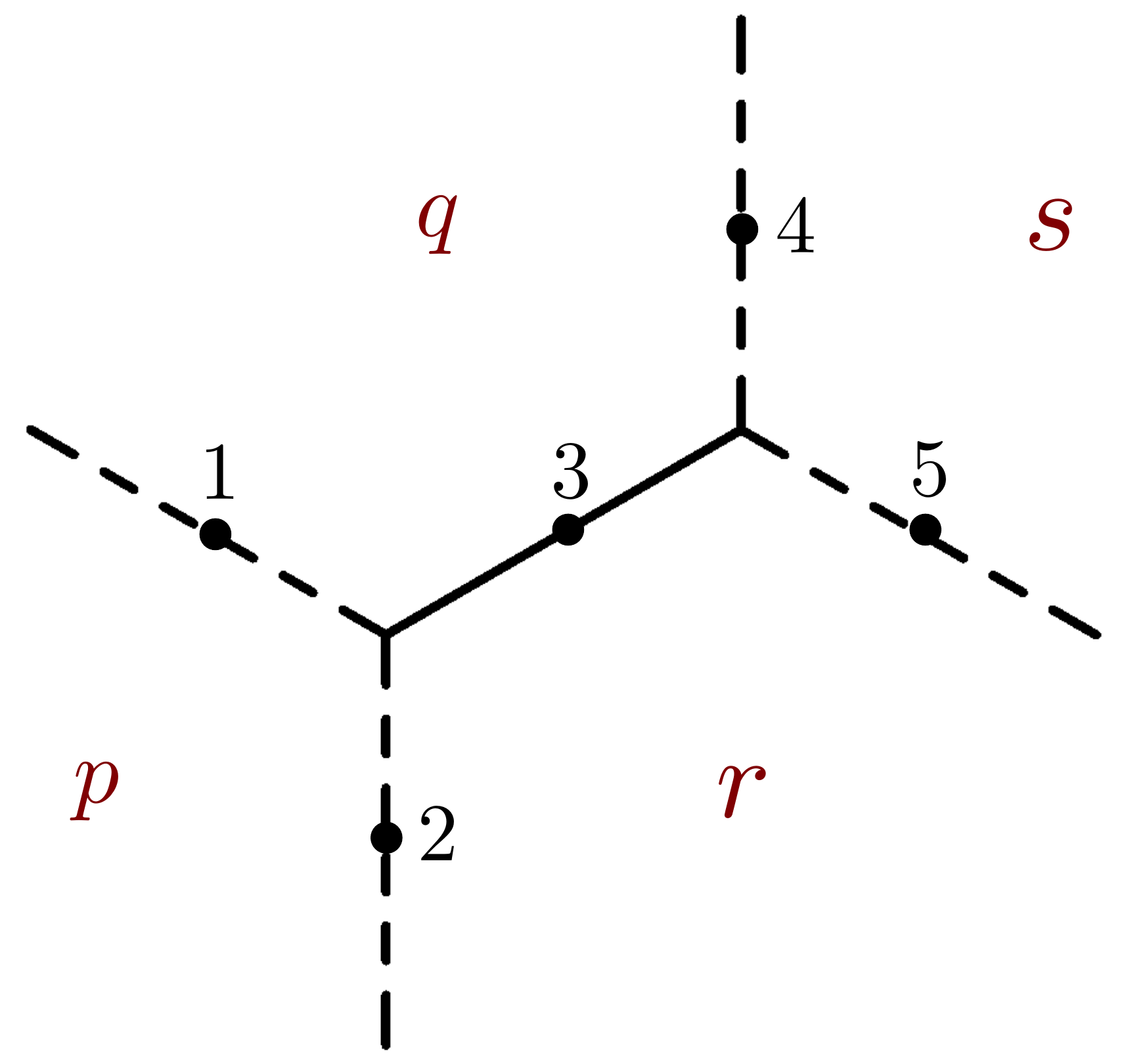}
    }    
    \subfloat[\label{fig:orientationb}]{
    \includegraphics[scale=0.175]{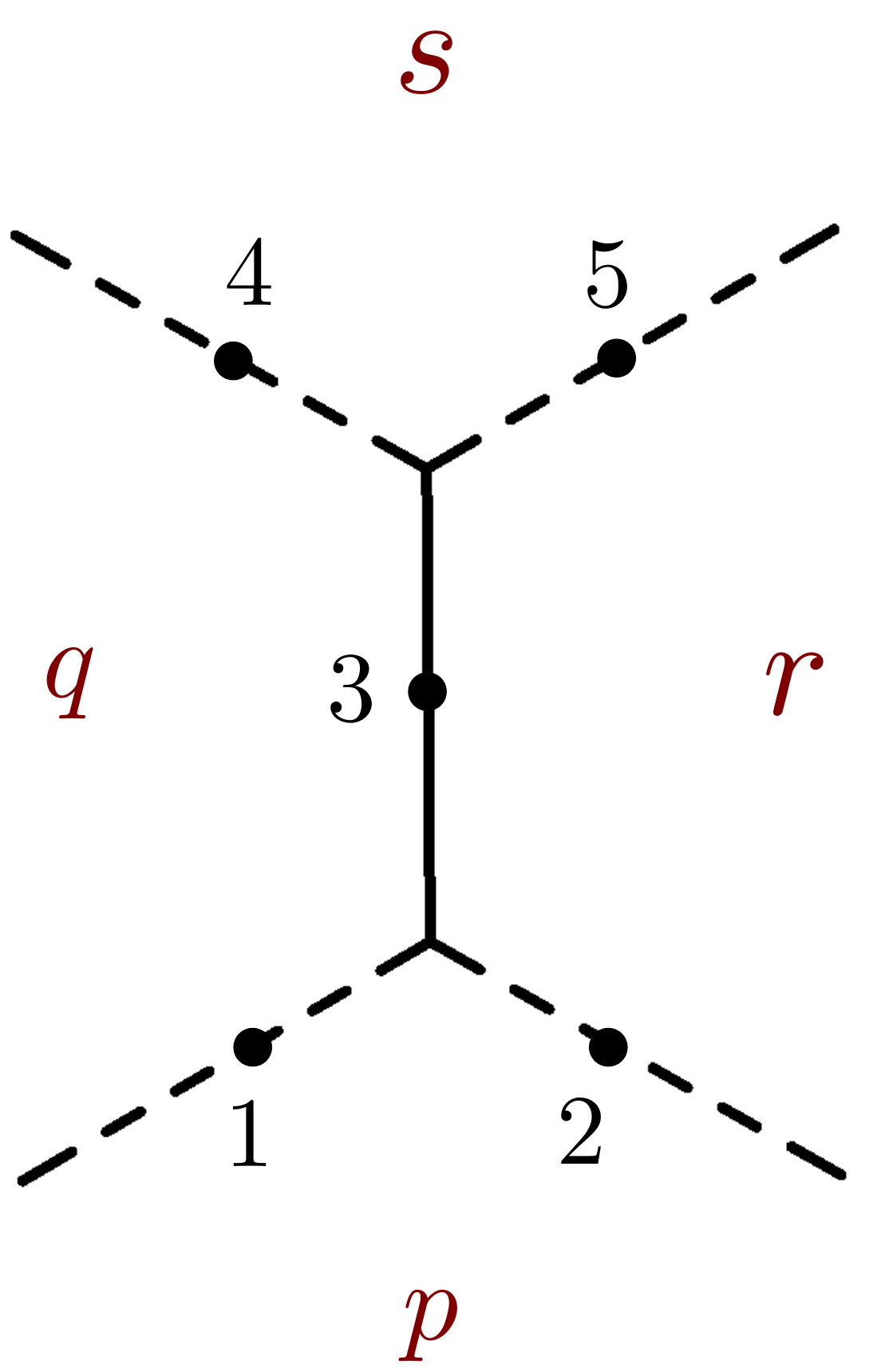}
    }
    \subfloat[\label{fig:orientationc}]{
    \includegraphics[scale=0.175]{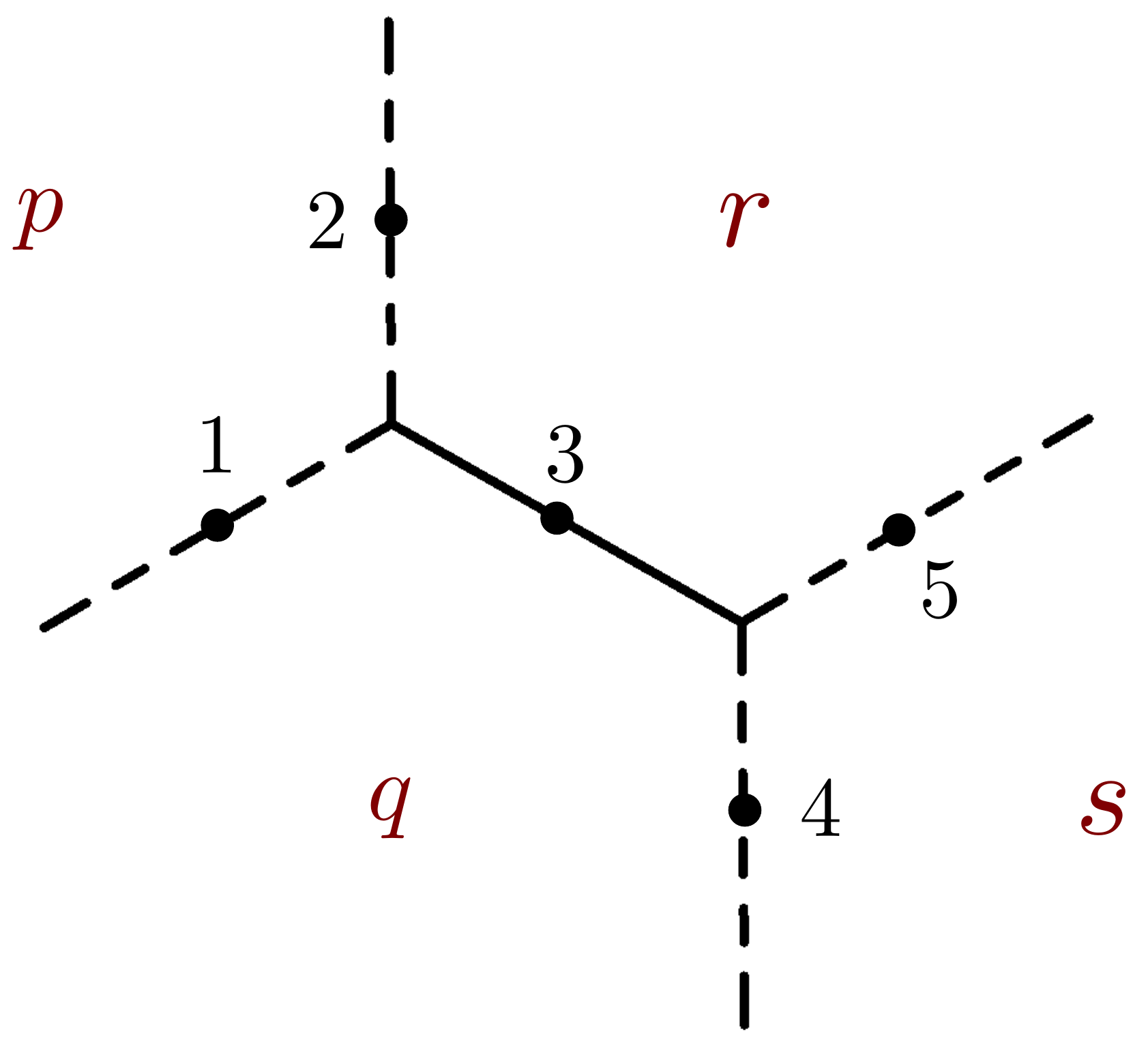}
    }
    \caption{The three possible edge orientations on which the $\sigma^x$ operator can be applied. Qubit $3$ is affected in all cases, and may leave flux excitations on the four surrounding plaquettes labeled by $p$, $q$, $r$ and $s$. The probabilities of measuring a given flux pattern are given in Tab.~\ref{tab:sigmaX_effect}.}
    \label{fig:sigmaX_effect} 
\end{figure}

Tab.~\ref{tab:sigmaX_effect} shows the probabilities of measuring a given flux configurations after applying a single $\sigma^x$ on the ground state for the three possible edge orientations shown in Fig.~\ref{fig:sigmaX_effect}. Note that as Eqs.\ \eqref{eqn:Pauli_form1} and~\eqref{eqn:Pauli_form2} suggest, the probabilities in Tab.\ \ref{tab:sigmaX_effect} do not depend of the phases used to initialize the $F_{\{ e \}}$ function in Algorithm \ref{alg:F_finding}. More details giving a deeper understanding on the structure of the string operators $S^+_{\mathcal P}$ can be found in App.\ \ref{app:completeness_strings}. 

A distinctive feature of the probability distributions in Tab.\ \ref{tab:sigmaX_effect} is that there is a directionality in the error pattern. A $\sigma^x$ error affecting a vertical edge (orientation (b)) is much more likely to leave flux excitations behind than for the other two orientations. This is clearly due to the specific structure of the plaquette operators, and could be used advantageously when dealing with asymmetric noise \cite{PhysRevX.2.021004,PhysRevLett.120.050505}.
Another major difference with the toric code is the fact that chains of $\sigma^x$ errors are likely to leave flux excitations along their path. This additional information could be used by the decoder and may lead to a higher threshold value.

\begin{table}
\centering
\begin{tabular}{| c | c | c | c |}
\cline{2-4}
\multicolumn{1}{c|}{}
& \multicolumn{3}{c |}{Probability} \\ \hline
$( b_p, b_q, b_r, b_s )$ & \footnotesize{Orientation} (a) & \footnotesize{Orientation} (b) & \footnotesize{Orientation} (c) \\ \hline
$\left( 0,0,0,0 \right)$ & $ 9/16 $ & $1/16$ & $9/16$ \\ \hline
$\left( 1,1,0,0 \right)$ & $ 1/16 $ & $ 1/16$ & $ 1/16$ \\ \hline 
$\left( 1,0,1,0 \right)$ & $1/16$ & $ 1/16$ & $ 1/16$ \\ \hline
$\left( 0,1,1,0 \right)$ & $1/16$ & $9/16$ & $ 1/16$ \\ \hline
$\left( 1,0,0,1 \right)$ & $1/16$ & $ 1/16$ & $ 1/16$ \\ \hline
$\left( 0,1,0,1 \right)$ & $1/16$ & $ 1/16$ & $ 1/16$ \\ \hline
$\left( 0,0,1,1 \right)$ & $1/16$ & $ 1/16$ & $ 1/16$ \\ \hline
$\left( 1,1,1,1 \right)$ & $1/16$ & $ 1/16$ & $ 1/16$ \\ \hline
\end{tabular}
\caption{The various probabilities of getting a given flux excitation configuration after the application of the operator $\sigma^x$ on a qubit, for the three possible orientations. The plaquettes label correspond to the ones in Fig.~\ref{fig:sigmaX_effect}.}
\label{tab:sigmaX_effect}
\end{table}

\section{Conclusions and outlook}\label{sec:conclusions}

One of the key features of the off-­shell DS model developed here for error correction is that it is a non­-CSS code \cite{PhysRevLett.77.198,PhysRevLett.77.3260}. This is not novel in the theory of quantum error correcting codes. In fact, the answer to the important question of what is the minimal  complete error correction code that is able to encode one logical qubit and correct for an arbitrary error was precisely a non-­CSS code of five qubits \cite{PhysRevLett.77.198,PhysRevLett.77.3260}. This is consistent with the quantum Hamming bound \cite{gottesman96}, and is in sharp contrast with the classical case where the solution is the repetition code of three bits. However, what is peculiar of the off­-shell DS code is that, to our knowledge, this is the first non-CSS topological quantum memory that being a stabilizer code, it is also a topological code throughout the whole Hilbert space. In a sense, this was a missing link in the theory of topological quantum error correction codes and we have filled this gap with the tools introduced in our work.

We notice that a previous study \cite{doi:10.1063/1.4920923} attempted to construct a quantum error correction code using the DS model as the starting point. The main difference with our work is that they construct a non­-commuting quantum correcting code, whereas we have succeeded in constructing an extension that belongs to the stabilizer formalism. As a consequence of this, the whole error correction procedure of the off­-shell DS code is topological. On the contrary, the topological nature of the non­-commuting code in \cite{doi:10.1063/1.4920923} is unproven. Both constructions share the feature of using non-­Pauli operators to construct the basic string operators of the model.

The outcome of our work is a complete characterization of the error correction procedure for a quantum memory based on a topological non-­CSS stabilizer code. This is a major step for the reason explained above. However, a fully­-fledged quantum computer will demand more, namely, a universal gate set and a fault­-tolerant procedure to battle errors dynamically \cite{preskill97}. With the tools deployed here, it is conceivable that this goal will be achieved elsewhere.

 A new way of constructing quantum codes opens up with this work. The tools introduced here for  models like the DS  based on Abelian lattice gauge theories can be generalized to other Levin-�Wen models \cite{levinwen,PhysRevB.86.115109,ORTIZ2016193}, like doubled Fibonacci models, or twisted versions of fracton models \cite{shong_delgado}. This is the subject of further study.

\begin{acknowledgements}
We thank Fiona Burnell and Juan Miguel Nieto for helpful discussions.
We acknowledge financial support from the Spanish MINECO grants FIS2012-33152, FIS2015-67411, and the CAM research consortium QUITEMAD+, Grant No. S2013/ICE-2801. The research of M.A.M.-D. has been supported in part by the U.S. Army Research Office through Grant No. W911N F-14-1-0103. S.V. thanks FPU MECD Grant. 
\end{acknowledgements}

\appendix

\section{Example of an open-string operator} \label{app:example_string}

\begin{figure}
    \centering
    \includegraphics[width=0.6\linewidth]{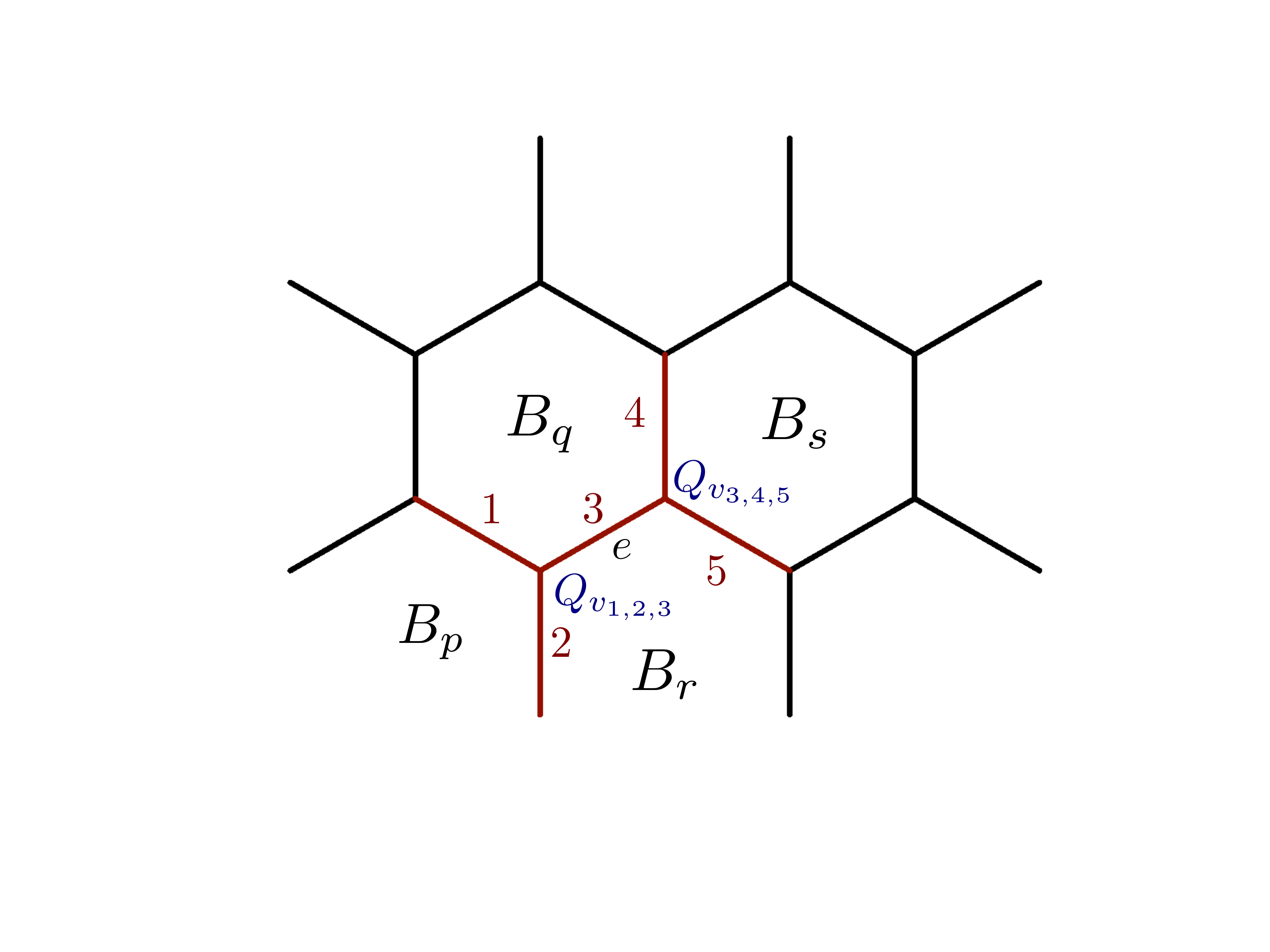}
       \caption{The layout we consider to find a possible string operator $S^{+}_{e}$ creating a pair of vertex excitations at $v_{1,2,3}$ and $v_{3,4,5}$.  The red edges correspond to the qubits in $\textnormal{Conn} (e)$, the ones on which $S^+_e$ acts non-trivially, with edge $e$ corresponding to qubit $3$. $S^+_{e}$ anti commutes with both vertex operators $Q_{v_{1,2,3}} = \sigma^z_1 \sigma^z_2 \sigma^z_3$ and $Q_{v_{3,4,5}} = \sigma^z_3 \sigma^z_4 \sigma^z_5$, while commuting with every other vertex operators as well as every generalized plaquette operators.}
\label{fig:E_1} 
\end{figure}

It is instructive to illustrate the workings of Algorithm \ref{alg:F_finding} to find string operators $S^{+}_{\mathcal P}$ in order to gain a more intuitive understanding. Consider the very simple path $\mathcal P = \{ e \}$, consisting only of edge $e$ shown in Fig.\ \ref{fig:E_1}. Any given configuration on the five edges included in $\textnormal{Conn} (\{ e \})$ is represented by a bit string of length $5$, for which a $0$  indicates the absence of a string, while a $1$ indicates that it is occupied. One can also interpret the bit string as a state in the computational basis, a $0$ indicating a $+1$ eigenstate of the corresponding $\sigma^z$, while a $1$ indicates a $-1$ eigenstate of $\sigma^z$. We compute the function $F^+_{\{ e\} }$ so that $S^+_{\{ e \}}$ is a canonical string operator, i.e., it also fulfills Eq.\ \eqref{eq:Hermicity_constraint}. Following Algorithm \ref{alg:F_finding}, the configuration $(0,0,0,0,0)$ is first chosen and we set $F^+_{\{ e \} } (0,0,0,0,0) = 1$. Noting that $\mathcal B_{\{ e \}}$ contains the $4$ plaquettes identified as $p, q, r$ and $s$ in Fig. \ref{fig:E_1}, we find the following values for $F^+_{\{ e \}}$:
\begin{eqnarray*} \label{eq:F_exemple_1}
\begin{aligned}
& F_{\{ e \}} (0,0,0,0,0) = 1, && F_{\{e \}} (1,1,0,0,0) = -i, \\ & F_{\{ e \} } (1,0,1,1,0) = -i, && F_{\{ e \}} (0,0,0,1,1) = -i, \\
& F_{\{ e \}} (0,0,1,0,0) = 1, && F_{\{ e \}}(1,1,1,0,0) = i , \\& F_{\{ e \}} (1,0,0,1,0) = i , && F_{\{ e \}} (0,0,1,1,1) = i, \\
& F_{\{ e \}} (0,1,1,0,1) = -i, && F_{\{ e \}} (1,1,1,1,1) = -1, \\ & F_{\{ e \}} (0,1,1,1,0) = -i , && F_{\{ e \}} (1,0,0,0,1) = i, \\
& F_{\{ e \}} (0,1,0,0,1) = i, && F_{\{ e \}} (1,1,0,1,1) = -1, \\& F_{\{ e \}} (0,1,0,1,0) = i, && F_{\{ e \}} (1,0,1,0,1) = -i.
\end{aligned}
\end{eqnarray*}
Here we are not only computing the values for configuration class $\mathcal{C}_{\{ e \}}(\vec 0)$, but also for configuration class $\mathcal{C}_{\{ e \}}(\vec{\alpha}^{\{ e \}})$, since these two are related by Eq.\ \eqref{eq:Hermicity_constraint}and therefore Algorithm\ \ref{alg:F_finding} makes the assignment $F_{\{ e \}} (i_1, i_2, i_3 \oplus 1, i_4, i_5) = \lbrack F_{\{ e \}} (i_1, i_2, i_3, i_4, i_5) \rbrack^{\ast}$.

Choosing next the configuration $(0,0,0,0,1)$ and setting $F_{\{ e \}} (0,0,0,0,1) = 1$, we can fix the following values of $F_{\{ e \}}$ :

\begin{eqnarray*} \label{eq:F_exemple_2}
\begin{aligned}
& F_{\{ e \}} (0,0,0,0,1) = 1, && F_{\{ e \}} (0,0,0,1,0) = 1, \\& F_{\{ e \}} (1,1,0,0,1) = -i, && F_{\{ e \}} (1,1,0,1,0) = -i, \\
& F_{\{ e \}} (0,0,1,0,1) = 1, && F_{\{ e \}} (0,0,1,1,0) = 1, \\& F_{\{ e \}} (1,1,1,0,1) = i, && F_{\{ e \}} (1,1,1,1,0) = i, \\
& F_{\{ e \}} (1,0,1,1,1) = -1, && F_{\{ e \}} (0,1,1,1,1) = -1, \\& F_{\{ e \}} (0,1,1,0,0) = i, && F_{\{ e \}} (1,0,1,0,0) = i, \\
& F_{\{ e \}} (1,0,0,1,1) = -1, && F_{\{ e \}} (0,1,0,1,1) = -1, \\& F_{\{ e \}} (0,1,0,0,0) = -i, && F_{\{ e \}} (1,0,0,0,0) = -i.
\end{aligned}
\end{eqnarray*}
Again, two class of configurations are fixed due to the fact that the string is canonical.
We keep doing this till all configuration classes have been fixed. As a result, we obtain  $S^+_{\{ e \}}$, which  commutes with the four neighbouring plaquette operators $B_p, B_q, B_r$ and $B_s$ shown in Fig. \ref{fig:E_1}, as well as all the other plaquette operators which are farther away.

\section{Proofs regarding string operators produced by Algorithm 1} \label{app:proof_thm_string_operators}

In order to prove Theorem \ref{thm:string_operators}, we begin by stating several technical results in the following section.

\subsection{Useful technical lemmas} \label{appsub:useful_lemmas}

\begin{lemma} \label{lem:theta_permutation}
Let $\{ p_1, p_2, \dots, p_n \}$ an ordered set of plaquettes and let $\{ q_1, q_2, \dots, q_n \}$ be a permutation of it. For any configuration $\vec i $, we have that $\theta_{\mathcal P} \left( \vec i , p_1, p_2, \dots, p_n \right) = \theta_{\mathcal P} \left( \vec i , q_1, q_2, \dots, q_n \right)$.
\begin{proof}
According to Eq.\ \eqref{eq:def_theta}, we have that
\begin{eqnarray}
\begin{aligned}
&\theta_{\mathcal P} (\vec i, p_1, \dots, p_m)  =  \frac{\langle \vec i \bigoplus\limits_{i=1}^m \vec{\alpha}^{p_i} \oplus \vec{\alpha}^{\mathcal P} \vert B_{p_m} \dots B_{p_1} \vert \vec i \oplus \vec{\alpha}^{\mathcal P} \rangle }{\langle \vec i \bigoplus\limits_{i=1}^m \vec{\alpha}^{p_i} \vert B_{p_m} \dots B_{p_1} \vert \vec i \rangle } \\
& =  \frac{ \langle \vec i \bigoplus\limits_{i=1}^{m} \vec{\alpha}^{q_i} \oplus \vec{\alpha}^{\mathcal P} \vert B_{q (p_m)} \dots B_{q (p_1)} \vert \vec i \oplus \vec{\alpha}^{\mathcal P} \rangle }{\langle \vec i \bigoplus\limits_{i=1}^m \vec{\alpha}^{q_i} \vert B_{q(p_m)} \dots B_{q(p_1)} \vert \vec i \rangle} \\
& = \theta_{\mathcal P} (q_1, \dots, q_m),
\end{aligned}
\end{eqnarray}
where $q (p_i)$ denotes the permutation of plaquettes that exchange $\{ p_1, \dots, p_m \}$ to $\{ q_1, \dots, q_m \}$, and where we used the fact that the plaquette operators all commute.
\end{proof}
\end{lemma}

\begin{lemma} \label{lem:theta_product}
Let $\{ p_1, p_2, \dots, p_m \}$ and $\{ q_1, q_2, \dots, q_k \}$ be two different set of plaquettes in $\mathcal B_{\mathcal P}$ such that $\bigoplus\limits_{i=1}^m \vec{\alpha}^{p_i} = \bigoplus\limits_{i=1}^{k} \vec{\alpha}^{q_i}$ on the string configuration of  $\;\textnormal{Conn} (\mathcal P)$. Then, for any configuration $\vec i$, we find that
\begin{equation} \label{eq:theta_product2}
\theta_{\mathcal P} \left( \vec i, p_1, p_2, \dots, p_m \right) = \theta_{\mathcal P} \left( \vec i, q_1, q_2, \dots, q_k \right).
\end{equation}
\begin{proof}
First notice that given the structure of the plaquette operators, $\bigoplus\limits_{i=1}^{m} \vec{\alpha}^{p_i} = \bigoplus\limits_{i=1}^{k} \vec{\alpha}^{q_i}$ on $\textnormal{Conn} (\mathcal P)$ implies that $\prod\limits_{i=1}^{m} B_{p_i} \prod\limits_{j=1}^{k} B_{q_j} \prod\limits_{l=1}^s B_{r_l} = 1$, where $\{ r_1, \dots, r_s \}$ are all the plaquettes outside of $\mathcal B_{\mathcal P}$. We also have that $\bigoplus_{i=1}^{m} \vec{\alpha}^{p_i} = \left( \bigoplus_{j=1}^{k} \vec{\alpha}^{q_j} \right) \oplus \left( \bigoplus_{l = 1}^{s} \vec{\alpha}^{r_l} \right)$, where the configurations are taken over the whole system.
Using those facts, we find
\begin{widetext}
\begin{eqnarray}
\begin{aligned}
&\theta_{\mathcal P} (\vec i, p_1, \dots, p_m)  = \frac{ \langle \vec i \bigoplus\limits_{i=1}^m \vec{\alpha}^{p_i} \oplus \vec{\alpha}^{\mathcal P} \vert B_{p_m} \dots B_{p_1} \prod_{i=1}^m B_{p_i} \prod_{j=1}^{k} B_{q_j}  \prod_{l=1}^s B_{r_l} \vert \vec i \oplus \vec{\alpha}^{\mathcal P} \rangle}{ \langle \vec i \bigoplus\limits_{i=1}^m \vec{\alpha}^{p_i} \vert B_{p_m} \dots B_{p_1} \prod_{i=1}^m B_{p_1} \prod_{j=1}^{k} B_{q_j} \prod_{l=1}^s B_{r_l}  \vert \vec i \rangle} \\
& =  \frac{ \langle \vec i \bigoplus\limits_{i=1}^m \vec{\alpha}^{p_i} \oplus \vec{\alpha}^{\mathcal P} \vert \prod_{j=1}^{k} B_{q_j} \prod_{l=1}^s B_{r_l} \vert \vec i \oplus \vec{\alpha}^{\mathcal P} \rangle}{ \langle \vec i \bigoplus\limits_{i=1}^m \vec{\alpha}^{p_i} \vert \prod_{j=1}^{k} B_{q_j} \prod_{l=1}^s B_{r_l} \vert \vec i \rangle}   =  \frac{ \langle \vec i \bigoplus\limits_{i=1}^m \vec{\alpha}^{p_i} \oplus \vec{\alpha}^{\mathcal P} \vert  \prod_{l=1}^s B_{r_l} \left( \sum_{\vec j} \vert \vec j \rangle \langle \vec j \vert \right)  \prod_{j=1}^{k} B_{q_j} \vert \vec i \oplus \vec{\alpha}^{\mathcal P} \rangle}{ \langle \vec i \bigoplus\limits_{i=1}^m \vec{\alpha}^{p_i} \vert \prod_{l=1}^s B_{r_l} \left( \sum_{\vec j} \vert \vec j \rangle \langle \vec j \vert \right) \prod_{j=1}^{k} B_{q_j}  \vert \vec i \rangle} \\
&  =  \frac{ \langle \vec i \bigoplus_{j=1}^{k} \vec{\alpha}^{q_j} \oplus \left( \bigoplus_{l = 1}^{s} \vec{\alpha}^{r_l} \right) \oplus \vec{\alpha}^{\mathcal P} \vert  \prod_{l=1}^s B_{r_l} \vert \vec i \bigoplus_{j=1}^{k} \vec{\alpha}^{q_j} \oplus \vec{\alpha}^{\mathcal P}  \rangle }{ \langle \vec i \bigoplus_{j=1}^{k} \vec{\alpha}^{q_j} \oplus \left( \bigoplus_{l = 1}^{s} \vec{\alpha}^{r_l} \right)   \vert  \prod_{l=1}^s B_{r_l} \vert \vec i \bigoplus_{j=1}^{k} \vec{\alpha}^{q_j}  \rangle } \frac{ \langle \vec i \bigoplus_{j = 1}^k \vec{\alpha}^{q_j} \oplus \vec{\alpha}^{\mathcal P} \vert \prod_{j=1}^{k} B_{q_j} \vert \vec i \oplus \vec{\alpha}^{\mathcal P} \rangle}{ \langle \vec i \bigoplus_{j = 1}^k \vec{\alpha}^{q_j} \vert \prod_{j=1}^{k} B_{q_j} \vert \vec i \rangle } \\
& = \frac{ \langle \vec i \bigoplus_{j=1}^{k} \vec{\alpha}^{q_j} \oplus \left( \bigoplus_{l = 1}^{s} \vec{\alpha}^{r_l} \right)  \vert X_{\mathcal P}  \prod_{l=1}^s B_{r_l} X_{\mathcal P} \vert \vec i \bigoplus_{j=1}^{k} \vec{\alpha}^{q_j}   \rangle }{ \langle \vec i \bigoplus_{j=1}^{k} \vec{\alpha}^{q_j} \oplus \left( \bigoplus_{l = 1}^{s} \vec{\alpha}^{r_l} \right)  \vert  \prod_{l=1}^s B_{r_l} \vert \vec i \bigoplus_{j=1}^{k} \vec{\alpha}^{q_j} \rangle } \theta_{\mathcal P} (\vec i, q_1, \dots, q_k), = \theta_{\mathcal P} (\vec i, q_1, \dots, q_k),
\end{aligned}
\end{eqnarray}
\end{widetext}
where we used the fact that $X_{\mathcal P}$ commutes with the plaquette operators $B_{r_l}$, since the plaquettes in $\{r_1, \dots, r_s \}$ are not in $\textnormal{Conn} (\mathcal P)$. $X_{\mathcal P}$ is the string of $\sigma_x$ corresponding to the string operator defined on $\mathcal P$.
\end{proof}
\end{lemma}
\begin{lemma} \label{lem:well_defined}
The functions $F_{\mathcal P}$ constructed by Algorithm \ref{alg:F_finding} are well-defined.
\begin{proof}
First note that Lemma \ref{lem:theta_permutation} states that the order in which the plaquettes $\{p_1, \dots, p_n \}$ appear in a specific subset of $\mathcal B_{\mathcal P}$ and the order into which the subsets are chosen do not affect its value.

Next, we show that if there are two different sets of plaquettes $\{p_1, \dots, p_m \} \subset \mathcal B_{\mathcal P}$ and $\{q_1, \dots, q_k \} \subset \mathcal B_{\mathcal P}$ such that $\vec{\alpha}^{p_1} \oplus \dots \oplus \vec{\alpha}^{p_m} = \vec{\alpha}^{q_1} \oplus \dots \oplus \vec{\alpha}^{q_k}$, where it is understood that the configurations are equal on $\textnormal{Conn} (\mathcal P)$ (as opposed to the whole lattice), then Algorithm \ref{alg:F_finding} ensures that $F_{\mathcal P} (\vec i \oplus \vec{\alpha}^{p_1} \oplus \dots \oplus \vec{\alpha}^{p_n}) = F_{\mathcal P} (\vec i \oplus \vec{\alpha}^{q_1} \oplus \dots \oplus \vec{\alpha}^{q_m})$, for any configurations $\vec i$ (and for configuration $\vec i \oplus \vec{\alpha}^{\mathcal P}$ as well). This is a simple consequence of Lemma \ref{eq:theta_product2}, which tells us that $\theta_{\mathcal P} (\vec i , p_1, \dots, p_m) = \theta_{\mathcal P} (\vec i, q_1, \dots, q_k)$, and of the way that the $F_{\mathcal P}$ functions are built;

\begin{eqnarray*}
\begin{aligned}
F_{\mathcal P} ( \vec i \oplus \vec{\alpha}^{p_1} \oplus \dots \oplus \vec{\alpha}^{p_m} )& = \theta_{\mathcal P} (\vec i, p_1, \dots, p_m ) F_{\mathcal P} ( \vec i ) \\
& = \theta_{\mathcal P} (\vec i , q_1, \dots, q_k) F_{\mathcal P} ( \vec i ) \\
& = F_{\mathcal P} ( \vec i \oplus \vec{\alpha}^{q_1} \oplus \dots \oplus \vec{\alpha}^{q_k} ).
\end{aligned}
\end{eqnarray*}
\end{proof}
\end{lemma}

\begin{lemma} \label{lem:alg_success}
Let $F_{\mathcal P}$ be a function determined by Algorithm \ref{alg:F_finding}. Then $F_{\mathcal P}$ simultaneously satisfies all the constraints (\ref{eq:commutation_string}).
\begin{proof}

Consider an arbitrary configuration $\vec i$ for which the value of $F_{\mathcal P} (\vec i)$ has been determined using Algorithm \ref{alg:F_finding}. Two possible cases need to be analyzed:
\begin{enumerate}
\item[ 1.] If $\vec i$ is one of the configuration picked to set an unknown value of $F_{\mathcal P} ( \vec i )$, then we find that for any plaquette $p$, 
\begin{equation}
F_{\mathcal P} (\vec i \oplus \vec{\alpha}^{p}) =  \frac{b_p (\vec i \oplus \vec{\alpha}^{\mathcal P})}{b_p (\vec i )} F_{\mathcal P} ( \vec i  ),
\end{equation}
by definition of $\theta_{\mathcal P}$.

\item[2.] If $\vec i$ is not one of the configurations picked, then there is another configuration $\vec i'$ which was picked as a class representative and a set of plaquettes $\{ p_1, \dots, p_m \} \subset \mathcal B_{\mathcal P}$ such that $\vec i = \vec i' \oplus \vec{\alpha}^{p_1} \oplus \dots \oplus \vec{\alpha}^{p_m}$. For any plaquette $p$, according to Algorithm \ref{alg:F_finding}, we find that
\begin{equation}
\frac{F_{\mathcal P} (\vec i \oplus \vec{\alpha}^{p}) }{ F_{\mathcal P} ( \vec i ) } = \frac{\theta_{\mathcal P} (\vec i' , p_1, \dots, p_m, p) }{ \theta_{\mathcal P} (\vec i' , p_1, \dots, p_m)}.
\end{equation}
By using definition (\ref{eq:def_theta}), we find
\begin{eqnarray*}
\begin{aligned}
\frac{\theta_{\mathcal P}(\vec i', p_1, \dots, p_m, p) }{\theta_{\mathcal P} (\vec i' , p_1, \dots, p_m)} &=  \frac{b_{p}( \vec i' \oplus \vec{\alpha}^{p_1} \oplus \dots \oplus \vec{\alpha}^{p_m}  \oplus \vec{\alpha}^{\mathcal P} ) }{b_{p} ( \vec i' \oplus \vec{\alpha}^{p_1} \oplus \dots \oplus \vec{\alpha}^{p_m} ) } \\
& = \frac{b_p ( \vec i  \oplus \vec{\alpha}^{\mathcal P} ) }{b_p ( \vec i  ) }.
\end{aligned}
\end{eqnarray*}
\end{enumerate}
We thus have that in both cases, all the constraints (\ref{eq:commutation_string}) are satisfied.

\end{proof}
\end{lemma}

\begin{lemma} \label{lem:square_identity}
(canonical strings) $F_{\mathcal P} ( \vec i \oplus \vec{\alpha}^{\mathcal P} ) = \lbrack F_{\mathcal P} ( \vec i ) \rbrack^{\ast}$ for any string configuration $\vec i$ if and only if $\lbrack S^{+}_{\mathcal P} \rbrack^2 = 1$.
\begin{proof}
Explicit calculation of $\lbrack S^{+}_{\mathcal P} \rbrack^2$ gives
\begin{eqnarray*} \label{eq:string_square1}
\begin{aligned}
\lbrack S^{+}_{\mathcal P} \rbrack^2  &= \left( \prod\limits_{i \in \, \mathcal P} \sigma^x_i \sum\limits_{\textnormal{String conf. } \vec i } F_{\mathcal P} ( \vec i ) \vert \vec i \rangle \langle \vec i \vert \right)\\ 
&\times \left( \prod\limits_{i' \in \, \mathcal P} \sigma^x_{i'} \sum\limits_{\textnormal{String conf. } \vec i'} F_{\mathcal P} ( \vec i' ) \vert \vec i' \rangle \langle \vec i' \vert \right) \\
& = \sum\limits_{\textnormal{String conf. } \vec i } F_{\mathcal P} ( \vec i ) \vert \vec i \oplus \vec{\alpha}^{\mathcal P} \rangle \langle \vec i \oplus \vec{\alpha}^{\mathcal P} \vert \\&\times\sum\limits_{\textnormal{String conf. }\vec i' } F_{\mathcal P} ( \vec i' ) \vert \vec i' \rangle \langle \vec i' \vert \\
& = \sum\limits_{\textnormal{String conf. }\vec i } F_{\mathcal P} ( \vec i \oplus \vec{\alpha}^{\mathcal P}) F_{\mathcal P} ( \vec i ) \vert \vec i \rangle \langle \vec i \vert.
\end{aligned}
\end{eqnarray*}
Clearly, $F_{\mathcal P} (\vec i \oplus \vec{\alpha}^{\mathcal P}) = \lbrack F_{\mathcal P} (\vec i ) \rbrack^{\ast}$ implies that $\lbrack S^{+}_{\mathcal P} \rbrack^2 = 1$, since $F_{\mathcal P} (\vec i)$ is a complex number lying on the unit circle.
On the other hand, $\lbrack S^{+}_{\mathcal P} \rbrack^2 = 1$ implies that $F_{\mathcal P} ( \vec i ) F_{\mathcal P} (\vec i \oplus \vec{\alpha}^{\mathcal P}) = 1$, which means that $F_{\mathcal P} (\vec i \oplus \vec{\alpha}^{\mathcal P}) = \lbrack F_{\mathcal P} (\vec i ) \rbrack^{\ast}$, once again because $F_{\mathcal P} (\vec i)$ is a complex number lying on the unit circle.

\end{proof}
\end{lemma}

\begin{lemma} \label{lem:hermiticity}
(canonical strings) $\lbrack S^{+}_{\mathcal P} \rbrack^{\dagger} = S^{+}_{\mathcal P}$ if and only if $F_{\mathcal P} ( \vec i  \oplus \vec{\alpha}^{\mathcal P}) = \lbrack F_{\mathcal P} ( \vec i ) \rbrack^{\ast}$ for any configuration $\vec i$.
\begin{proof}
Explicit calculation of $\lbrack S^{+}_{\mathcal P} \rbrack^{\dagger}$ gives
\begin{eqnarray*} \label{eq:string_hermitian}
\begin{aligned}
&\lbrack S^{+}_{\mathcal P} \rbrack^{\dagger} = \sum\limits_{\textnormal{String conf. } \vec i } \lbrack F_{\mathcal P} ( \vec i ) \rbrack^{\ast} \vert \vec i \rangle \langle \vec i \vert \prod\limits_{i \in \mathcal P} \sigma^x_{i} \\
& = \sum\limits_{\textnormal{String conf. } \vec i } \lbrack F_{\mathcal P} ( \vec i ) \rbrack^{\ast} \left( \prod\limits_{j \in \mathcal P} \sigma^x_j \right)^2 \vert \vec i \rangle \langle \vec i \vert \prod\limits_{i \in \mathcal P} \sigma^x_i \\
& = \prod\limits_{i \in \mathcal P} \sigma^x_i \sum\limits_{\textnormal{String conf. }\vec i } \lbrack F_{\mathcal P} (\vec i ) \rbrack^{\ast} \vert \vec i \oplus \vec{\alpha}^{\mathcal P} \rangle \langle \vec i \oplus \vec{\alpha}^{\mathcal P} \vert,
\end{aligned}
\end{eqnarray*}
Suppose that $\lbrack S^{+}_{\mathcal P} \rbrack^{\dagger} = S^{+}_{\mathcal P}$. In that case, it is clear that $\lbrack F_{\mathcal P} (\vec i) \rbrack^{\ast} = F_{\mathcal P} (\vec i \oplus \vec{\alpha}^{\mathcal P})$.

It is also clear that $\lbrack F_{\mathcal P} (\vec i) \rbrack^{\ast} = F_{\mathcal P} (\vec i \oplus \vec{\alpha}^{\mathcal P})$ implies that $\lbrack S^{+}_{\mathcal P} \rbrack^{\dagger} = S^{+}_{\mathcal P}$. 
\end{proof}
\end{lemma}

\begin{lemma} \label{lem:theta_consistency}
(canonical strings) Suppose that for any set of plaquettes $\{ p_1, p_2, \dots, p_m \} \subset \mathcal B_{\mathcal P}$ and for a string configuration $\vec i$, we have that $F_{\mathcal P} ( \vec i \oplus \vec{\alpha}^{p_1} \oplus \dots \oplus \vec{\alpha}^{p_m} ) = \theta_{\mathcal P} ( \vec i, p_1, \dots, p_m ) F_{\mathcal P} ( \vec i )$ and $F_{\mathcal P} ( \vec i \oplus \vec{\alpha}^{\mathcal P} \oplus \vec{\alpha}^{p_1} \oplus \dots \oplus \vec{\alpha}^{p_m} ) = \theta_{\mathcal P} ( \vec i \oplus \vec{\alpha}^{\mathcal P}, p_1, \dots, p_m ) F_{\mathcal P} ( \vec i \oplus \vec{\alpha}^{\mathcal P})$. If $F_{\mathcal P} ( \vec i \oplus \vec{\alpha}^{\mathcal P} ) = \lbrack F_{\mathcal P} ( \vec i ) \rbrack^{\ast}$, then for any set of plaquettes $\{ q_1, \dots, q_k \} \subset \mathcal B_{\mathcal P}$, we have that $F_{\mathcal P} ( \vec i \oplus \vec{\alpha}^{\mathcal P} \oplus \vec{\alpha}^{q_1} \oplus \dots \oplus \vec{\alpha}^{q_k} ) = \lbrack F_{\mathcal P} ( \vec i \oplus \vec{\alpha}^{q_1} \oplus \dots \oplus \vec{\alpha}^{q_k} ) \rbrack^{\ast}$.
\begin{proof}
By hypothesis, we have that
\begin{equation} \label{eq:consistency1}
F_{\mathcal P} (\vec i \oplus \vec{\alpha}^{q_1} ) = \theta_{\mathcal P} (\vec i, q_1) F_{\mathcal P} (\vec i),
\end{equation}
as well as
\begin{equation}
F_{\mathcal P} (\vec i \oplus \vec{\alpha}^{\mathcal P} \oplus \vec{\alpha}^{q_1} ) = \lbrack \theta_{\mathcal P} (\vec i, q_1) \rbrack^{-1} F_{\mathcal P} (\vec i \oplus \vec{\alpha}^{\mathcal P}),
\end{equation}
where we took advantage of the fact that $\theta_{\mathcal P} (\vec i \oplus \vec{\alpha}^{\mathcal P}, q_1) = \lbrack \theta_{\mathcal P} ( \vec i , q_1) \rbrack^{-1}$, as is clear from Eq.\ \eqref{eq:Hermicity_constraint}.

Using the fact that $F_{\mathcal P} (\vec i \oplus \vec{\alpha}^{\mathcal P}) = \lbrack F_{\mathcal P} (\vec i ) \rbrack^{\ast}$ in Eq.\ \eqref{eq:consistency1}, we find
\begin{eqnarray}
\begin{aligned}
\lbrack F_{\mathcal P} (\vec i \oplus \vec{\alpha}^{\mathcal P} ) \rbrack^{\ast} & = \lbrack \theta_{\mathcal P} (\vec i, q_1) \rbrack^{-1} F_{\mathcal P} (\vec i \oplus \vec{\alpha}^{q_1} ) \\
& = \frac{F_{\mathcal P} (\vec i \oplus \vec{\alpha}^{\mathcal P} \oplus \vec{\alpha}^{q_1} ) }{ F_{\mathcal P} (\vec i \oplus \vec{\alpha}^{\mathcal P} ) } F_{\mathcal P} (\vec i \oplus \vec{\alpha}^{q_1} ).
\end{aligned}
\end{eqnarray}
Since $F_{\mathcal P}$ lies on the unit circle, we find that
\begin{equation}
F_{\mathcal P} (\vec i \oplus \vec{\alpha}^{\mathcal P} \oplus \vec{\alpha}^{q_1} ) F_{\mathcal P} (\vec i \oplus \vec{\alpha}^{q_1}) = 1,
\end{equation}
which in turn implies that
\begin{equation}
F_{\mathcal P} (\vec i \oplus \vec{\alpha}^{\mathcal P} \oplus \vec{\alpha}^{q_1} ) = \lbrack F_{\mathcal P} (\vec i \oplus \vec{\alpha}^{q_1} ) \rbrack^{\ast}.
\end{equation}
The same reasoning can be recursively employed to show that
\begin{equation}
F_{\mathcal P} (\vec i \oplus \vec{\alpha}^{\mathcal P} \oplus \vec{\alpha}^{q_1} \oplus \dots \oplus \vec{\alpha}^{q_k}) = \lbrack F_{\mathcal P} (\vec i \oplus \vec{\alpha}^{q_1} \oplus \dots \oplus \vec{\alpha}^{q_k} ) \rbrack^{\ast}.
\end{equation}

\end{proof}
\end{lemma}

\begin{lemma} \label{lem:string_conjugation}
(canonical strings) Let $F_{\mathcal P}$ be a function determined by Algorithm \ref{alg:F_finding}. If the phases $e^{i \phi}$ are assigned to  the class representatives $\vec i$ in such a way that $F_{\mathcal P} (\vec i) = \lbrack F_{\mathcal P} (\vec i \oplus \vec{\alpha}^{\mathcal P}) \rbrack^{\ast}$, then $F_{\mathcal P}$ simultaneously satisfies all the constraints  in Eq.\ \eqref{eq:Hermicity_constraint}.
\begin{proof}
As it was done before, two possible cases are considered.
\begin{enumerate}
\item[1.] If $\vec i$ is one of the configuration picked as a class representative to set the value of $F_{\mathcal P} (\vec i)$, then we trivially have that $F_{\mathcal P} (\vec i \oplus \vec{\alpha}^{\mathcal P}) = \lbrack F_{\mathcal P} (\vec i ) \rbrack^{\ast}$.

\item[2.] If $\vec i$ is not one of the configurations picked, then once again we find that $\vec i = \vec i' \oplus \vec{\alpha}^{p_1} \oplus \dots \oplus \vec{\alpha}^{p_m}$, as mentioned above.

We find
\begin{eqnarray}
\begin{aligned}
F_{\mathcal P} (\vec i \oplus \vec{\alpha}^{\mathcal P} ) & = F_{\mathcal P} (\vec i' \oplus \vec{\alpha}^{\mathcal P} \oplus \vec{\alpha}^{p_1} \oplus \dots \oplus \vec{\alpha}^{p_m}).
\end{aligned}
\end{eqnarray}

Using Algorithm \ref{alg:F_finding}, we find that
\begin{eqnarray*} \label{eq:lemma_conditions}
\begin{aligned}
& F_{\mathcal P} (\vec i' \bigoplus\limits_{j=1}^{m} \vec{\alpha}^{p_j} ) = \theta_{\mathcal P}  (\vec i' , p_1, \dots, p_m) F^{\pm}_{\mathcal P} ( \vec i' ), \\
& F_{\mathcal P} (\vec i' \oplus \vec{\alpha}^{\mathcal P} \bigoplus\limits_{j=1}^{m} \vec{\alpha}^{p_j}) = \\
& \quad \quad \theta_{\mathcal P} (\vec i'  \oplus \vec{\alpha}^{\mathcal P}, p_1, \dots, p_m) F^{+}_{\mathcal P} (\vec i' \oplus \vec{\alpha}^{\mathcal P}).
\end{aligned}
\end{eqnarray*}

Since all the conditions of Lemma \ref{lem:theta_consistency} are satisfied, we have that
\begin{eqnarray}
\begin{aligned}
& F_{\mathcal P} (\vec i'  \oplus \vec{\alpha}^{\mathcal P} \oplus \vec{\alpha}^{p_1} \oplus \dots \oplus \vec{\alpha}^{p_m}) \\
& = \lbrack F_{\mathcal P} ( \vec i' \oplus \vec{\alpha}^{p_1} \oplus \dots \oplus \vec{\alpha}^{p_m} ) \rbrack^{\ast} \\
& = \lbrack F_{\mathcal P} (\vec i )  \rbrack^{\ast}.
\end{aligned}
\end{eqnarray}

\end{enumerate}
\end{proof}
\end{lemma}

\subsection{Proof of Theorem \ref{thm:string_operators}} \label{appsub:proof_thm}

Given all the previous technical results, it is straightforward to give the proof of Theorem \ref{thm:string_operators}, which we restate here for ease of reading:

\begin{reptheorem}{thm:string_operators}
Let $\mathcal P$ be a path. Any function $F_{\mathcal P}$ defined by Algorithm \ref{alg:F_finding} is such that $S^{\pm}_{\mathcal P}$ is a string operator. Furthermore, it is possible to choose the phases $e^{i \phi(i)}$ such that the string operator is canonical.

\begin{proof}
First note that according to Lemma \ref{lem:well_defined}, $F_{\mathcal P}$ is well-defined. By construction, $F_{\mathcal P}$ has non-trivial support only in $\textnormal{Conn} (\mathcal P)$, thus any operator $S^{\pm}_{\mathcal P}$ built from it satisfies condition \ref{property_2}. Furthermore, Lemma \ref{lem:alg_success} states that $S^{+}_{\mathcal P}$ satisfies conditions \eqref{eq:commutation_string}. We thus have that condition \ref{property_1} is satisfied as well, proving that $S^{\pm}_{\mathcal P}$ is a string operator.

In order to show that it is always possible to choose the phases $e^{i \phi(\vec i)}$'s so that $S^{\pm}_{\mathcal P}$ is canonical, we must consider two cases, depending on whether $\mathcal P$ is open or close.

Suppose first that the path $\mathcal P$ is open. In that case, we have that configurations $\vec i$ and $\vec i \oplus \vec{\alpha}^{\mathcal P}$ are in two distinct configuration classes. Given the class representative $\vec i$ for which we set $F_{\mathcal P} (\vec i) = e^{i \phi}$, we simply pick class representative $\vec i \oplus \vec{\alpha}^{\mathcal P}$ as representative for its corresponding class, and set $F_{\mathcal P} (\vec i \oplus \vec{\alpha}^{\mathcal P} ) = e^{-i \phi}$.

If $\mathcal P$ is close, then we find that $\vec i$ and $\vec i \oplus \vec{\alpha}^{\mathcal P}$ belong to the same class of configurations, since there exists a set of plaquettes $\{ p_1, \dots, p_m \} \subseteq \mathcal B_{\mathcal P}$ such that on $\textnormal{Conn} (\mathcal P)$, $\vec i \oplus \vec{\alpha}^{\mathcal P} = \vec i \oplus \vec{\alpha}^{p_1} \oplus \dots \oplus \vec{\alpha}^{p_m} $. Setting $F_{\mathcal P} ( \vec i ) = \lbrack \theta_{\mathcal P} (\vec i, p_1, \dots, p_m ) \rbrack^{-\frac{1}{2}}$, we find that $F_{\mathcal P} (\vec i \oplus \vec{\alpha}^{\mathcal P} ) = \lbrack F_{\mathcal P} \rbrack^{\ast}$.

In both cases, we can use Lemma \ref{lem:string_conjugation} to find that all constraints of Eq.\ \eqref{eq:Hermicity_constraint} are fulfilled and therefore,  conditions \ref{property_3} and \ref{property_4} are also satisfied.

\end{proof}
\end{reptheorem}

\subsection{Consistency of the probability of measuring an excitation configuration} \label{app:completeness_strings}

The decomposition of $X_{\mathcal P}$ in terms of string operators given by Eq.\ \eqref{eqn:Pauli_form2} is not unique given that in Algorithm~\ref{alg:F_finding}, we are free to choose different initial phases for the various class representatives. However, we show here that the probabilities associated with finding a given excitation pattern after the application $X_{\mathcal P}$ are insensitive to those initial phases. For simplicity, we assume that $\mathcal P$ is a single  non-overlapping and non-crossing open path. The arguments below generalize in a straightforward manner to the case where we need to consider $\mathcal P = \mathcal P_1 \# \dots \# \mathcal P_m$ for some $m > 1$.
	First note that 
	\begin{equation} \label{eq:Pauli_decomp}
	\vert \vec i \rangle \langle \vec i \vert = \prod_i \frac{1}{2} (1 + (-1)^{\alpha_i} \sigma^z_i),
	\end{equation}
	where $\vec i = (\alpha_1, \dots, \alpha_n)$. We can thus write $F_{\mathcal P} (\vec i) \vert \vec i \rangle \langle \vec i \vert = \sum_{P_z} c_{P_z} (\vec i) P_z$, where $P_z$ are all the possible Pauli operators acting on the qubits in $\textnormal{Conn} (\mathcal P)$ and composed of $\sigma^z$ and identities only, and where $c_{P_z} (\vec i)$ are complex coefficients given by Eq.\ \eqref{eq:Pauli_decomp} multiplied by $F_{\mathcal P} (\vec i)$. Given $\tilde{F}_{\mathcal P}$ differing of $F_{\mathcal P}$ by the choice of phases associated with the different class  representatives, we have that $\tilde{F}_{\mathcal P} (\vec i) \vert \vec i \rangle \langle \vec i \vert = \sum_{P_z} c_{P_z} (\vec i) e^{i \varphi(\vec j)} P_z$, where $\vec j$ is the representative of the configuration class into which $\vec i$ belongs, and $\varphi(\vec j)$ is the phase difference used between $F_{\mathcal P}$ and $\tilde{F}_{\mathcal P}$ to initialize Algorithm~\ref{alg:F_finding}.
	
	We define the orthonormal basis $\{ \vert L, C \rangle \}$ where $C$ labels the vertex and flux excitations configuration while $L$ is a label for the $4^g$ degenerate states corresponding to a given configuration. The probability of the transition $\vert L, C \rangle \rightarrow \vert L', C' \rangle $ caused by the application of $X_{\mathcal P}$ is given by
	\begin{widetext}
	\begin{eqnarray}
	\begin{aligned}
	P (\vert L, C \rangle \rightarrow \vert L', C' \rangle) & = \vert \langle L', C' \vert S^+_{\mathcal P} \sum_{\vec j} \sum_{\vec i \in \mathcal C_{\mathcal P} (\vec j) } \sum_{P_z} c_{P_z}^{\ast} (\vec i) e^{-i \varphi (\vec j)} P_z \vert L, C \rangle \vert^2 \\
	& = \sum_{\vec j, \vec j'} \sum_{\substack{\vec i \in \mathcal C_{\mathcal P} (\vec j) \\ \vec i' \in \mathcal C_{\mathcal P} (\vec j')}} \sum_{\substack{P_z  \textnormal{ s.t. }\vert \langle L', C' \vert S^+_{\mathcal P} P_z \vert L, C \rangle \vert = 1 \\ P_z' \textnormal{ s.t. } \vert \langle L', C' \vert S^+_{\mathcal P} P_z' \vert L, C \rangle \vert = 1 }} c_{P_z}^{\ast} (\vec i) c_{P_z'} (\vec i') e^{i (\varphi (\vec j) - \varphi (\vec j'))}. 
	\end{aligned}
	\end{eqnarray}
Clearly, $P_z$ and $P_z'$ share the same endpoints and belong to the same homological class, \emph{i.e.} $P_z P_z'$ form a trivial closed loop. Given the decomposition in Eq.\ \eqref{eq:Pauli_decomp}, we find that for $\vec j \neq \vec j'$
	\begin{equation} \label{eqn:ortho}
	\sum_{\substack{ P_z  \textnormal{ s.t. } \vert \langle L', C' \vert S^+_{\mathcal P} P_z \vert L, C \rangle \vert = 1 \\ P_z' \textnormal{ s.t. } \vert \langle L', C' \vert S^+_{\mathcal P} P_z' \vert L, C \rangle \vert = 1 }} c_{P_z}^{\ast} (\vec i) c_{P_z'} (\vec i') = 0.
	\end{equation}
	
	To see this, we rewrite Eq.~\eqref{eqn:ortho} as
	\begin{equation}
	\sum_{P_z  \textnormal{ s.t. } \vert \langle L', C' \vert S^+_{\mathcal P} P_z \vert L, C \rangle \vert = 1} \sum_{Q_z \subseteq \{ Q_{\nu} \}}  c_{P_z}^{\ast} (\vec i) c_{P_z \cdot Q_z} (\vec i'),
	\end{equation}
	where $\{ Q_{\nu} \}$ denotes the set of all subsets of products of charge operators associated with the vertices in path $\mathcal P$. Noting that since $\vec i $ and $\vec i'$ belongs to different configuration classes, there exists a vertex $\nu'$ such that $c_{P_z} (\vec i') = -c_{P_z \cdot Q_{\nu'}} (\vec i')$. Using this last fact, we get
	\begin{equation}
	\sum_{P_z  \textnormal{ s.t. } \vert \langle L', C' \vert S^+_{\mathcal P} P_z \vert L, C \rangle \vert = 1} \sum_{Q_z \subseteq \{ Q_{\nu} \} \backslash Q_{\nu'}}  \big[ c_{P_z}^{\ast} (\vec i) c_{P_z \cdot Q_z} (\vec i') + c_{P_z}^{\ast} (\vec i) c_{P_z \cdot Q_z Q_{\nu'}} (\vec i') \big],
	\end{equation}
	which clearly equals $0$.
	
	The probability transition is thus given by
	
	\begin{equation}
	P ( \vert L, C \rangle \rightarrow \vert L', C' \rangle ) = \sum_{j} \sum_{i, i' \in \mathcal C_{\mathcal P} (\vec j) } \sum_{\substack{ P_z \textnormal{ s.t. } \vert \langle L', C' \vert P_z \vert L, C \rangle \vert = 1 \\ P_z' \textnormal{ s.t. } \vert \langle L', C' \vert P_z' \vert L, C \rangle \vert = 1 }} c_{P_z}^{\ast} (\vec i') c_{P_z'} (\vec i) \langle L, C \vert P_z P_z' \vert L, C \rangle,
	\end{equation}
which is independent of the phases $e^{i \varphi ( \vec j )}$.
\end{widetext}

\onecolumngrid
\section{ Topological properties of strings operators} \label{app:openstrings_commutation}
Before presenting various technical results, we begin by precisely defining what we mean by \emph{crossing paths}.

\begin{definition} \label{def:path}
Let a path $\mathcal P = \{ e_1, e_2, \dots, e_n \}$ be a sequence of edges such that edge $e_i$ connects vertices $v_{i-1}$ to $v_i$.  If $v_0 = v_n$, we say that the path $\mathcal P$ is \emph{closed}; otherwise it is \emph{open}. If there exists $0 < i < n$ and $j \neq i$, $0 \leq j \leq n$ such that $v_i = v_j$ in the sequence of vertices it contains, $v_{\mathcal P} = \{ v_0, v_1, \dots, v_n \}$, then $\mathcal P$ is said to be \emph{self-crossing}. If there exists a plaquette containing two or more vertices of $v_{\mathcal P}$ that cannot form a single consecutive sequence, then path $\mathcal P$ is said to be \emph{self-overlapping}. Note that the notions of self-overlapping and self-crossing do not imply each other (see Fig.~\ref{fig:self_cross_overlap}).
\end{definition}

\begin{figure}
\subfloat[\label{fig:crossing}]{
\includegraphics[width=.2\textwidth]{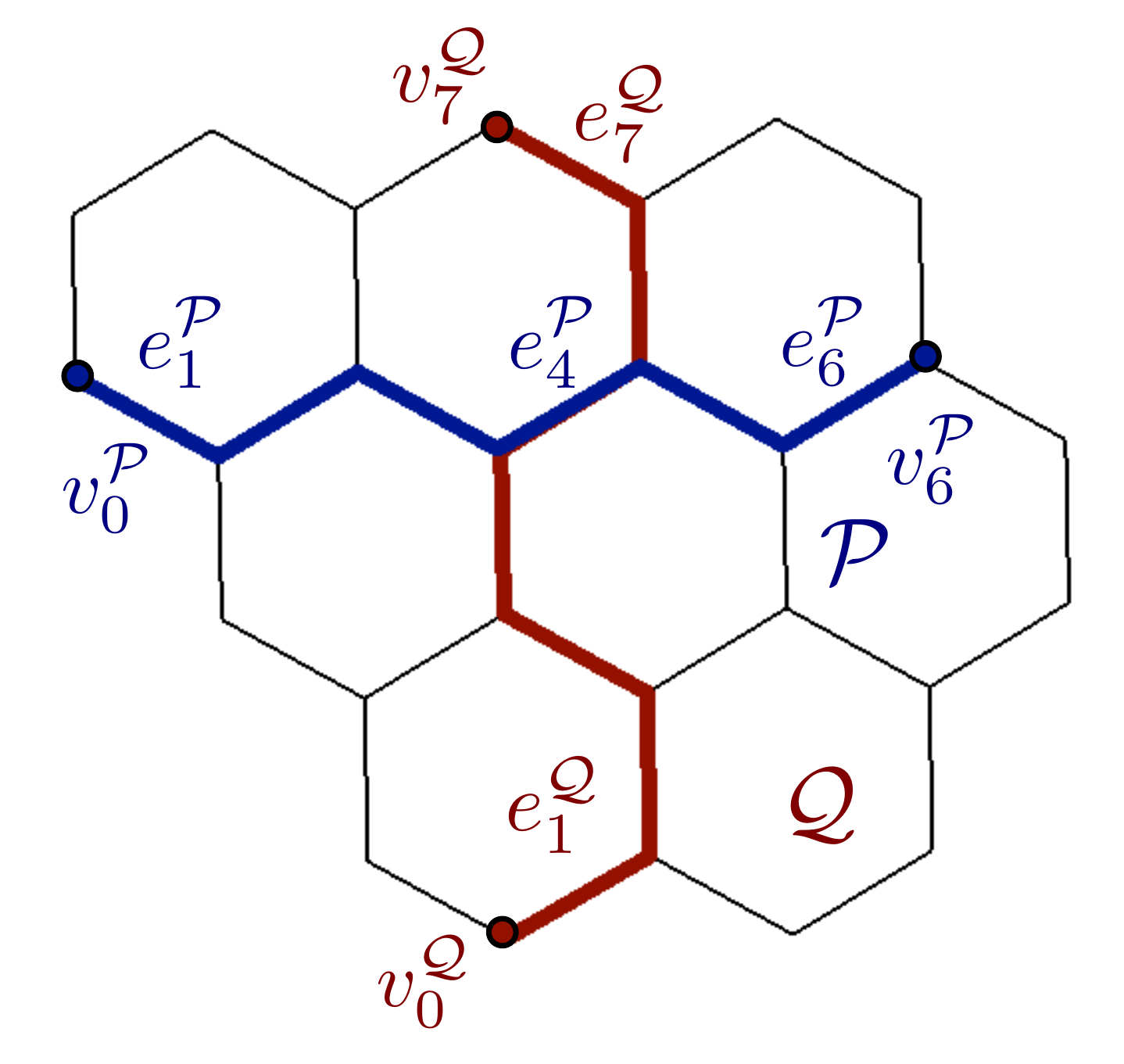}}
\hspace{.5cm}
\subfloat[\label{fig:not_crossing}]{
\includegraphics[width=.25\textwidth]{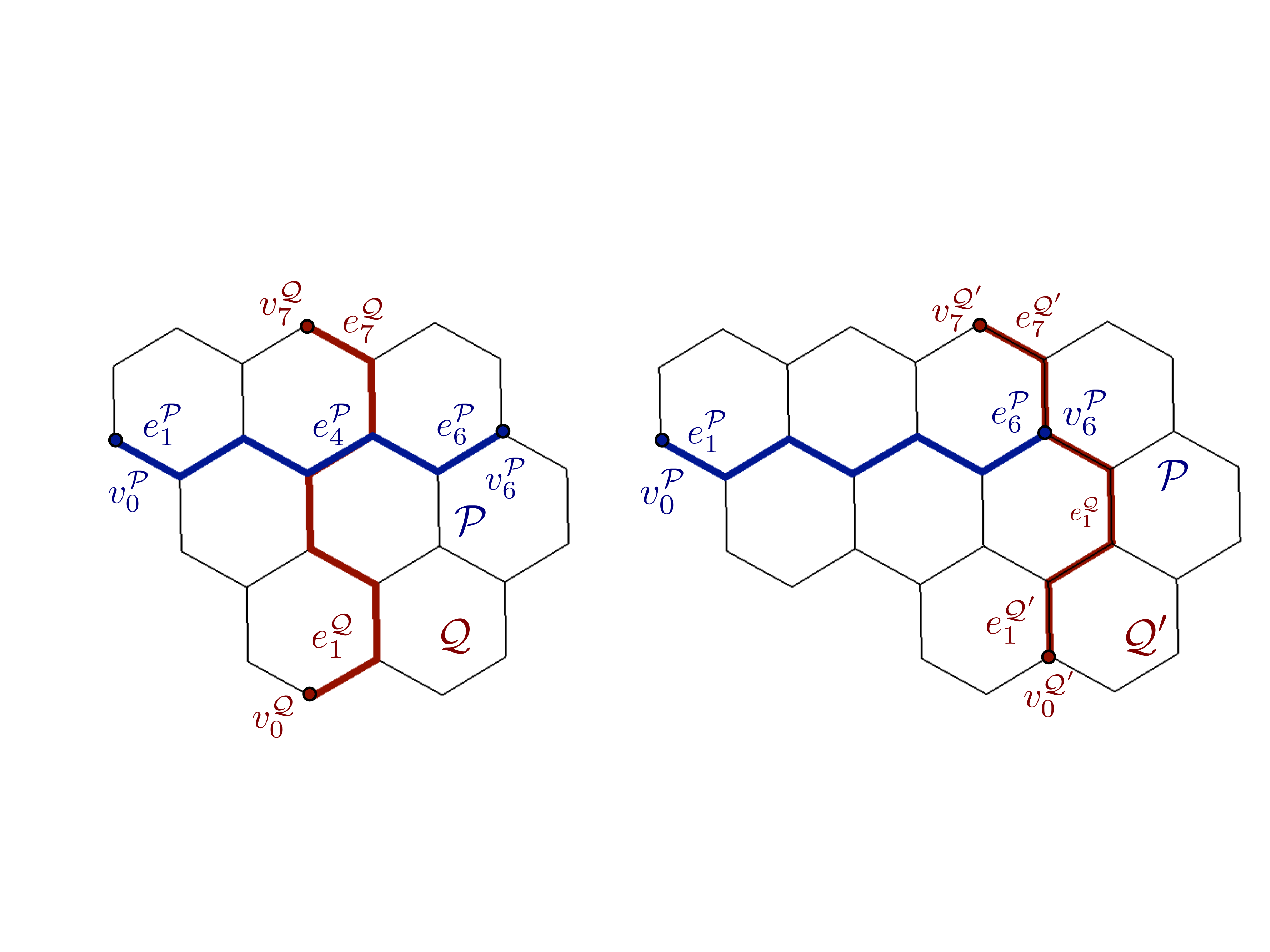}}
\hspace{.5cm}
\subfloat[\label{fig:crossing_cero}]{
\includegraphics[width=.2\textwidth]{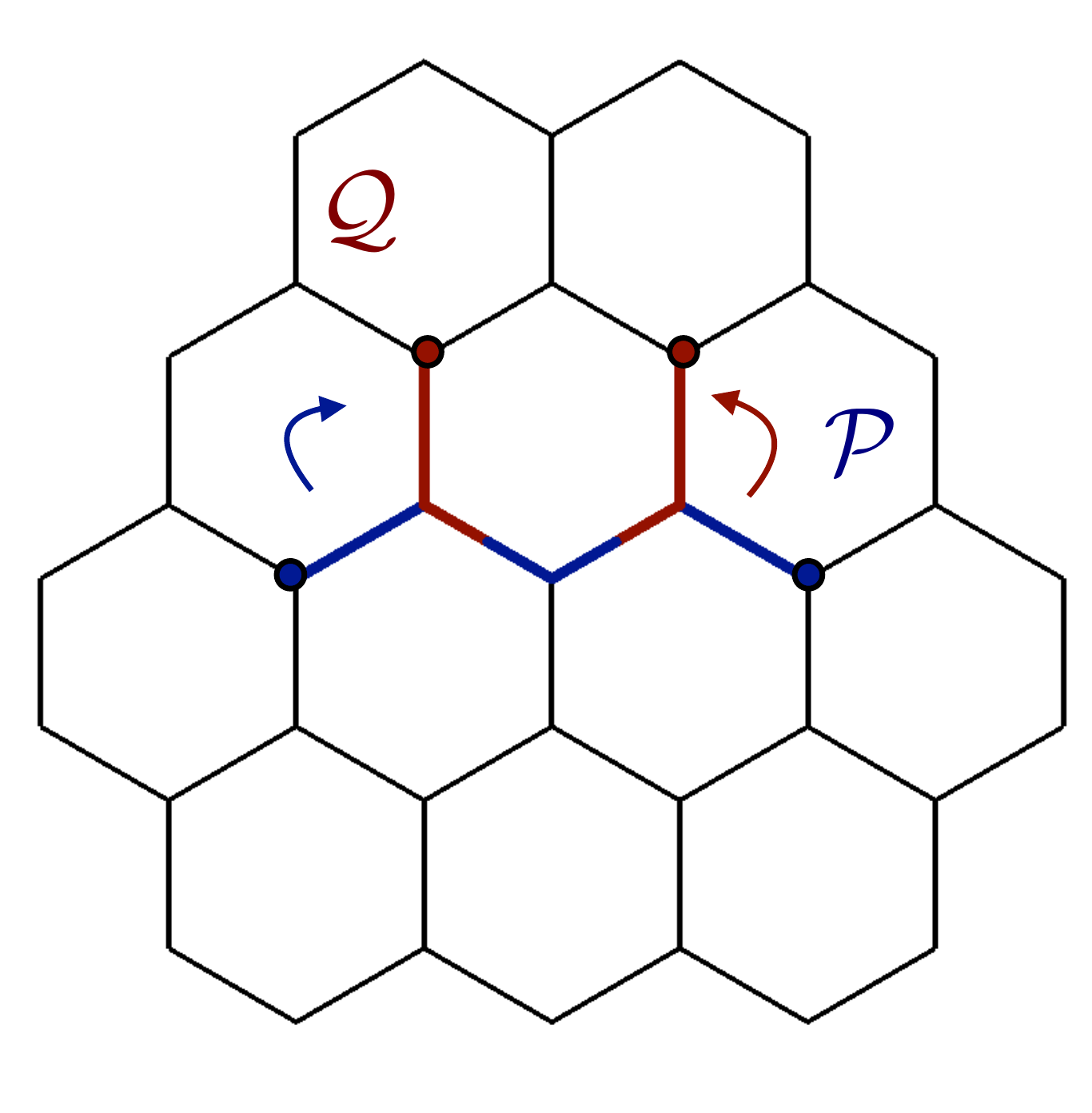}}
\hspace{.5cm}
\subfloat[\label{fig:crossing_once}]{
\includegraphics[width=.2\textwidth]{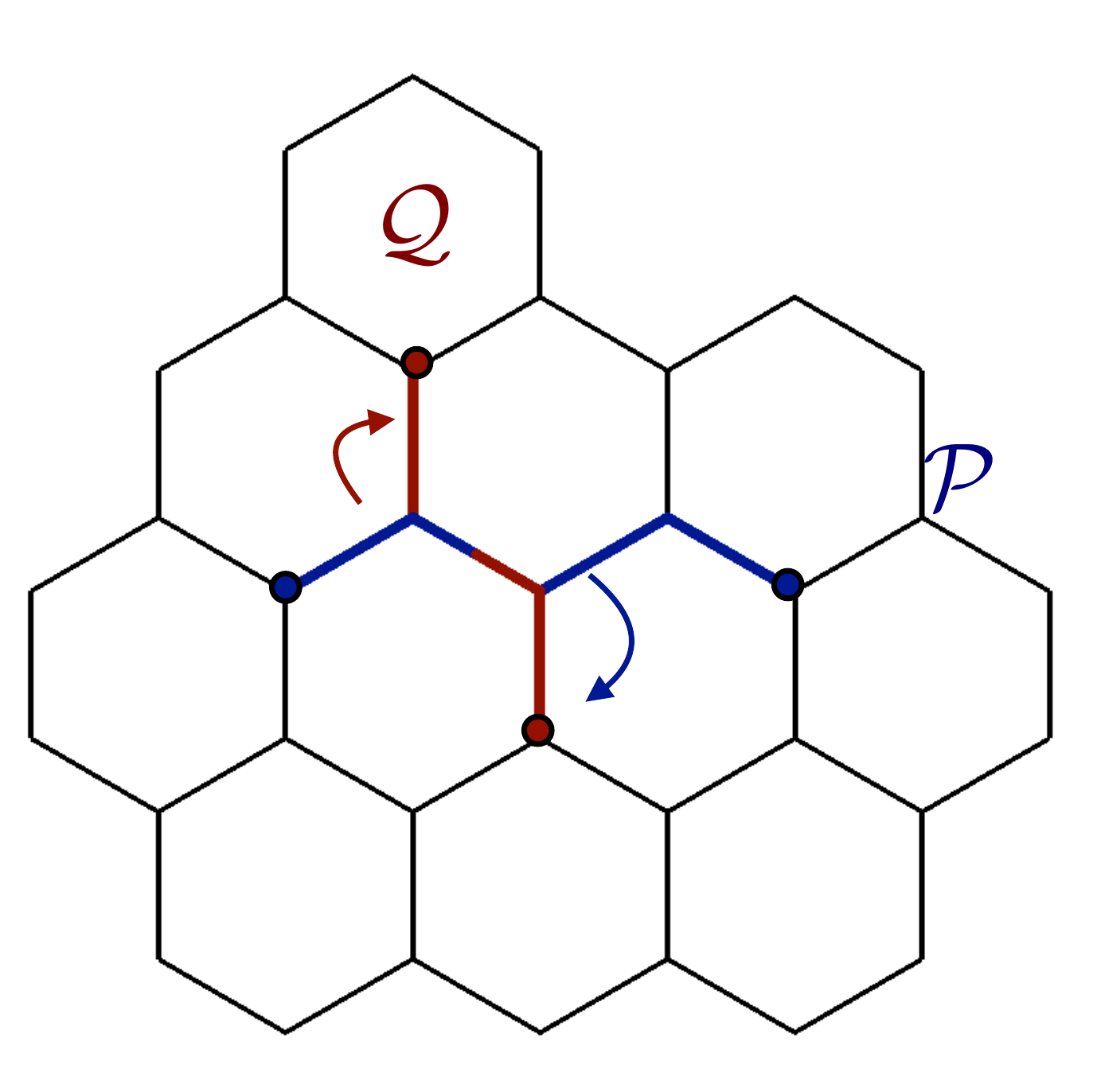}}
\caption{Example of two open paths $\mathcal P$ and $\mathcal Q$ which, according to criteria $i)$ of Definition \ref{def:crossing_strings} are crossing over the edge $E_{\mathcal P, \mathcal Q} = \{ e_4^{\mathcal P} = e_4^{\mathcal Q} \}$(a), as well as paths $\mathcal P$ and $\mathcal Q'$ which are not crossing (b), since path $\mathcal Q$ contains vertex $v_6^{\mathcal P}$. The two open paths $\mathcal P$ and $\mathcal Q$ cross $0$ times in (c), according to criteria $ii)$, while they cross once in (d). The arrows indicate the relative orientation among the strings.}
\label{fig:crossing_exemples}
\end{figure}

\begin{definition} \label{def:crossing_strings}
Consider two paths $\mathcal P = \{e^{\mathcal P}_1, \dots, e^{\mathcal P}_n \}$ and $\mathcal Q = \{e^{\mathcal Q}_1, \dots, e^{\mathcal Q}_{n'} \}$ connecting vertices $\{ v^{\mathcal P}_{0}, \dots, v^{\mathcal P}_{n} \}$ and $\{ v^{\mathcal Q}_{0}, \dots, v^{\mathcal Q}_{n'} \}$ respectively. Consider a sequence of edges in common of both paths $\mathcal P$ and $\mathcal Q$, and consider the largest (possibly empty) such sequence (supposing for now that it is unique), $E_{\mathcal P, \mathcal Q} = \{ e^{\mathcal P}_i = e^{\mathcal Q}_j, \dots, e^{\mathcal P}_{i'} = e^{\mathcal Q}_{j'} \}$ (with both sequences ordered in increasing order of simplicity, \emph{i.e.} $i' > i$ and $j' > j$) connecting the common vertices in $\mathcal P$ and $\mathcal Q$, denoted by $\Lambda_{\mathcal P, \mathcal Q} = \{ v^{\mathcal P}_{i-1} = v^{\mathcal Q}_{j-1}, \dots, v^{\mathcal P}_{i'} = v^{\mathcal Q}_{j'} \}$. Consider the following properties :
\begin{itemize}
\item[i)] (in the case where both paths open) none of the vertices $v^{\mathcal P}_0$, $v^{\mathcal P}_n$, $v^{\mathcal Q}_0$ and $v^{\mathcal Q}_{n'}$ are in $\Lambda_{\mathcal P, \mathcal Q}$,
\item[i)] (in the case where one path is open ($\mathcal P$), the other is closed ($\mathcal Q$) ) none of the vertices $v^{\mathcal P}_{0}$ and $v^{\mathcal P}_n$ are in $\Lambda_{\mathcal P, \mathcal Q}$,
\item[ii)] both pairs of edges $(e^{\mathcal P}_{i-1}, e^{\mathcal Q}_{j-1})$ and $(e^{\mathcal P}_{i'+1}, e^{\mathcal Q}_{j'+1})$ have the same relative orientation, i.e., clockwise or counter-clockwise.
\end{itemize}
If condition $i)$ is satisfied, we say that paths $\mathcal P$ and $\mathcal Q$ cross over the edges $E_{\mathcal P, \mathcal Q}$. Note that for the case of two closed paths, we always say that they cross.
If condition (ii) is not satisfied (including the case where $E_{\mathcal P, \mathcal Q}$ is the empty set), we say that $\mathcal P$ and $\mathcal Q$ cross $0$ times, otherwise we say that paths $\mathcal P$ and $\mathcal Q$ cross once.
Finally, if there is more than one pair of sequences $E^i_{\mathcal P, \mathcal Q}$ and $\Lambda^i_{\mathcal P, \mathcal Q}$ where $i$ runs from $1$ through $m$ which \emph{all} satisfy conditions $i)$, we say that paths $\mathcal P$ and $\mathcal Q$ cross. For all the regions $E^{i_1}_{\mathcal P, \mathcal Q}, \dots, E^{i_{m'}}_{\mathcal P, \mathcal Q}$ and $\Lambda^{i_1}_{\mathcal P, \mathcal Q}, \dots, \Lambda^{i_{m'}}_{\mathcal P, \mathcal Q}$ for which $ii)$ is satisfied, with $m' \leq m$, we say that paths $\mathcal P$ and $\mathcal Q$ cross over the relevant region, and we additionally say that paths $\mathcal P$ and $\mathcal Q$ cross $m'$ times. See Fig. \ref{fig:crossing_exemples} for explicit examples.
\end{definition}

Note that in the previous definition, paths $\mathcal P$ and $\mathcal Q$ can be formed of smaller paths, i.e., $\mathcal P = \mathcal P_1 \# \dots \# \mathcal P_m$ and $\mathcal Q = \mathcal Q_1 \# \dots \# \mathcal Q_n$.
Notice that when $\mathcal P$ and $\mathcal Q$ cross, we may define a reference frame such that one of the paths plays the role of the horizontal and the other the vertical. In the following, we consider that path $\mathcal P$ is the horizontal and $\mathcal Q$ the vertical.

 It is useful to define $\mathcal B_{\mathcal P}^{\mathcal Q, \textnormal{left}} = \{p^{\textnormal{left}}_1, \dots, p^{\textnormal{left}}_m \}$, the set of plaquettes in $\mathcal B_{\mathcal P}$ such that $\vec{\alpha}^{p^{\textnormal{left}}_1} \oplus \dots \oplus \vec{\alpha}^{p^{\textnormal{left}}_m} = \vec{\alpha}^{\mathcal Q}$ when restricted to $\textnormal{Conn} (\mathcal P)$ and containing the left-most plaquette of $\mathcal B_{\mathcal P}$. In a complementary way, we define $\mathcal B_{\mathcal P}^{\mathcal Q, \textnormal{right}} = \mathcal B_{\mathcal P} \backslash \mathcal B_{\mathcal P}^{\mathcal Q, \textnormal{left}}$. In a similar way, $\mathcal B_{\mathcal Q}^{\mathcal P, \textnormal{up}}$ and $\mathcal B_{\mathcal Q}^{\mathcal P, \textnormal{down}}$ can be defined for a suitable path $\mathcal Q$. The nomenclature of \emph{left} vs \emph{right} is an arbitrary choice (just as is the case for \emph{up} vs \emph{down}). 

Note that we implicitly used the fact that paths $\mathcal P$ and $\mathcal Q$ cross, are open, and that they are not self-overlapping nor self-crossing in the above definition. If it were not the case, then it would not be  possible to find a set of plaquettes such that the associated configuration corresponds to the configuration of the other path on its connected region.

Furthermore, for a general path $\mathcal P = \mathcal P_1 \# \dots \# \mathcal P_m$ with every path $\{ \mathcal P_i \}$ open, not self-crossing nor self-overlapping, but for which it is not necessarily true for the \emph{whole} path $\mathcal P$, and for a path $\mathcal Q$ such that $\mathcal P$ and $\mathcal Q$ are crossing, it is always possible to similarly define $\mathcal B^{\mathcal Q, \textnormal{left (right)}}_{\mathcal P_i}$ for $i \in \{ 1, \dots, m \}$.

Consider two crossing paths $\mathcal P = \mathcal P_1 \# \dots \# \mathcal P_m$ and $\mathcal Q = \mathcal Q_1 \# \dots \# \mathcal Q_n$ such that every individual path $\mathcal P_i$ and $\mathcal Q_j$ is open, is not self-crossing nor self-overlapping, and we are interested in computing the commutation relations between $S^+_{\mathcal P_1 \# \dots \# \mathcal P_m}$ and $S^+_{\mathcal Q_1 \# \dots \# \mathcal Q_n}$. Explicit calculations give

\begin{eqnarray} \label{eqn:crossing_strings_1}
\begin{aligned}
S^+_{\mathcal P_1 \# \dots \# \mathcal P_m}  S^+_{\mathcal Q_1 \# \dots \# \mathcal Q_n}  =  \prod_{i \in \mathcal P} \sigma^x_i \prod_{j \in \mathcal Q} \sigma^x_j \sum_{\textnormal{string conf. } \vec i} \prod_{i = 1}^m F_{\mathcal P_i} (\vec i \oplus \vec{\alpha}^{\mathcal Q} \bigoplus_{j = 1}^{i-1} \vec{\alpha}^{\mathcal P_j}) \prod_{i'=1}^n F_{\mathcal Q_{i'}} (\vec i \bigoplus_{j'=1}^{i'-1} \vec{\alpha}^{\mathcal Q_{j'}}) \vert \vec i \rangle \langle \vec i \vert,
\end{aligned}
\end{eqnarray}
where the product over the small strings are taken in the reverse order, and where the string configurations are considered  over the whole system.

Similarly computing the product of the string operators in the reverse order, we get

\begin{eqnarray} \label{eqn:crossing_strings_2}
\begin{aligned}
S^+_{\mathcal Q_1 \# \dots \# \mathcal Q_n} S^+_{\mathcal P_1 \# \dots \# \mathcal P_m} = \prod_{i \in \mathcal P} \sigma^x_i \prod_{j \in \mathcal Q} \sigma^x_j \sum_{\textnormal{string conf. } \vec i} \prod_{i = 1}^m F_{\mathcal P_i} (\vec i \bigoplus_{j = 1}^{i-1} \vec{\alpha}^{\mathcal P_j}) \prod_{i'=1}^n F_{\mathcal Q_{i'}} (\vec i \oplus \vec{\alpha}^{\mathcal P} \bigoplus_{j'=1}^{i'-1} \vec{\alpha}^{\mathcal Q_{j'}}) \vert \vec i \rangle \langle \vec i \vert.
\end{aligned}
\end{eqnarray}

Considering Eq.\ \eqref{eqn:crossing_strings_1} and \eqref{eqn:crossing_strings_2}, we define the quantity
\begin{equation} \label{eqn:R_ratio_def}
\mathcal R_{\mathcal P_1 \# \dots \# \mathcal P_m, \mathcal Q_1 \# \dots \# \mathcal Q_n} (\vec i ) = \frac{\prod_{i = 1}^m F_{\mathcal P_i} (\vec i \oplus \vec{\alpha}^{\mathcal Q} \bigoplus_{j = 1}^{i-1} \vec{\alpha}^{\mathcal P_j} ) \prod_{i'=1}^n F_{\mathcal Q_{i'}} (\vec i \bigoplus_{j'=1}^{i'-1} \vec{\alpha}^{\mathcal Q_{j'}} ) }{ \prod_{i = 1}^{m} F_{\mathcal P_i} (\vec i \bigoplus_{j=1}^{i-1} \vec{\alpha}^{\mathcal P_j}) \prod_{i'=1}^n F_{\mathcal Q_{i'}} (\vec i \oplus \vec{\alpha}^{\mathcal P} \bigoplus_{j'=1}^{i'-1} \vec{\alpha}^{\mathcal Q_{j'}}) }.
\end{equation}
which gives the commutation relations between $S^{+}_{\mathcal P_1 \# \dots \# \mathcal P_m}$ and $S^{+}_{\mathcal Q_1 \# \dots \# \mathcal Q_n}$.The quantity defined in Eq.\ \eqref{eqn:R_ratio_def} is independent on the specific string configuration $\vec i$, which is shown in Lemma \ref{lem:R_form} and \ref{lem:R_configuration_independence}, and that it does not depend on the specific way that the paths $\mathcal P$ and $\mathcal Q$ are partitioned in the smaller paths $\{ \mathcal P_i \}$ and $\{ \mathcal Q_j \}$, as long as those are not self-crossing nor self-overlapping.
It can also be shown that if paths $\mathcal P$ and $\mathcal Q$ are transformed to paths $\mathcal P'$ and $\mathcal Q'$ using a series of elongations, reductions and valid deformations such that none of the elementary step makes a path crossing an endpoint of the other path, then we find $\mathcal R_{\mathcal P, \mathcal Q} = \mathcal R_{\mathcal P', \mathcal Q'}$. This is done in Lemmas \ref{lem:path_p_ind} through \ref{lem:odd_crossing} as well as in Corollary \ref{cor:path_cut_ind}.
If paths $\mathcal P$ and $\mathcal Q$ cross an odd number of times, one can then proceed to transforms paths $\mathcal P$ and $\mathcal Q$ to minimal configurations $\mathcal P_{\textnormal{min}}$ and $\mathcal Q_{\textnormal{min}}$ such that $\mathcal R_{\mathcal P, \mathcal Q} = \mathcal R_{\mathcal P_{\textnormal{min}}, \mathcal Q_{\textnormal{min}}}$ (see Fig. \ref{fig:ratio_computation}). Explicitly computing this last quantity for a given string configuration yields $\mathcal R_{\mathcal P, \mathcal Q} = -1$.
If, on the other hand, $\mathcal P$ and $\mathcal Q$ cross an even number of times, one can consider the deformed paths $\mathcal P_{0}$ and $\mathcal Q_{0}$ such that $\mathcal R_{\mathcal P, \mathcal Q} = \mathcal R_{\mathcal P_{0}, \mathcal Q_{0}}$ and such that $\mathcal P_0$ and $\mathcal Q_0$ supports are disjoint, showing that $\mathcal R_{\mathcal P, \mathcal Q} = +1$.
Remarkably, all those previous results are essentially due to the fact that the plaquette operators commute and square to the identity.

\subsection{Useful technical lemmas} \label{appsub:useful_lemmas}

Lemmas mentioned before are introduced here. They are necessary to show Theorem \ref{thm:crossing_strings}. 

\begin{lemma} \label{lem:R_form}
Let $\{ p_1^i, \dots, p_{m^i}^i \} = \mathcal B_{\mathcal P_i}^{\mathcal Q, a_i}$ and $\{ q_1^j, \dots, q_{n^j}^j \} = \mathcal B_{\mathcal Q_j}^{\mathcal P, b_j}$ with $a_i = \textnormal{left}$ or $a_i = \textnormal{right}$ and $b_j = \textnormal{up}$ or $b_j = \textnormal{down}$, as it is shown in Fig. \ref{fig:crossing_strings1}. Then, $\mathcal{R}_{\mathcal P_1 \# \dots \# \mathcal P_m, \mathcal Q_1 \# \dots \# \mathcal Q_n} (\vec i) $ can be written as: 

\begin{eqnarray} \label{eqn:ratio_form}
\begin{aligned}
\mathcal R_{\mathcal P_1 \# \dots \# \mathcal P_m, \mathcal Q_1 \# \dots \# \mathcal Q_n} (\vec i) &= \prod_{i =1}^{m} \frac{\langle \vec i \bigoplus_{j = 1}^{i} \vec{\alpha}^{\mathcal P_j} \bigoplus_{k = 1}^{m^i} \vec{\alpha}^{p_{k}^i} \vert \prod_{j = 1}^{m^i} B_{p_j^i} \vert \vec i \bigoplus_{j = 1}^{i} \vec{\alpha}^{\mathcal P_j} \rangle }{ \langle \vec i \bigoplus_{j = 1}^{i-1} \vec{\alpha}^{\mathcal P_j} \bigoplus_{k = 1}^{m^i} \vec{\alpha}^{p_{k}^i} \vert \prod_{j = 1}^{m^i} B_{p_j^i} \vert \vec i \bigoplus_{j = 1}^{i-1} \vec{\alpha}^{\mathcal P_j} \rangle  } \\
 \times & \prod_{i =1}^{n} \frac{\langle \vec i \bigoplus_{j = 1}^{i-1} \vec{\alpha}^{\mathcal Q_j} \bigoplus_{k = 1}^{n^i} \vec{\alpha}^{q_{k}^i} \vert \prod_{j = 1}^{n^i} B_{q_j^i} \vert \vec i \bigoplus_{j = 1}^{i-1} \vec{\alpha}^{\mathcal Q_j} \rangle }{ \langle \vec i \bigoplus_{j = 1}^{i} \vec{\alpha}^{\mathcal Q_j} \bigoplus_{k = 1}^{n^i} \vec{\alpha}^{q_{k}^i} \vert \prod_{j = 1}^{n^i} B_{q_j^i} \vert \vec i \bigoplus_{j = 1}^{i} \vec{\alpha}^{\mathcal Q_j} \rangle  }.
 \end{aligned}
\end{eqnarray}
\begin{figure}[h!]
\center
\includegraphics[scale=0.50]{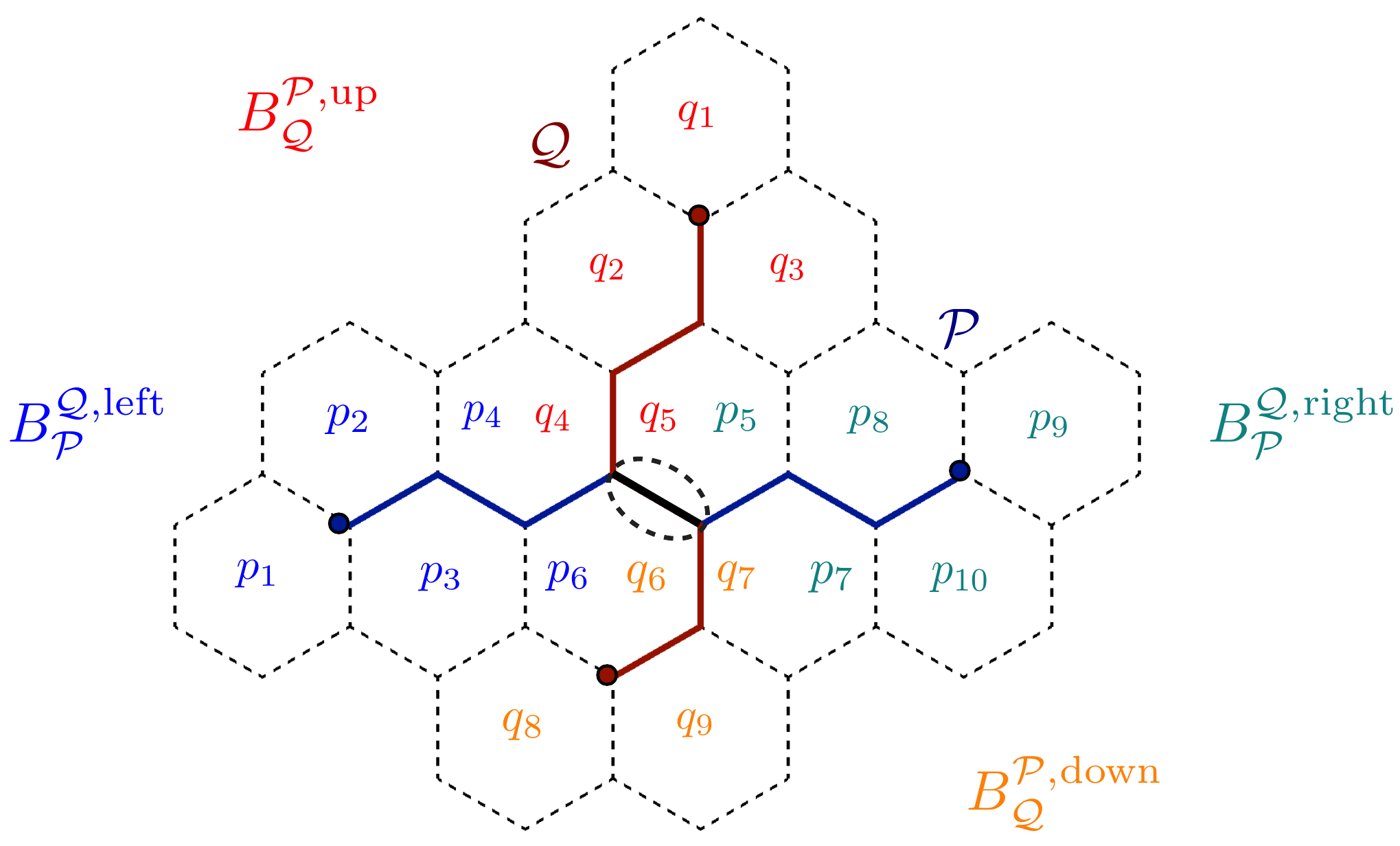}
\caption{An example of the structure of string operators $S^+_{\mathcal P}$ and $S^+_{\mathcal Q}$ on crossing paths $\mathcal P$ and $\mathcal Q$, shown in dark blue and red, respectively. The various sets of plaquettes $\mathcal B_{\mathcal P}^{\mathcal Q, \textnormal{left (right)}}$ and $\mathcal B_{\mathcal Q}^{\mathcal P, \textnormal{up (down)}}$ are presented in various matching colours. The edge over which both paths $\mathcal P$ and $\mathcal Q$ are overlapping is circled by a dashed line.}
\label{fig:crossing_strings1}
\end{figure}
\begin{proof}

We begin by considering the quantity

\begin{equation}
\frac{ F_{\mathcal P_i} (\vec i \oplus \vec{\alpha}^{\mathcal Q} \bigoplus_{j = 1}^{i-1} \vec{\alpha}^{\mathcal P_j} )}{ F_{\mathcal P_i} (\vec i \bigoplus_{j=1}^{i-1} \vec{\alpha}^{\mathcal P_j})},
\end{equation}
which appears in Eq.\ \eqref{eqn:R_ratio_def}. Using the definition of $\mathcal B_{\mathcal P_i}^{\mathcal Q, a_i}$, we find

\begin{equation}
\frac{ F_{\mathcal P_i} (\vec i \oplus \vec{\alpha}^{\mathcal Q} \bigoplus_{j = 1}^{i-1} \vec{\alpha}^{\mathcal P_j} )}{ F_{\mathcal P_i} (\vec i \bigoplus_{j=1}^{i-1} \vec{\alpha}^{\mathcal P_j})} = \frac{ F_{\mathcal P_i} (\vec i \bigoplus_{j = 1}^{m^i} \vec{\alpha}^{p_j^i}  \bigoplus_{j = 1}^{i-1} \vec{\alpha}^{\mathcal P_j} )}{ F_{\mathcal P_i} (\vec i \bigoplus_{j=1}^{i-1} \vec{\alpha}^{\mathcal P_j})}.
\end{equation}

Using the structure of the $F_{\mathcal P_i}$ as defined by Algorithm \ref{alg:F_finding}, we find that there exists a configuration $\vec i'$ and a set of plaquettes $\{p'_1, \dots, p'_{m'} \}$ (possibly empty) such that $\vec i \bigoplus_{j = 1}^{i-1} \vec{\alpha}^{\mathcal P_j}  = \vec i' \bigoplus_{j = 1}^{m'} \vec{\alpha}^{p'_j}$ when restricted to $\textnormal{Conn} (\mathcal P_i)$. We can thus write

\begin{eqnarray} \label{eqn:R_theta}
\begin{aligned}
\frac{ F_{\mathcal P_i} (\vec i \oplus \vec{\alpha}^{\mathcal Q} \bigoplus_{j = 1}^{i-1} \vec{\alpha}^{\mathcal P_j} )}{ F_{\mathcal P_i} (\vec i \bigoplus_{j=1}^{i-1} \vec{\alpha}^{\mathcal P_j})} &= \frac{\theta_{\mathcal P_i} (\vec i', p'_1, \dots, p'_{m'}, p_1^i, \dots, p_{m^i}^i) }{ \theta_{\mathcal P_i} (\vec i', p'_1, \dots, p'_{m'}) } \\
&=  \theta_{\mathcal P_i} ( \vec i, p_1^i, \dots, p_{m^i}^i ),
\end{aligned}
\end{eqnarray}
where we made use of Eq.\ \eqref{eq:def_theta}. Note that using Lemma \ref{lem:theta_product}, we are free to choose the values for $a_i$=left/right and $b_j$=up/down, as we please. This will turn out to be very useful later on.

Using Eq.\ \eqref{eq:def_theta}, we find

\begin{eqnarray}
\begin{aligned}
\frac{ F^+_{\mathcal P_i} (\vec i \oplus \vec{\alpha}^{\mathcal Q} \bigoplus_{j = 1}^{i-1} \vec{\alpha}^{\mathcal P_j} )}{ F^+_{\mathcal P_i} (\vec i \bigoplus_{j=1}^{i-1} \vec{\alpha}^{\mathcal P_j})} &= \frac{\langle \vec i \bigoplus_{j = 1}^{i} \vec{\alpha}^{\mathcal P_j} \bigoplus_{k = 1}^{m^i} \vec{\alpha}^{p_{k}^i} \vert \prod_{j = 1}^{m^i} B_{p_j^i} \vert \vec i \bigoplus_{j = 1}^{i} \vec{\alpha}^{\mathcal P_j} \rangle }{ \langle \vec i \bigoplus_{j = 1}^{i-1} \vec{\alpha}^{\mathcal P_j} \bigoplus_{k = 1}^{m^i} \vec{\alpha}^{p_{k}^i} \vert \prod_{j = 1}^{m^i} B_{p_j^i} \vert \vec i \bigoplus_{j = 1}^{i-1} \vec{\alpha}^{\mathcal P_j} \rangle  }.
\end{aligned}
\end{eqnarray}

A similar reasoning holding for the quantity

\begin{equation}
\left( \frac{F^+_{\mathcal Q_{i'}} (\vec i \bigoplus_{j'=1}^{i'-1} \vec{\alpha}^{\mathcal Q_{j'}} )}{ F^+_{\mathcal Q_{i'}} (\vec i \oplus \vec{\alpha}^{\mathcal P} \bigoplus_{j'=1}^{i'-1} \vec{\alpha}^{\mathcal Q_{j'}}) } \right)^{-1},
\end{equation}
this completes the proof.
\end{proof}
\end{lemma}
The precedent lemma stipulates that $\mathcal R_{\mathcal P, \mathcal Q} ( \vec i)$ can be written in terms of the phases acquired by the product of the plaquette operators of the plaquettes contained in $\mathcal B_{\mathcal P}^{\mathcal Q, a}$ and $\mathcal B_{\mathcal Q}^{\mathcal P, b}$, on the appropriate string configurations.

\begin{lemma} \label{lem:R_configuration_independence}
The quantity $\mathcal R_{\mathcal P_1 \# \dots \# \mathcal P_m, \mathcal Q_1 \# \dots \# \mathcal Q_n} ( \vec i )$ is independent of the string configuration $\vec i$.
\begin{proof}

Consider an arbitrary edge $e$ and its associated canonical string operator $S^+_{e}$, which we can always find according to Theorem \ref{thm:string_operators}. Further let $\{ p_1^i, \dots, p_{m^i}^i \} = \mathcal B_{\mathcal P_i}^{\mathcal Q, a_i}$ and $\{ q_1^j, \dots, q_{n^j}^j \} = \mathcal B_{\mathcal Q_j}^{\mathcal P, b_j}$ with $a_i = \textnormal{left}$ or $a_i = \textnormal{right}$ and $b_j = \textnormal{up}$ or $b_j = \textnormal{down}$, and which we are always free to choose according to Lemma \ref{lem:theta_product}. Using Lemma \ref{lem:R_form} and the fact that $S^+_{e}$ is canonical and therefore squares to one, the quantity $\mathcal R_{\mathcal P_1 \# \dots \# \mathcal P_m, \mathcal Q_1 \# \dots \# Q_m} (\vec i )$ is given by

\begin{eqnarray} \label{eqn:ratio_path_cut_ind}
\begin{aligned}
\mathcal R_{\mathcal P_1 \# \dots \# \mathcal P_m, \mathcal Q_1 \# \dots \# \mathcal Q_n} (\vec i ) &= \prod_{i =1}^{m} \left( \frac{\langle \vec i \bigoplus_{j = 1}^{i} \vec{\alpha}^{\mathcal P_j} \bigoplus_{k = 1}^{m^i} \vec{\alpha}^{p_{k}^i} \vert S^+_{e} \prod_{j = 1}^{m^i} B_{p_j^i} S^+_e \vert \vec i \bigoplus_{j = 1}^{i} \vec{\alpha}^{\mathcal P_j} \rangle }{ \langle \vec i \bigoplus_{j = 1}^{i-1} \vec{\alpha}^{\mathcal P_j} \bigoplus_{k = 1}^{m^i} \vec{\alpha}^{p_{k}^i} \vert S^+_e \prod_{j = 1}^{m^i} B_{p_j^i} S^+_e \vert \vec i \bigoplus_{j = 1}^{i-1} \vec{\alpha}^{\mathcal P_j} \rangle  } \right) \\
 \times & \prod_{i =1}^{n} \left( \frac{\langle \vec i \bigoplus_{j = 1}^{i-1} \vec{\alpha}^{\mathcal Q_j} \bigoplus_{k = 1}^{n^i} \vec{\alpha}^{q_{k}^i} \vert S^+_e \prod_{j = 1}^{n^i} B_{q_j^i} S^+_e \vert \vec i \bigoplus_{j = 1}^{i-1} \vec{\alpha}^{\mathcal Q_j} \rangle }{ \langle \vec i \bigoplus_{j = 1}^{i} \vec{\alpha}^{\mathcal Q_j} \bigoplus_{k = 1}^{n^i} \vec{\alpha}^{q_{k}^i} \vert S^+_e \prod_{j = 1}^{n^i} B_{q_j^i} S^+_e \vert \vec i \bigoplus_{j = 1}^{i} \vec{\alpha}^{\mathcal Q_j} \rangle  } \right) \\
 & = \prod_{i =1}^{m} \left( \frac{\langle \vec i \bigoplus_{j = 1}^{i} \vec{\alpha}^{\mathcal P_j} \bigoplus_{k = 1}^{m^i} \vec{\alpha}^{p_{k}^i} \oplus \vec{\alpha}^e \vert  \prod_{j = 1}^{m^i} B_{p_j^i}  \vert \vec i \bigoplus_{j = 1}^{i} \vec{\alpha}^{\mathcal P_j} \oplus \vec{\alpha}^e \rangle }{ \langle \vec i \bigoplus_{j = 1}^{i-1} \vec{\alpha}^{\mathcal P_j} \bigoplus_{k = 1}^{m^i} \vec{\alpha}^{p_{k}^i} \oplus \vec{\alpha}^e \vert  \prod_{j = 1}^{m^i} B_{p_j^i}  \vert \vec i \bigoplus_{j = 1}^{i-1} \vec{\alpha}^{\mathcal P_j} \oplus \vec{\alpha}^e \rangle  } \right) \\
& \times \prod_{i =1}^{n} \left( \frac{\langle \vec i \bigoplus_{j = 1}^{i-1} \vec{\alpha}^{\mathcal Q_j} \bigoplus_{k = 1}^{n^i} \vec{\alpha}^{q_{k}^i} \oplus \vec{\alpha}^e \vert \prod_{j = 1}^{n^i} B_{q_j^i} \vert \vec i \bigoplus_{j = 1}^{i-1} \vec{\alpha}^{\mathcal Q_j} \oplus \vec{\alpha}^e \rangle }{ \langle \vec i \bigoplus_{j = 1}^{i} \vec{\alpha}^{\mathcal Q_j} \bigoplus_{k = 1}^{n^i} \vec{\alpha}^{q_{k}^i} \oplus \vec{\alpha}^e \vert \prod_{j = 1}^{n^i} B_{q_j^i} \vert \vec i \bigoplus_{j = 1}^{i} \vec{\alpha}^{\mathcal Q_j} \oplus \vec{\alpha}^e \rangle } \right) \\
& \times \prod_{i = 1}^m \left( \frac{ F_e ( \vec i \bigoplus_{j = 1}^{i} \vec{\alpha}^{\mathcal P_j} \bigoplus_{k = 1}^{m^i} \vec{\alpha}^{p_{k}^i} \oplus \vec{\alpha}^e) F_e (\vec i \bigoplus_{j = 1}^{i} \vec{\alpha}^{\mathcal P_j} ) }{  F_e ( \vec i \bigoplus_{j = 1}^{i-1} \vec{\alpha}^{\mathcal P_j}) F_e (\vec i \bigoplus_{j = 1}^{i-1} \vec{\alpha}^{\mathcal P_j} \bigoplus_{k = 1}^{m^i} \vec{\alpha}^{p_{k}^i} \oplus \vec{\alpha}^e ) } \right) \\
& \times \prod_{i = 1}^n \left( \frac{F_e ( \vec i \bigoplus_{j = 1}^{i-1} \vec{\alpha}^{\mathcal Q_j} \bigoplus_{k = 1}^{n^i} \vec{\alpha}^{q_{k}^i} \oplus \vec{\alpha}^e) F_e (\vec i \bigoplus_{j = 1}^{i-1} \vec{\alpha}^{\mathcal Q_j}) }{ F_e ( \vec i \bigoplus_{j = 1}^{i} \vec{\alpha}^{\mathcal Q_j} ) F_e (\vec i \bigoplus_{j = 1}^{i} \vec{\alpha}^{\mathcal Q_j} \bigoplus_{k = 1}^{n^i} \vec{\alpha}^{q_{k}^i} \oplus \vec{\alpha}^e)   } \right).
\end{aligned}
\end{eqnarray}

Carefully looking at the right-hand side of the equality, we find that

\begin{eqnarray}
\begin{aligned}
\mathcal R_{\mathcal P_1 \# \dots \# \mathcal P_m, \mathcal Q_1 \# \dots \# \mathcal Q_n} ( \vec i ) &= \mathcal R_{\mathcal P_1 \# \dots \# \mathcal P_m, \mathcal Q_1 \# \dots \# \mathcal Q_n} ( \vec i \oplus \vec{\alpha}^e ) \\
& \times \prod_{i = 1}^m \left( \frac{ F_e ( \vec i \bigoplus_{j = 1}^{i} \vec{\alpha}^{\mathcal P_j} \bigoplus_{k = 1}^{m^i} \vec{\alpha}^{p_{k}^i} \oplus \vec{\alpha}^e) F_e (\vec i \bigoplus_{j = 1}^{i} \vec{\alpha}^{\mathcal P_j} ) }{  F_e ( \vec i \bigoplus_{j = 1}^{i-1} \vec{\alpha}^{\mathcal P_j}) F_e (\vec i \bigoplus_{j = 1}^{i-1} \vec{\alpha}^{\mathcal P_j} \bigoplus_{k = 1}^{m^i} \vec{\alpha}^{p_{k}^i} \oplus \vec{\alpha}^e ) } \right) \\
& \times \prod_{i = 1}^n \left( \frac{F_e ( \vec i \bigoplus_{j = 1}^{i-1} \vec{\alpha}^{\mathcal Q_j} \bigoplus_{k = 1}^{n^i} \vec{\alpha}^{q_{k}^i} \oplus \vec{\alpha}^e) F_e (\vec i \bigoplus_{j = 1}^{i-1} \vec{\alpha}^{\mathcal Q_j}) }{ F_e ( \vec i \bigoplus_{j = 1}^{i} \vec{\alpha}^{\mathcal Q_j} ) F_e (\vec i \bigoplus_{j = 1}^{i} \vec{\alpha}^{\mathcal Q_j} \bigoplus_{k = 1}^{n^i} \vec{\alpha}^{q_{k}^i} \oplus \vec{\alpha}^e)   } \right).
\end{aligned}
\end{eqnarray}

Note that it is always possible to choose the $ a_i $'s and the $ b_j$'s so that, in case of need, we can add some additional plaquettes in $\mathcal B_{\mathcal P_i}^{\mathcal Q, a_i}$ and $\mathcal B_{\mathcal Q_j}^{\mathcal P, b_j}$ respectively without affecting $\mathcal R_{\mathcal P_1 \# \dots \# \mathcal P_m, \mathcal Q_1 \# \dots \# \mathcal Q_n} ( \vec i )$, in order to have that on $\textnormal{Conn} (\mathcal \{e \})$, we find $\vec{\alpha}^{p^i_{1}} \oplus \dots \oplus \vec{\alpha}^{p^i_{m^i}} = \vec{\alpha}^{\mathcal Q}$ for any $i$, and $\vec{\alpha}^{q^j_1} \oplus \dots \oplus \vec{\alpha}^{q^j_{n^j}} = \vec{\alpha}^{\mathcal P}$ for any $j$. Using this, we find

\begin{eqnarray}
\begin{aligned}
& \prod_{i = 1}^m \left( \frac{ F_e ( \vec i \bigoplus_{j = 1}^{i} \vec{\alpha}^{\mathcal P_j} \bigoplus_{k = 1}^{m^i} \vec{\alpha}^{p_{k}^i} \oplus \vec{\alpha}^e) F_e (\vec i \bigoplus_{j = 1}^{i} \vec{\alpha}^{\mathcal P_j} ) }{  F_e ( \vec i \bigoplus_{j = 1}^{i-1} \vec{\alpha}^{\mathcal P_j}) F_e (\vec i \bigoplus_{j = 1}^{i-1} \vec{\alpha}^{\mathcal P_j} \bigoplus_{k = 1}^{m^i} \vec{\alpha}^{p_{k}^i} \oplus \vec{\alpha}^e ) } \right) \\
& \times \prod_{i = 1}^n \left( \frac{F_e ( \vec i \bigoplus_{j = 1}^{i-1} \vec{\alpha}^{\mathcal Q_j} \bigoplus_{k = 1}^{n^i} \vec{\alpha}^{q_{k}^i} \oplus \vec{\alpha}^e) F_e (\vec i \bigoplus_{j = 1}^{i-1} \vec{\alpha}^{\mathcal Q_j}) }{ F_e ( \vec i \bigoplus_{j = 1}^{i} \vec{\alpha}^{\mathcal Q_j} ) F_e (\vec i \bigoplus_{j = 1}^{i} \vec{\alpha}^{\mathcal Q_j} \bigoplus_{k = 1}^{n^i} \vec{\alpha}^{q_{k}^i} \oplus \vec{\alpha}^e)   } \right) \\
& = \frac{F_e (\vec i \oplus \vec{\alpha}^{\mathcal P} \oplus \vec{\alpha}^{\mathcal Q} \oplus \vec{\alpha}^e ) F_e (\vec i \oplus \vec{\alpha}^{\mathcal P})}{F_e (\vec i ) F_e (\vec i \oplus \vec{\alpha}^{\mathcal Q} \oplus \vec{\alpha}^e)}  \times \frac{F_e (\vec i \oplus \vec{\alpha}^{\mathcal P} \oplus \vec{\alpha}^e) F_e (\vec i)}{ F_e (\vec i \oplus \vec{\alpha}^{\mathcal Q} ) F_e (\vec i \oplus \vec{\alpha}^{\mathcal Q} \oplus \vec{\alpha}^{\mathcal P} \oplus \vec{\alpha}^e ) } \\
& = 1,
\end{aligned}
\end{eqnarray}
where we used that $F_e ( \vec i' \oplus \vec{\alpha}^e ) = \lbrack F_e (\vec i') \rbrack^{-1}$ for any string configuration $\vec i'$, since $S^+_{ e}$ is canonical.

\end{proof}
\end{lemma}

\begin{lemma} \label{lem:path_p_ind}
Let $\mathcal P = \mathcal P_1 \# \dots, \# \mathcal P_i \# \mathcal P_{i+1} \# \dots \# \mathcal P_m$ and $\mathcal Q = \mathcal Q_1 \# \dots \# \mathcal Q_n$ be two crossing paths. For any $i \in \{ 1, \dots, m-1 \}$, we denote $\mathcal P_i = \{ e^{\mathcal P_i}_1 , \dots, e^{\mathcal P_i}_{\vert \mathcal P_i \vert} \}$, we define $\mathcal P_i^+ = \{ e^{\mathcal P_i}_1 , \dots, e^{\mathcal P_i}_{\vert \mathcal P_i \vert}, e^{\mathcal P_{i+1}}_1\}$, ${}^+ \mathcal P_i = \{ e^{\mathcal P_{i-1}}_{\vert \mathcal P_{i-1} \vert}, e^{\mathcal P_i}_1 , \dots, e^{\mathcal P_i}_{\vert \mathcal P_i \vert}\}$,  and similarly for $\mathcal P_i^-$ and $ {}^- \mathcal P_i$, this time \emph{removing} the first or last edge, depending on the case. If the paths $\mathcal P^{\pm}_i$ and ${}^{\pm}\mathcal P_{i+1}^{\phantom{\pm}}$ are not self-overlapping nor self-crossing, we have that
\begin{eqnarray}
\begin{aligned}
\mathcal R_{\mathcal P_1 \# \dots \# \mathcal P_i \# \mathcal P_{i+1} \# \dots \# \mathcal P_m, \mathcal Q_1 \# \dots \# \mathcal Q_n} & = \mathcal R_{\mathcal P_1 \# \dots \# \mathcal P_i^+ \# {}^- \mathcal P_{i+1} \# \dots \# \mathcal P_m, \mathcal Q_1 \# \dots \# \mathcal Q_n} \\
& = \mathcal R_{\mathcal P_1 \# \dots \# \mathcal P_i^- \# {}^+ \mathcal P_{i+1} \# \dots \# \mathcal P_m, \mathcal Q_1 \# \dots \# \mathcal Q_n}.
\end{aligned}
\end{eqnarray}
\begin{proof}
Consider the edge $e^{\mathcal P_{x+1}}_{1}$ and the corresponding \emph{canonical} string operator, written  as  $S^+_e$. We then have that
\begin{eqnarray}
\begin{aligned}
\mathcal R_{\mathcal P_1 \# \dots \# \mathcal P_x \# \mathcal P_{x+1} \# \dots \mathcal P_m, \mathcal Q_1 \# \dots \# \mathcal Q_n} &= \prod_{i =1}^{m} \frac{\langle \vec i \bigoplus_{j = 1}^{i} \vec{\alpha}^{\mathcal P_j} \bigoplus_{k = 1}^{m^i} \vec{\alpha}^{p_{k}^i} \vert \lbrack S^+_e \rbrack^{\delta_{x,i}} \prod_{j = 1}^{m^i} B_{p_j^i} \lbrack S^+_e \rbrack ^{\delta_{x,i}} \vert \vec i \bigoplus_{j = 1}^{i} \vec{\alpha}^{\mathcal P_j} \rangle }{ \langle \vec i \bigoplus_{j = 1}^{i-1} \vec{\alpha}^{\mathcal P_j} \bigoplus_{k = 1}^{m^i} \vec{\alpha}^{p_{k}^i} \vert  \lbrack S^+_e \rbrack^{\delta_{x,i+1}} \prod_{j = 1}^{m^i} B_{p_j^i}  \lbrack S^+_e \rbrack^{\delta_{x,i+1}} \vert \vec i \bigoplus_{j = 1}^{i-1} \vec{\alpha}^{\mathcal P_j} \rangle  } \\
 \times & \prod_{i =1}^{n} \frac{\langle \vec i \bigoplus_{j = 1}^{i-1} \vec{\alpha}^{\mathcal Q_j} \bigoplus_{k = 1}^{n^i} \vec{\alpha}^{q_{k}^i} \vert \prod_{j = 1}^{n^i} B_{q_j^i} \vert \vec i \bigoplus_{j = 1}^{i-1} \vec{\alpha}^{\mathcal Q_j} \rangle }{ \langle \vec i \bigoplus_{j = 1}^{i} \vec{\alpha}^{\mathcal Q_j} \bigoplus_{k = 1}^{n^i} \vec{\alpha}^{q_{k}^i} \vert \prod_{j = 1}^{n^i} B_{q_j^i} \vert \vec i \bigoplus_{j = 1}^{i} \vec{\alpha}^{\mathcal Q_j} \rangle  },
\end{aligned}
\end{eqnarray}

We thus get
\begin{eqnarray}
\begin{aligned}
\mathcal R_{\mathcal P_1 \# \dots \# \mathcal P_x \# \mathcal P_{x+1} \# \dots \mathcal \# P_m, \mathcal Q_1 \# \dots \# \mathcal Q_n} = & \mathcal R_{\mathcal P_1 \# \dots \# \mathcal P_x^+ \# \mathcal {}^-P_{x+1} \# \dots \mathcal \# P_m, \mathcal Q_1 \# \dots \# \mathcal Q_n} \\
& \times \frac{F_e (\vec i \bigoplus_{j = 1}^{x} \vec{\alpha}^{\mathcal P_j}) F_e (\vec i \bigoplus_{j}^{x} \vec{\alpha}^{\mathcal P_j} \bigoplus_{k = 1}^{m^{x}} \vec{\alpha}^{p^{x}_k} \oplus \vec{\alpha}^{e} ) }{F_e (\vec i \bigoplus_{j = 1}^{x} \vec{\alpha}^{\mathcal P_j}) F_e (\vec i \bigoplus_{j}^{x} \vec{\alpha}^{\mathcal P_j} \bigoplus_{k = 1}^{m^{x+1}} \vec{\alpha}^{p^{x+1}_k} \oplus \vec{\alpha}^{e} )}.
\end{aligned}
\end{eqnarray}

As in the reasoning of the proof of Lemma \ref{lem:R_configuration_independence}, we used our freedom to add some plaquettes to $\mathcal B_{\mathcal P_x}^{\mathcal Q, a_x}$ and $\mathcal B_{\mathcal P_{x+1}}^{\mathcal Q, a_{x+1}}$, so that we have $\mathcal B_{\mathcal P^+_x}^{\mathcal Q, a_x} \subset \mathcal B_{\mathcal P_x}^{\mathcal Q, a_x}$ and $\mathcal B_{{}^{-}\mathcal P_{x+1}}^{\mathcal Q, a_{x+1}} \subset \mathcal B_{\mathcal P_{x+1}}^{\mathcal Q, a_{x+1}}$. Additionally, when can choose them such that when restricted to $\textnormal{Conn} (\mathcal \{ e \})$, we have that $\bigoplus_{k = 1}^{m^x} \vec{\alpha}^{p^{x}_k} = \bigoplus_{k = 1}^{m^{x+1}} \vec{\alpha}^{p^{x+1}_k} = \mathcal Q$. We thus find that 

\begin{equation}
\mathcal R_{\mathcal P_1 \# \dots \# \mathcal P_x \# \mathcal P_{x+1} \# \dots \mathcal P_m, \mathcal Q_1 \# \dots \# \mathcal Q_n} =  \mathcal R_{\mathcal P_1 \# \dots \# \mathcal P_x^+ \# \mathcal {}^-P_{x+1} \# \dots \# \mathcal P_m, \mathcal Q_1 \# \dots \# \mathcal Q_n}.
\end{equation}

A similar reasoning shows that

\begin{equation}
\mathcal R_{\mathcal P_1 \# \dots \# \mathcal P_x \# \mathcal P_{x+1} \# \dots \mathcal P_m, \mathcal Q_1 \# \dots \# \mathcal Q_n} = \mathcal R_{\mathcal P_1 \# \dots \# \mathcal P_x^- \# \mathcal {}^+P_{x+1} \# \dots \# \mathcal P_m, \mathcal Q_1 \# \dots \# \mathcal Q_n}.
\end{equation}

\end{proof}
\end{lemma}
\begin{corollary} \label{cor:path_cut_ind}
Let $\mathcal P = \mathcal P_1 \# \dots \# \mathcal P_m$ and $\mathcal Q = \mathcal Q_1 \# \dots \# \mathcal Q_n$ be two crossing paths. For any $i \in \{ 1, \dots, n-1\}$, we have that
\begin{eqnarray} \label{eqn:q_path_ind}
\begin{aligned}
\mathcal R_{\mathcal P_1 \# \dots \# \mathcal P_m, \mathcal Q_1 \# \dots \# \mathcal Q_i \# \mathcal Q_{i+1} \# \dots \# \mathcal Q_n} & = \mathcal R_{\mathcal P_1 \# \dots \# \mathcal P_m, \mathcal Q_1 \# \dots \# \mathcal Q_i^+ \# {}^- \mathcal Q_{i+1} \# \dots \# \mathcal Q_n} \\
&  = \mathcal R_{\mathcal P_1 \# \dots \# \mathcal P_m, \mathcal Q_1 \# \dots \# \mathcal Q_i^- \# {}^+ \mathcal Q_{i+1} \# \dots \# \mathcal Q_n}.
\end{aligned}
\end{eqnarray}
\begin{proof}
To see that Eq.\ \eqref{eqn:q_path_ind} holds, it suffices to use the same reasoning than in Lemma \ref{lem:path_p_ind}, this time for the appropriate path $\mathcal Q_x$.
\end{proof}
\end{corollary}

Given the fact that the quantity $\mathcal R_{\mathcal P_1 \# \dots \# \mathcal P_m, \mathcal Q_1 \# \dots \# \mathcal Q_n}$ is insensitive to the specific decomposition of the paths $\mathcal P$ and $\mathcal Q$ as long as they are composed of simple paths which are not self-crossing nor self-overlapping, from now on we will simply write $\mathcal R_{\mathcal P, \mathcal Q}$.

\begin{lemma} \label{lem:path_elongation}
Consider the paths $\mathcal P$, $\mathcal P' = \mathcal P_0 \# \mathcal P \# \mathcal P_{m+1}$, $\mathcal Q$ and $\mathcal Q' = \mathcal Q_0 \# \mathcal Q \# \mathcal Q_{n+1}$, such that $\mathcal P$ and $\mathcal Q$ are crossing, and such that $\mathcal P_0$, $\mathcal P_{m+1}$ do not contain edges in $\textnormal{Conn} (\mathcal Q')$, as well as $\mathcal Q_0$, $\mathcal Q_{n+1}$ do not contain edges in $\textnormal{Conn} (\mathcal P')$. Then, we have that
\begin{equation}
\mathcal R_{\mathcal P, \mathcal Q} = \mathcal R_{\mathcal P', \mathcal Q'}.
\end{equation}
\begin{proof}
We begin by considering $\mathcal R_{\mathcal P_0 \# \mathcal P, \mathcal Q}$, given by
\begin{eqnarray}
\begin{aligned}
\mathcal R_{\mathcal P_0 \# \mathcal P, \mathcal Q} = & \prod_{i = 0}^{m} \frac{\langle \vec i \bigoplus_{j = 0}^{i} \vec{\alpha}^{\mathcal P_j} \bigoplus_{k = 1}^{m^i} \vec{\alpha}^{p_{k}^i} \vert \prod_{j = 1}^{m^i} B_{p_j^i} \vert \vec i \bigoplus_{j = 0}^{i} \vec{\alpha}^{\mathcal P_j} \rangle }{ \langle \vec i \bigoplus_{j = 0}^{i-1} \vec{\alpha}^{\mathcal P_j} \bigoplus_{k = 1}^{m^i} \vec{\alpha}^{p_{k}^i} \vert \prod_{j = 1}^{m^i} B_{p_j^i} \vert \vec i \bigoplus_{j = 0}^{i-1} \vec{\alpha}^{\mathcal P_j} \rangle  } \\
 \times & \prod_{i =1}^{n} \frac{\langle \vec i \bigoplus_{j = 1}^{i-1} \vec{\alpha}^{\mathcal Q_j} \bigoplus_{k = 1}^{n^i} \vec{\alpha}^{q_{k}^i} \vert \prod_{j = 1}^{n^i} B_{q_j^i} \vert \vec i \bigoplus_{j = 1}^{i-1} \vec{\alpha}^{\mathcal Q_j} \rangle }{ \langle \vec i \bigoplus_{j = 1}^{i} \vec{\alpha}^{\mathcal Q_j} \bigoplus_{k = 1}^{n^i} \vec{\alpha}^{q_{k}^i} \vert \prod_{j = 1}^{n^i} B_{q_j^i} \vert \vec i \bigoplus_{j = 1}^{i} \vec{\alpha}^{\mathcal Q_j} \rangle  }.
\end{aligned}
\end{eqnarray}
Inserting $\lbrack S^+_{\mathcal P_0} \rbrack^2$ at various appropriate locations, with $S^+_{\mathcal P_0}$ a canonical string operator, and rearranging the terms, we find
\begin{eqnarray}
\begin{aligned}
\mathcal R_{\mathcal P_0 \# \mathcal P, \mathcal Q} = & \prod_{i = 1}^{m} \frac{\langle \vec i \oplus \vec{\alpha}^{\mathcal P_0} \bigoplus_{j = 1}^{i} \vec{\alpha}^{\mathcal P_j} \bigoplus_{k = 1}^{m^i} \vec{\alpha}^{p_{k}^i} \vert \prod_{j = 1}^{m^i} B_{p_j^i} \vert \vec i \oplus \vec{\alpha}^{\mathcal P_0} \bigoplus_{j = 1}^{i} \vec{\alpha}^{\mathcal P_j} \rangle }{ \langle \vec i \oplus \vec{\alpha}^{\mathcal P_0} \bigoplus_{j = 1}^{i-1} \vec{\alpha}^{\mathcal P_j} \bigoplus_{k = 1}^{m^i} \vec{\alpha}^{p_{k}^i} \vert \prod_{j = 1}^{m^i} B_{p_j^i} \vert \vec i \oplus \vec{\alpha}^{\mathcal P_0} \bigoplus_{j = 1}^{i-1} \vec{\alpha}^{\mathcal P_j} \rangle  } \\
 \times & \prod_{i =1}^{n} \frac{\langle \vec i \bigoplus_{j = 1}^{i-1} \vec{\alpha}^{\mathcal Q_j} \bigoplus_{k = 1}^{n^i} \vec{\alpha}^{q_{k}^i} \vert S^+_{\mathcal P_0} \prod_{j = 1}^{n^i} B_{q_j^i} S^+_{\mathcal P_0} \vert \vec i \bigoplus_{j = 1}^{i-1} \vec{\alpha}^{\mathcal Q_j} \rangle }{ \langle \vec i \bigoplus_{j = 1}^{i} \vec{\alpha}^{\mathcal Q_j} \bigoplus_{k = 1}^{n^i} \vec{\alpha}^{q_{k}^i} \vert S^+_{\mathcal P_0} \prod_{j = 1}^{n^i} B_{q_j^i} S^+_{\mathcal P_0}  \vert \vec i \bigoplus_{j = 1}^{i} \vec{\alpha}^{\mathcal Q_j} \rangle  } \\
 & \times \frac{\langle \vec i \oplus \vec{\alpha}^{\mathcal P_0} \bigoplus_{k = 1}^{m^0} \vec{\alpha}^{p_{k}^0} \vert S^+_{\mathcal P_0} \prod_{j = 1}^{m^0} B_{p_j^0} S^+_{\mathcal P_0} \vert  \vec i \oplus \vec{\alpha}^{\mathcal P_0} \rangle }{ \langle \vec i \bigoplus_{k = 1}^{m^0} \vec{\alpha}^{p_{k}^0} \vert \prod_{j = 1}^{m^0} B_{p_j^0} \vert \vec i \rangle  }.
\end{aligned}
\end{eqnarray}
Using the structure of $S^+_{\mathcal P_0}$, we find that

\begin{eqnarray}
\begin{aligned}
\mathcal R_{\mathcal P_0 \# \mathcal P, \mathcal Q} = & \prod_{i = 1}^{m} \frac{\langle \vec i \oplus \vec{\alpha}^{\mathcal P_0} \bigoplus_{j = 1}^{i} \vec{\alpha}^{\mathcal P_j} \bigoplus_{k = 1}^{m^i} \vec{\alpha}^{p_{k}^i} \vert \prod_{j = 1}^{m^i} B_{p_j^i} \vert \vec i \oplus \vec{\alpha}^{\mathcal P_0} \bigoplus_{j = 1}^{i} \vec{\alpha}^{\mathcal P_j} \rangle }{ \langle \vec i \oplus \vec{\alpha}^{\mathcal P_0} \bigoplus_{j = 1}^{i-1} \vec{\alpha}^{\mathcal P_j} \bigoplus_{k = 1}^{m^i} \vec{\alpha}^{p_{k}^i} \vert \prod_{j = 1}^{m^i} B_{p_j^i} \vert \vec i \oplus \vec{\alpha}^{\mathcal P_0} \bigoplus_{j = 1}^{i-1} \vec{\alpha}^{\mathcal P_j} \rangle  } \\
 \times & \prod_{i =1}^{n} \frac{\langle \vec i \oplus \vec{\alpha}^{\mathcal P_0} \bigoplus_{j = 1}^{i-1} \vec{\alpha}^{\mathcal Q_j} \bigoplus_{k = 1}^{n^i} \vec{\alpha}^{q_{k}^i} \vert  \prod_{j = 1}^{n^i} B_{q_j^i} \vert \vec i \oplus \vec{\alpha}^{\mathcal P_0} \bigoplus_{j = 1}^{i-1} \vec{\alpha}^{\mathcal Q_j} \rangle }{ \langle \vec i \oplus \vec{\alpha}^{\mathcal P_0} \bigoplus_{j = 1}^{i} \vec{\alpha}^{\mathcal Q_j} \bigoplus_{k = 1}^{n^i} \vec{\alpha}^{q_{k}^i} \vert  \prod_{j = 1}^{n^i} B_{q_j^i} \vert \vec i \oplus \vec{\alpha}^{\mathcal P_0} \bigoplus_{j = 1}^{i} \vec{\alpha}^{\mathcal Q_j} \rangle  } \\
 & \times \frac{ F_{\mathcal P_0} (\vec i \bigoplus_{k = 1}^{n^i} \vec{\alpha}^{q_{k}^i} \oplus \vec{\alpha}^{\mathcal P_0 } ) F_{\mathcal P_0} (\vec i ) }{ F_{\mathcal P_0} (\vec i \oplus \vec{\alpha}^{\mathcal Q} \bigoplus_{k = 1}^{n^i} \vec{\alpha}^{q_{k}^i} \oplus \vec{\alpha}^{\mathcal P_0 } ) F_{\mathcal P_0} (\vec i \oplus \vec{\alpha}^{\mathcal Q}) }  F_{\mathcal P_0} (\vec i \oplus \vec{\alpha}^{\mathcal P_0}) F_{\mathcal P_0} (\vec i \bigoplus_{k = 1}^{m^0} \vec{\alpha}^{p_{k}^0}).
\end{aligned}
\end{eqnarray}
Since we have that $\mathcal P_0 \cap \textnormal{Conn}(\mathcal Q) = \emptyset$ and since paths $\mathcal P$ and $\mathcal Q$ are crossing, we find that for any string configuration $\vec i'$, $F_{\mathcal P_0} (\vec i' \oplus \vec{\alpha}^{\mathcal Q} ) = F_{\mathcal P_0} (\vec i')$. We thus conclude that
\begin{equation}
\frac{ F_{\mathcal P_0} (\vec i \bigoplus_{k = 1}^{n^i} \vec{\alpha}^{q_{k}^i} \oplus \vec{\alpha}^{\mathcal P_0 } ) F_{\mathcal P_0} (\vec i ) }{ F_{\mathcal P_0} (\vec i \oplus \vec{\alpha}^{\mathcal Q} \bigoplus_{k = 1}^{n^i} \vec{\alpha}^{q_{k}^i} \oplus \vec{\alpha}^{\mathcal P_0 } ) F_{\mathcal P_0} (\vec i \oplus \vec{\alpha}^{\mathcal Q}) }  F_{\mathcal P_0} (\vec i \oplus \vec{\alpha}^{\mathcal P_0}) F_{\mathcal P_0} (\vec i \bigoplus_{k = 1}^{m^0} \vec{\alpha}^{p_{k}^0}) = 1.
\end{equation}

Furthermore, using Lemma \ref{lem:R_configuration_independence}, we see that 
\begin{eqnarray}
\begin{aligned}
& \prod_{i = 1}^{m} \frac{\langle \vec i \oplus \vec{\alpha}^{\mathcal P_0} \bigoplus_{j = 1}^{i} \vec{\alpha}^{\mathcal P_j} \bigoplus_{k = 1}^{m^i} \vec{\alpha}^{p_{k}^i} \vert \prod_{j = 1}^{m^i} B_{p_j^i} \vert \vec i \oplus \vec{\alpha}^{\mathcal P_0} \bigoplus_{j = 1}^{i} \vec{\alpha}^{\mathcal P_j} \rangle }{ \langle \vec i \oplus \vec{\alpha}^{\mathcal P_0} \bigoplus_{j = 1}^{i-1} \vec{\alpha}^{\mathcal P_j} \bigoplus_{k = 1}^{m^i} \vec{\alpha}^{p_{k}^i} \vert \prod_{j = 1}^{m^i} B_{p_j^i} \vert \vec i \oplus \vec{\alpha}^{\mathcal P_0} \bigoplus_{j = 1}^{i-1} \vec{\alpha}^{\mathcal P_j} \rangle  } \\
& \times \prod_{i =1}^{n} \frac{\langle \vec i \oplus \vec{\alpha}^{\mathcal P_0} \bigoplus_{j = 1}^{i-1} \vec{\alpha}^{\mathcal Q_j} \bigoplus_{k = 1}^{n^i} \vec{\alpha}^{q_{k}^i} \vert  \prod_{j = 1}^{n^i} B_{q_j^i} \vert \vec i \oplus \vec{\alpha}^{\mathcal P_0} \bigoplus_{j = 1}^{i-1} \vec{\alpha}^{\mathcal Q_j} \rangle }{ \langle \vec i \oplus \vec{\alpha}^{\mathcal P_0} \bigoplus_{j = 1}^{i} \vec{\alpha}^{\mathcal Q_j} \bigoplus_{k = 1}^{n^i} \vec{\alpha}^{q_{k}^i} \vert  \prod_{j = 1}^{n^i} B_{q_j^i} \vert \vec i \oplus \vec{\alpha}^{\mathcal P_0} \bigoplus_{j = 1}^{i} \vec{\alpha}^{\mathcal Q_j} \rangle  } \\
& = \mathcal R_{\mathcal P, \mathcal Q}.
\end{aligned}
\end{eqnarray}
A similar reasoning allows one to add the remaining paths $\mathcal P_{m+1}$, $\mathcal Q_{0}$ and $\mathcal Q_{n+1}$, to finally find that
\begin{equation}
\mathcal R_{\mathcal P', \mathcal Q'} = \mathcal R_{\mathcal P, \mathcal Q}.
\end{equation}
\end{proof}
\end{lemma}

\begin{lemma} \label{lem:path_deformation}
Let $\mathcal P = \mathcal P_1 \# \dots \# \mathcal P_m$ and $\mathcal Q = \mathcal Q_1 \# \dots \# \mathcal Q_n$ be two crossing paths. Assume further that path $\mathcal P$ is not self-crossing. Consider a set of distinct paths $\{ \mathcal P^p_{\beta} \}$ differing from path $\mathcal P$ by a plaquette $p$, i.e., $\vec{\alpha}^{\mathcal P} = \bigoplus_{\beta} \vec{\alpha}^{\mathcal P^p_{\beta}} \oplus \vec{\alpha}^{p}$, with the plaquette $p$ not containing the vertices at the endpoint of path $\mathcal Q$ and such that every $\mathcal P^p_{\beta}$ is composed of non self-overlapping nor self-crossing individual open paths. Then we have that $\mathcal R_{\mathcal P, \mathcal Q} = \prod_{\beta} \mathcal R_{\mathcal P^p_{\beta}, \mathcal Q}$.

\begin{figure}[h!]
\center
\includegraphics[scale=0.45]{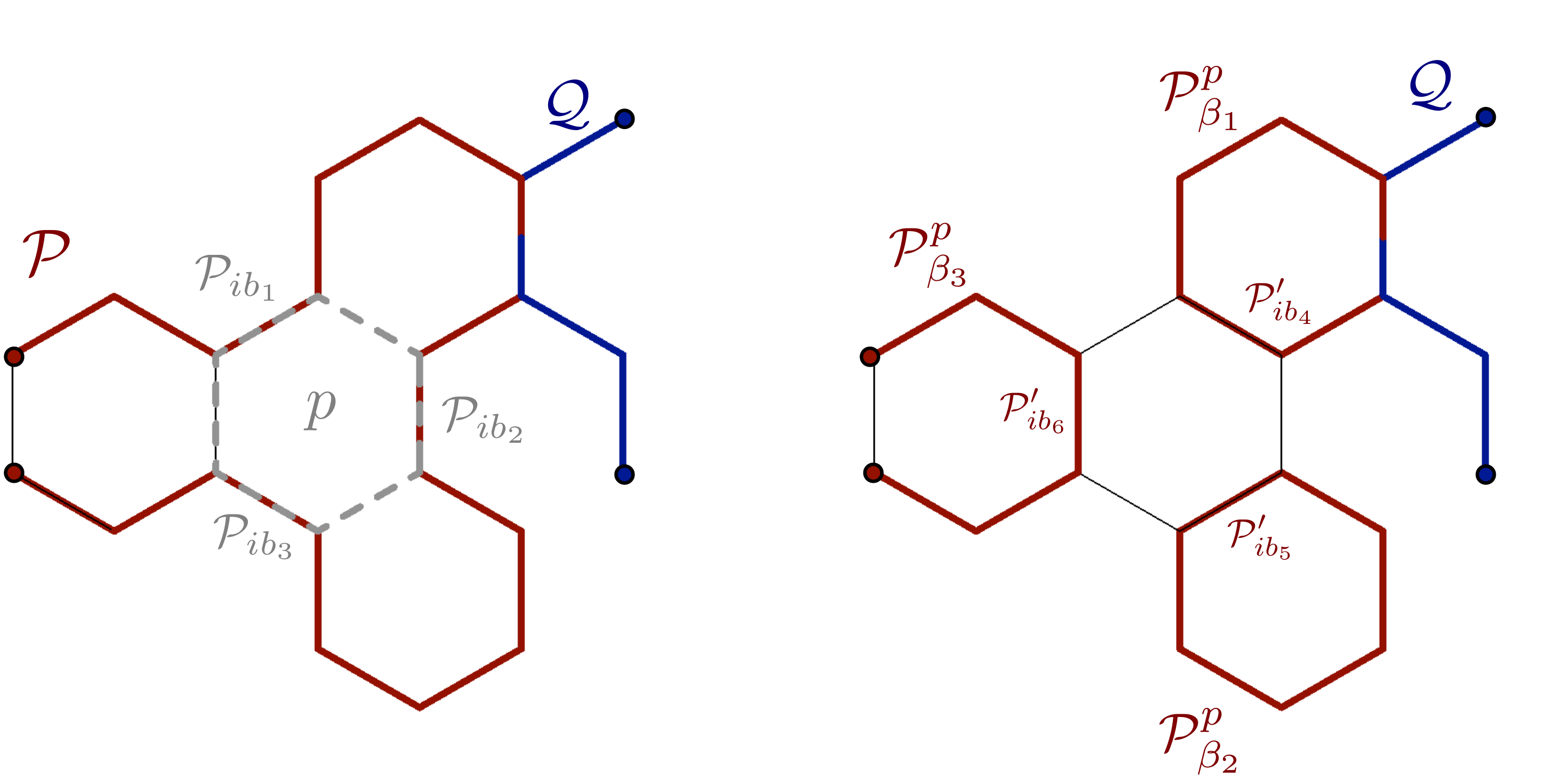}
\caption{This illustration shows a specific example on path deformation. Three distinct paths ($\{\mathcal{P}_{\beta_1},\mathcal{P}_{\beta_2},\mathcal{P}_{\beta_3}\}$, in dark red) are created by applying a plaquette $p$ (in dark blue) to the original path $\mathcal P$(dark red). $\mathcal{P}_{ib_1}, \mathcal{P}_{ib_2},\mathcal{P}_{ib_3}$ are individual paths which give the parts  of path $\mathcal P$ that overlaps with plaquette $p$. $\mathcal{P}'_{ib_4}, \mathcal{P}'_{ib_5},\mathcal{P}'_{ib_6}$ are the individual paths which complete plaquette $p$ and are contained in the newly formed paths.   }. 
\label{fig:lemma_13}
\end{figure}

\begin{proof}
First notice that since paths $\mathcal P$ and $\mathcal Q$ cross, we have that paths $\mathcal P^p_{\beta}$ and $\mathcal Q$ also cross, since the set of paths $\{ \mathcal P^p_{\beta} \}$ and $\mathcal P$ differs only by a plaquette, which cannot change its endpoints, and since the plaquette $p$ does not contain the endpoints of path $\mathcal Q$.

We can decompose the path $\mathcal P_p$ of the plaquette in a series of small paths $\mathcal P_{b_1} \# \dots \# \mathcal P_{b_k}$ with $k \leq 6$ which are not self-overlapping nor self-crossing. Furthermore, using Lemma \ref{lem:path_p_ind}, we can assume without loss of generality that the parts of path $\mathcal P$ that overlap with plaquette $p$ are given by individuals paths $\mathcal P_{i_{b_1}}, \dots, \mathcal P_{i_{b_l}}$, while the rest of the plaquette is given by individual paths $\mathcal P'_{b_{l+1}}, \dots, \mathcal P'_{b_k}$ (see Fig.\ \ref{fig:lemma_13}). Note that the various individual paths $\mathcal P_{i_{b_1}}, \dots, P_{i_{b_l}}$ (as well as $\mathcal P'_{i_{b_{l+1}}}, \dots, \mathcal P'_{i_{b_k}}$) need not be adjacent to each other. We note that in order for $\{ \mathcal P^p_{\beta} \}$ to contain more than one path, the path $\mathcal P$ must have at least two different sequences of individual paths in $\{ \mathcal P_{i_{b_{1}}}, \dots, \mathcal P_{i_{b_l}} \}$ such that they are separated by some paths in $\{ \mathcal P'_{i_{b_{l+1}}}, \dots, \mathcal P'_{i_{b_k}} \}$. Given the structure of a plaquette and of the connected region of a path, and using the fact that $\mathcal P$ is not self-crossing, we find that the resulting set of paths $\{ \mathcal P^p_{\beta} \}$ is such that any path in it does not contain any edge which is in the connected regions of the other ones.

For convenience, we denote the newly formed paths $\mathcal{P}^p_{\beta} = \mathcal P_{\beta, 1} \# \dots \# \mathcal P_{\beta, m^{\beta}}$, and the corresponding paths $\mathcal P'_{b_{l+1}}, \dots, \mathcal P'_{b_k}$ that are in $\mathcal P^p_{\beta}$ as
	$\{ \mathcal P_{\beta, i_{\beta_1}}, \dots, \mathcal P_{\beta, i_{\beta_{l'}}} \}$, appearing in order. Note that $l' \leq k - l$, and its dependance on $\beta$ has been omitted for the sake of clarity.
	
We begin by considering $\mathcal R^{+,+}_{\mathcal P, \mathcal Q}$ which is given by
\begin{eqnarray} \label{eqn:ratio_intact}
\begin{aligned}
\mathcal R_{\mathcal P, \mathcal Q} &= \prod_{ i =1 }^{m} \frac{\langle \vec i \bigoplus_{j = 1}^{i} \vec{\alpha}^{\mathcal P_j} \bigoplus_{k = 1}^{m^i} \vec{\alpha}^{p_{k}^i} \vert S^+_{\mathcal P \lbrack i \rbrack} \prod_{j = 1}^{m^i} B_{p_j^i} \lbrack S^+_{\mathcal P \lbrack i \rbrack} \rbrack^{\dagger} \vert \vec i \bigoplus_{j = 1}^{i} \vec{\alpha}^{\mathcal P_j} \rangle }{ \langle \vec i \bigoplus_{j = 1}^{i-1} \vec{\alpha}^{\mathcal P_j} \bigoplus_{k = 1}^{m^i} \vec{\alpha}^{p_{k}^i} \vert S^+_{\mathcal P \lbrack i-1 \rbrack} \prod_{j = 1}^{m^i} B_{p_j^i} \lbrack  S^+_{\mathcal P \lbrack i-1 \rbrack} \rbrack^{\dagger} \vert \vec i \bigoplus_{j = 1}^{i-1} \vec{\alpha}^{\mathcal P_j} \rangle  } \\
 	& \times \prod_{i =1}^{n} \frac{\langle \vec i \bigoplus_{j = 1}^{i-1} \vec{\alpha}^{\mathcal Q_j} \bigoplus_{k = 1}^{n^i} \vec{\alpha}^{q_{k}^i} \vert B_p \left( \prod_{j = 1}^{n^i} B_{q_j^i} B_p \right) \vert \vec i \bigoplus_{j = 1}^{i-1} \vec{\alpha}^{\mathcal Q_j} \rangle }{ \langle \vec i \bigoplus_{j = 1}^{i} \vec{\alpha}^{\mathcal Q_j} \bigoplus_{k = 1}^{n^i} \vec{\alpha}^{q_{k}^i} \vert B_p \left( \prod_{j = 1}^{n^i} B_{q_j^i} B_p \right) \vert \vec i \bigoplus_{j = 1}^{i} \vec{\alpha}^{\mathcal Q_j} \rangle  },
\end{aligned}
\end{eqnarray}
where we have defined
\begin{equation} \label{eqn:def_composite_string}
S^+_{\mathcal P \lbrack i \rbrack} =  \begin{cases}
1\quad & \textnormal{ for } \quad i < i_{b_1}, \\
S^+_{\mathcal P_{i_{b_1}} } \dots S^+_{\mathcal P_{i_{b_j}}} \quad & \textnormal{ for } \quad i_{b_j} \leq i < i_{b_{j+1}}, j < l \\
S^+_{\mathcal P_{i_{b_1}} } \dots S^+_{\mathcal P_{i_{b_l}}} \quad & \textnormal{ for } \quad i \geq i_{b_l},
\end{cases}
\end{equation}
and where $B_p$ is the plaquette operator associated to plaquette $p$.
We thus get
\begin{eqnarray}
\begin{aligned}
\mathcal R_{\mathcal P, \mathcal Q} &= \prod_{ \substack{ i =1 \\ i \notin \{ i_{b_1}, \dots, i_{b_l} \} }}^{m} \frac{\langle \vec i \bigoplus_{\substack{j = 1 \\  j \notin \{ i_{b_1}, \dots, i_{b_l} \} } }^{i} \vec{\alpha}^{\mathcal P_j} \bigoplus_{k = 1}^{m^i} \vec{\alpha}^{p_{k}^i} \vert  \prod_{j = 1}^{m^i} B_{p_j^i} \vert \vec i \bigoplus_{\substack{j = 1 \\  j \notin \{ i_{b_1}, \dots, i_{b_l} \} } }^{i} \vec{\alpha}^{\mathcal P_j} \rangle }{ \langle \vec i \bigoplus_{\substack{j = 1 \\  j \notin \{ i_{b_1}, \dots, i_{b_l} \} } }^{i-1} \vec{\alpha}^{\mathcal P_j} \bigoplus_{k = 1}^{m^i} \vec{\alpha}^{p_{k}^i} \vert \prod_{j = 1}^{m^i} B_{p_j^i} \vert \vec i \bigoplus_{\substack{j = 1 \\  j \notin \{ i_{b_1}, \dots, i_{b_l} \} } }^{i-1} \vec{\alpha}^{\mathcal P_j} \rangle  } \\
 	& \times \prod_{i =1}^{n} \frac{\langle \vec i \bigoplus_{}^{i-1} \vec{\alpha}^{\mathcal Q_j} \bigoplus_{k = 1}^{n^i} \vec{\alpha}^{q_{k}^i} \oplus \vec{\alpha}^p \vert \prod_{j = 1}^{n^i} B_{q_j^i} B_p \vert \vec i \bigoplus_{j = 1}^{i-1} \vec{\alpha}^{\mathcal Q_j} \rangle }{ \langle \vec i \bigoplus_{j = 1}^{i} \vec{\alpha}^{\mathcal Q_j} \bigoplus_{k = 1}^{n^i} \vec{\alpha}^{q_{k}^i} \oplus \vec{\alpha}^p \vert \prod_{j = 1}^{n^i} B_{q_j^i} B_p \vert \vec i \bigoplus_{j = 1}^{i} \vec{\alpha}^{\mathcal Q_j} \rangle } \\
	& \times \prod_{j =1}^{l} F_{\mathcal P_{i_{b_j}} } ( \vec i \oplus \vec{\alpha}^{\mathcal P} \oplus \vec{\alpha}^{\mathcal Q} \bigoplus_{i = 1}^{j-1} \vec{\alpha}^{\mathcal P_{i_{b_i}}}  ) F_{\mathcal P_{i_{b_j}} } (\vec i \oplus \vec{\alpha}^{\mathcal P} \bigoplus_{i = 1}^{j} \vec{\alpha}^{\mathcal P_{i_{b_i}}} ) \\
	& \times \frac{b_p (\vec i \oplus \vec{\alpha}^{\mathcal P})}{ b_p (\vec i \oplus \vec{\alpha}^{\mathcal P} \oplus \vec{\alpha}^{\mathcal Q})},
\end{aligned}
\end{eqnarray}

where we have implicitly used our freedom in choosing the sets of plaquettes $\mathcal B_{\mathcal P_i}^{\mathcal Q, a_i}$, $\mathcal B_{\mathcal Q_j}^{\mathcal P, b_j}$ and of adding plaquettes if necessary without affecting the value of $\mathcal R_{\mathcal P, \mathcal Q}$ in order to have that on $\textnormal{Conn} (\mathcal P_{i_{b_j}})$, we have that $\bigoplus_{k=1}^{m^i} \vec{\alpha}^{p^i_k} = \vec{\alpha}^{\mathcal Q}$ for any $i \in \{1, \dots, m \}$ and for any $j \in \{ 1, \dots, l \}$, and where we also have modified the set of plaquettes $\mathcal B_{\mathcal Q_j}^{\mathcal P, b_j}$ so that on $\textnormal{Conn} (\mathcal P_p)$, $\bigoplus_{k = 1}^{n^j} \vec{\alpha}^{q^j_k} = \vec{\alpha}^{\mathcal P}$ for any $j \in \{1, \dots, n \}$ and such that $\bigcup_{\beta} \mathcal B_{\mathcal Q_j}^{\mathcal P_{\beta}^p, b_{\beta, j}} = \mathcal B_{\mathcal Q_j}^{\mathcal P, b_j} \cup \{ p \}$.

We next introduce the various individual paths $\mathcal P'_{i_{b_{l+1}}}, \dots, \mathcal P'_{i_{b_k}}$ and rearrange the terms in an appropriate order so as to explicitly make appear the various $\mathcal P^p_{\beta}$'s. Note that in order to do so, we used the fact that the various paths do not have support on the other path's connected regions. We get

\begin{eqnarray} \label{eqn:adding_missing_parts}
\begin{aligned}
\mathcal R_{\mathcal P, \mathcal Q} = & \prod_{\beta} \left( \prod_{i =1}^{m^{\beta}} \frac{\langle \vec i \bigoplus_{\substack{j = 1 } }^{i} \vec{\alpha}^{\mathcal P^p_{\beta, j}} \bigoplus_{k = 1}^{m^{\beta, i}} \vec{\alpha}^{p_{k}^{\beta, i}} \vert  S^+_{\mathcal P^p_{\beta} \lbrack i \rbrack } \prod_{j = 1}^{m^{\beta, i}} B_{p_j^{\beta, i}} \lbrack S^+_{\mathcal P^p_{\beta} \lbrack i \rbrack } \rbrack^{\dagger} \vert \vec i \bigoplus_{\substack{j = 1 } }^{i} \vec{\alpha}^{\mathcal P^p_{\beta, j}} \rangle }{ \langle \vec i \bigoplus_{\substack{j = 1 } }^{i-1} \vec{\alpha}^{\mathcal P^p_{\beta, j}} \bigoplus_{k = 1}^{m^{\beta, i}} \vec{\alpha}^{p_{k}^{\beta, i}} \vert S^+_{\mathcal P^p_{\beta} \lbrack i-1 \rbrack } \prod_{j = 1}^{m^i} B_{p_j^{\beta, i}} \lbrack S^+_{\mathcal P^p_{\beta} \lbrack i-1 \rbrack } \rbrack^{\dagger} \vert \vec i \bigoplus_{\substack{j = 1 } }^{i-1} \vec{\alpha}^{\mathcal P^p_{\beta, j}} \rangle  } \right) \\
 	& \times \prod_{\beta} \left( \prod_{i =1}^{n^{\beta}} \frac{\langle \vec i \bigoplus_{}^{i-1} \vec{\alpha}^{\mathcal Q_j} \bigoplus_{k = 1}^{n^{\beta, i}} \vec{\alpha}^{q_{k}^{\beta,i}} \vert \prod_{j = 1}^{n^{\beta, i}} B_{q_j^{\beta, i}} \vert \vec i \bigoplus_{j = 1}^{i-1} \vec{\alpha}^{\mathcal Q_j} \rangle }{ \langle \vec i \bigoplus_{j = 1}^{i} \vec{\alpha}^{\mathcal Q_j} \bigoplus_{k = 1}^{n^{\beta, i}} \vec{\alpha}^{q_{k}^{\beta, i}} \vert \prod_{j = 1}^{n^{\beta, i}} B_{q_j^{\beta, i}} \vert \vec i \bigoplus_{j = 1}^{i} \vec{\alpha}^{\mathcal Q_j} \rangle } \right) \\
	& \times \left( \prod_{j =1}^{l} \frac{ F_{\mathcal P_{i_{b_j}} } ( \vec i \oplus \vec{\alpha}^{\mathcal P} \oplus \vec{\alpha}^{\mathcal Q} \bigoplus_{i = 1}^{j-1} \vec{\alpha}^{\mathcal P_{i_{b_i}}} )}{ F_{\mathcal P_{i_{b_j}} } (\vec i \oplus \vec{\alpha}^{\mathcal P} \bigoplus_{i = 1}^{j-1} \vec{\alpha}^{\mathcal P_{i_{b_i}}})}  \right) \frac{b_p (\vec i \oplus \vec{\alpha}^{\mathcal P})}{ b_p (\vec i \oplus \vec{\alpha}^{\mathcal P} \oplus \vec{\alpha}^{\mathcal Q})},
\end{aligned}
\end{eqnarray}
where we have defined
\begin{equation}
S^+_{\mathcal P^p_{\beta} \lbrack i \rbrack } = \begin{cases}
1 \quad & \textnormal{ for } i < i_{\beta_1} \\
S^+_{\mathcal P_{\beta, i_{\beta_1}}} \dots S^+_{\mathcal P_{\beta, i_{\beta_j}}} \quad & \textnormal{ for } i_{\beta_j} \leq i < i_{\beta_{j+1}}, j < l' \\
S^+_{\mathcal P_{\beta, i_{\beta_1}}} \dots S^+_{\mathcal P_{\beta, i_{\beta_l}}} \quad & \textnormal{ for } i \geq i_{\beta_{l'}},
\end{cases}
\end{equation}
and where we have $\{ p_j^{\beta, i }\} = \mathcal B_{\mathcal P_{\beta, i}}^{\mathcal Q, a_{\beta, i}}$, $\{ q_j^{\beta, i}\} = \mathcal B_{\mathcal Q}^{\mathcal P_{\beta, i},b_{\beta, i}}$.

Under close inspection, it thus becomes clear that
\begin{eqnarray} \label{eqn:additionnal_phase}
\begin{aligned}
\mathcal R_{\mathcal P, \mathcal Q} =& \prod_{\beta} \mathcal R_{\mathcal P^p_{\beta}} \times \left( \prod_{j =1}^{k} \frac{ F_{\mathcal P_{i_{b_j}} } ( \vec i \oplus \vec{\alpha}^{\mathcal P} \oplus \vec{\alpha}^{\mathcal Q} \bigoplus_{i = 1}^{j-1} \vec{\alpha}^{\mathcal P_{i_{b_i}}} )}{ F_{\mathcal P_{i_{b_j}} } (\vec i \oplus \vec{\alpha}^{\mathcal P} \bigoplus_{i = 1}^{j-1} \vec{\alpha}^{\mathcal P_{i_{b_i}}})}\right) \frac{b_p (\vec i \oplus \vec{\alpha}^{\mathcal P})}{ b_p (\vec i \oplus \vec{\alpha}^{\mathcal P} \oplus \vec{\alpha}^{\mathcal Q})}.
\end{aligned}
\end{eqnarray}

We first notice that since all the paths $\mathcal P^p_{\beta}$ and $\mathcal P$ cross with path $\mathcal Q$, we can define $\mathcal B_{\mathcal P_p}^{\mathcal Q} = \{p_{\mathcal Q, 1}, \dots, p_{\mathcal Q, o}\}$, a set of plaquettes in $\mathcal B_{\mathcal P_p}$ such that on $\textnormal{Conn} (\mathcal P_b)$, we have that $\bigoplus_{i = 1}^o \vec{\alpha}^{p_{\mathcal Q, i}} = \vec{\alpha}^{\mathcal Q}$. Using the same reasoning as in Lemma \ref{lem:R_form}, we find that
\begin{equation} \label{eqn:phase_canceling_1}
\frac{b_p ( \vec i \oplus \vec{\alpha}^{\mathcal P}) }{b_p ( \vec i \oplus \vec{\alpha}^{\mathcal P} \oplus \vec{\alpha}^{\mathcal Q} ) } = \frac {\langle \vec i \oplus \vec{\alpha}^{\mathcal P} \bigoplus_{i = 1}^{o} \vec{\alpha}^{p_{\mathcal Q, i}} \vert \prod_{i = 1}^o B_{p_{\mathcal Q, i}} \vert \vec i \oplus \vec{\alpha}^{\mathcal P} \rangle } {\langle \vec i \oplus \vec{\alpha}^{\mathcal P} \oplus \vec{\alpha}^{\mathcal P_p} \bigoplus_{i = 1}^{o} \vec{\alpha}^{p_{\mathcal Q, i}} \vert \prod_{i = 1}^o B_{p_{\mathcal Q, i}} \vert \vec i \oplus \vec{\alpha}^{\mathcal P} \oplus \vec{\alpha}^{\mathcal P_p} \rangle }.
\end{equation}
On the other hand, we find that
\begin{eqnarray} \label{eqn:phase_canceling_2}
\begin{aligned}
\prod_{j =1}^{k} & \frac{ F_{\mathcal P_{i_{b_j}} } ( \vec i \oplus \vec{\alpha}^{\mathcal P} \oplus \vec{\alpha}^{\mathcal Q} \bigoplus_{i = 1}^{j-1} \vec{\alpha}^{\mathcal P_{i_{b_i}}} )}{ F_{\mathcal P_{i_{b_j}} } (\vec i \oplus \vec{\alpha}^{\mathcal P} \bigoplus_{i = 1}^{j-1} \vec{\alpha}^{\mathcal P_{i_{b_i}}})} = \\
& \prod_{j = 1}^{k} \frac {\langle \vec i \oplus \vec{\alpha}^{\mathcal P} \bigoplus_{i = 1}^{o} \vec{\alpha}^{p_{\mathcal Q, i}} \bigoplus_{i = 1}^{j} \vec{\alpha}^{\mathcal P_{i_{b_i}}} \vert \prod_{i = 1}^o B_{p_{\mathcal Q, i}} \vert \vec i \oplus \vec{\alpha}^{\mathcal P} \bigoplus_{i = 1}^{j} \vec{\alpha}^{\mathcal P_{i_{b_i}}} \rangle } {\langle \vec i \oplus \vec{\alpha}^{\mathcal P} \bigoplus_{i = 1}^{o} \vec{\alpha}^{p_{\mathcal Q, i}} \bigoplus_{i = 1}^{j-1} \vec{\alpha}^{\mathcal P_{i_{b_i}}} \vert \prod_{i = 1}^o B_{p_{\mathcal Q, i}} \vert \vec i \oplus \vec{\alpha}^{\mathcal P} \bigoplus_{i = 1}^{j-1} \vec{\alpha}^{\mathcal P_{i_{b_i}}} \rangle } \\
& = \frac{\langle \vec i \oplus \vec{\alpha}^{\mathcal P} \oplus \vec{\alpha}^{\mathcal P_p} \bigoplus_{i = 1}^{o} \vec{\alpha}^{p_{\mathcal Q, i}} \vert \prod_{i = 1}^o B_{p_{\mathcal Q, i}} \vert \vec i \oplus \vec{\alpha}^{\mathcal P} \oplus \vec{\alpha}^{\mathcal P_p} \rangle } {\langle \vec i \oplus \vec{\alpha}^{\mathcal P} \bigoplus_{i = 1}^{o} \vec{\alpha}^{p_{\mathcal Q, i}} \vert \prod_{i = 1}^o B_{p_{\mathcal Q, i}} \vert \vec i \oplus \vec{\alpha}^{\mathcal P} \rangle }.
\end{aligned}
\end{eqnarray}
Putting Eqs.\ \eqref{eqn:additionnal_phase}, \eqref{eqn:phase_canceling_1} and \eqref{eqn:phase_canceling_2} together, we conclude that

\begin{equation}
\mathcal R_{\mathcal P, \mathcal Q} = \prod_{\beta} \mathcal R_{\mathcal P^p_{\beta}}.
\end{equation}
To finally conclude the proof, we remark that according to Lemma \ref{lem:path_p_ind} we are free to modify the composition in terms of individual paths of the various $\{ \mathcal P_{\beta}^p \}$'s as long as the their individual paths remain non self-overlapping nor self-crossing.
\end{proof}
\end{lemma}
\begin{lemma} \label{lem:even_crossing}
Let $\mathcal P$ and $\mathcal Q$ be two crossing paths such that each of them is made of simple non self-overlapping nor self-crossing open paths, and such that $\mathcal P$ is not self-crossing. If they cross an \emph{even} number of times, then $\mathcal R_{\mathcal P, \mathcal Q} = 1$.

\begin{proof}
The idea of the proof is to deform the path $\mathcal P$ using the results of Lemma \ref{lem:path_deformation}, as well as path $\mathcal Q$ by some elongation and reductions using Lemma \ref{lem:path_elongation}, in order to get a set of paths $\{ \mathcal P_{\beta} \}$ and path $\mathcal Q'$ such that $\mathcal R_{\mathcal P, \mathcal Q} = \prod_{\beta} \mathcal R_{\mathcal P_{\beta}, \mathcal Q'}$, and such that all of them are outside of $\textnormal{Conn} (\mathcal Q')$, thus implying that $R_{\mathcal P_{\beta}, \mathcal Q} = 1$.

We first note that if $\mathcal P$ and $\mathcal Q$ do not have any edges in common, then they trivially commute, since by supposition they are crossing each other. This implies that $\mathcal R_{\mathcal P, \mathcal Q} = 1$.

Consider the case where $\mathcal P$ and $\mathcal Q$ have some edges in common. Suppose first that $\mathcal P$ and $\mathcal Q$ have a single contiguous set of common edges, $\mathcal E$. In that case, the path $\mathcal P$ can be sequentially deformed along the set of contiguous plaquettes in $\textnormal{Conn} (\mathcal Q)$ containing the edges in $\mathcal E$ as well as the two edges in $\mathcal P \backslash \mathcal E$ sharing vertices with the edges in $\mathcal E$ in order to give a new sets of paths $\{ \mathcal P_{\beta} \}$. Note that since $\mathcal P$ and $\mathcal Q$ cross $0$ times, we can assume that none of the paths in $\{ \mathcal P_{\beta}\}$ contains edges in $\textnormal{Conn} (\mathcal Q)$. If it is not the case, then we can use Lemma \ref{lem:path_elongation} to first find a shorter path $\mathcal Q'$ such that $\mathcal R_{\mathcal P, \mathcal Q} = \mathcal R_{\mathcal P, \mathcal Q'}$ and for which it is true. By suitably choosing a decomposition of those plaquettes such that they are non self-overlapping nor self-crossing, we can use Lemma \ref{lem:path_deformation}, to find that $\mathcal R_{\mathcal P, \mathcal Q} = \prod_{\beta} \mathcal R_{\mathcal P_{\beta}, \mathcal Q} = 1$.

\begin{figure}[h!]
\center
\includegraphics[scale=0.25]{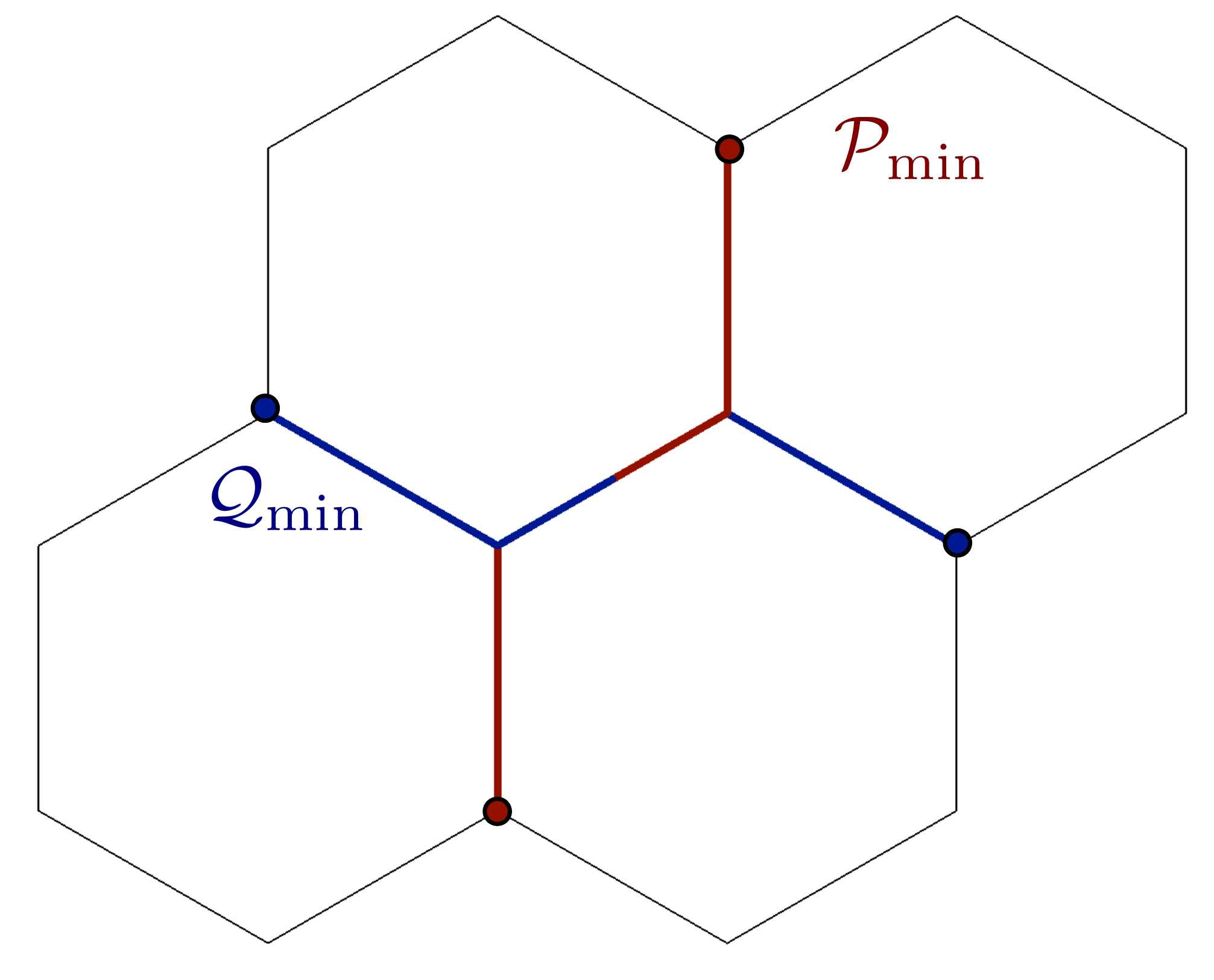}
\caption{The quantity $\mathcal R_{\mathcal P, \mathcal Q}$ for $2$ paths $\mathcal P$ and $\mathcal Q$ crossing an odd number of times is equal to $\mathcal R_{\mathcal P_{\textnormal{min}}, \mathcal Q_{\textnormal{min}}}$, for the minimal paths $\mathcal P_{\textnormal{min}}$ and $\mathcal Q_{\textnormal{min}}$ shown above.}
\label{fig:ratio_computation}
\end{figure}
Suppose next that $\mathcal P$ and $\mathcal Q$ have two or more different contiguous sets of common edges $\{ \mathcal E_i \}$. In that case, we can deform $\mathcal P$ by a subset of plaquettes in $\mathcal B_{\mathcal Q}$ so as to form a single contiguous set of common edges $\mathcal E$ such that $\mathcal E \cup \delta \mathcal E \supset \bigcup_{i} \mathcal E_i$, where $\delta \mathcal E$ denotes the first and last edges in $\mathcal Q$ which are also in $\mathcal E$, and such that all newly formed paths $\{ \mathcal P_{\beta} \}$ cross $\mathcal Q$, where we have again chosen a suitable decomposition of the paths along the plaquettes. Again using Lemma \ref{lem:path_deformation} and considering a shortened path $\mathcal Q'$ using Lemma \ref{lem:path_elongation} if necessary, we find that $\mathcal R_{\mathcal P, \mathcal Q} = \prod_{\beta} \mathcal R_{\mathcal P_{\beta}, \mathcal Q'}$, where one of the $\mathcal P_{\beta}$'s have a single set of edges ($\mathcal E$) in common with $\mathcal Q'$, and where the other paths cross $0$ times with $\mathcal Q'$. Since modifying the path $\mathcal P$ by a set of plaquettes cannot change the parity of the number of crossing, we have that the former path crosses $0$ times with $\mathcal Q'$ as well. By the previous reasoning, we thus find that $\mathcal R_{\mathcal P, \mathcal Q} = 1$.

\end{proof}
\end{lemma}

\begin{lemma} \label{lem:odd_crossing}
Let $\mathcal P$ and $\mathcal Q$ be two crossing paths such that each of them is made of simple non self-overlapping nor self-crossing open paths, and such that $\mathcal P$ is not self-crossing. If they cross an \emph{odd} number of times, then $\mathcal R_{\mathcal P, \mathcal Q} = -1$.
\begin{proof}
The idea of the proof closely follows that of Lemma \ref{lem:even_crossing}. We begin by sequentially deforming the path $\mathcal P$ into a set of paths $\{ \mathcal P_{\beta} \}$ such that all of them are crossing with path $\mathcal Q$, and such that there is a single set of contiguous edges between one of the path $\mathcal P_{o} \in \{ \mathcal P_{\beta} \}$ and $\mathcal Q$, and such that $\mathcal R_{\mathcal P, \mathcal Q} = \prod_ {\beta} \mathcal R_{\mathcal P_{\beta}, \mathcal Q'}$, where $\mathcal Q'$ may be a shortened path of $\mathcal Q$, as described in the proof of Lemma \ref{lem:even_crossing}. Since any path $\mathcal P_{\beta} \neq \mathcal P_o$ crosses path $\mathcal Q'$ $0$ times, we have that $\mathcal R_{\mathcal P_{\beta}, \mathcal Q} = 1$ for $\beta \neq o$. We thus have that $\mathcal R_{\mathcal P, \mathcal Q} = \mathcal R_{\mathcal P_o, \mathcal Q'}$.

	In order to calculate $\mathcal R_{\mathcal P_o, \mathcal Q'}$, we note that we can first sequentially deform path $\mathcal P_o$ as described previously so that there is a single common edge between the two paths, and we can use Lemma \ref{lem:path_elongation} to bring the endpoints of the two paths as close as possible in order to minimize the length of the paths. We thus find that computing $\mathcal R_{\mathcal P, \mathcal Q}$ reduces to computing this quantity for a single minimal path configurations $\mathcal P_{\textnormal{min}}$ and $\mathcal Q_{\textnormal{min}}$ illustrated in Fig.\ \ref{fig:ratio_computation}. Explicit calculations using Eq.\ \eqref{eqn:R_ratio_def} for a single underlying string configuration (Lemma \ref{lem:R_configuration_independence} ensures that its value is independent of the configuration), gives that $\mathcal R_{\mathcal P_{\textnormal{min}}, \mathcal Q_{\textnormal{min}}} = -1$.

\end{proof}
\end{lemma}

\subsection{Proof of Theorem \ref{thm:crossing_strings}} \label{appsub:proof_thm}
Having introduced all previous technical lemmas, we are in a position to complete the demonstration of Theorem~\ref{thm:crossing_strings}.
 \begin{reptheorem}{thm:crossing_strings}
Let $\mathcal P$ and $\mathcal Q$ be two paths crossing $n$ times, composed of non self-overlapping nor self-crossing individual open paths. We have that
\begin{eqnarray}
\begin{aligned}
\lbrack S^+_{\mathcal P}, S^+_{\mathcal Q} \rbrack &= 0 \textnormal{ if } n \textnormal{ is even}, \\
\{ S^+_{\mathcal P}, S^+_{\mathcal Q} \} &= 0 \textnormal{ if } n \textnormal{ is odd}.
\end{aligned}
\end{eqnarray}
\begin{proof}
Consider first the case where path $\mathcal P$ is not self-crossing. Given the  definition in Eq.\ \eqref{eqn:R_ratio_def} of $\mathcal R_{\mathcal P, \mathcal Q}$, Lemma \ref{lem:even_crossing} shows that $\lbrack S^+_{\mathcal P}, S^+_{\mathcal Q} \rbrack = 0$ if $n$ is even, while Lemma \ref{lem:odd_crossing} shows that $\{ S^+_{\mathcal P}, S^+_{\mathcal Q} \} = 0$ is $n$ is odd.

Consider next the case where $\mathcal P$ is self-crossing. Using Lemma~\ref{lem:path_p_ind}, we can always find paths $\mathcal P_1, \dots, \mathcal P_m$ for which $\mathcal P = \mathcal P_1 \# \dots \# \mathcal P_m$, such that none of those are self-crossing, with $\mathcal R_{\mathcal P_1 \# \dots \# \mathcal P_m, \mathcal Q} = \mathcal R_{\mathcal P, \mathcal Q}$. Suppose that all those paths cross with $\mathcal Q$. Since $S^+_{\mathcal P_1 \# \dots \# \mathcal P_m} = S^+_{\mathcal P_m} \dots S^+_{\mathcal P_1}$, it suffices to know the commutation relations between every of the operators $S^+_{\mathcal P_1}, \dots, S^+_{\mathcal P_m}$ and $S^+_{\mathcal Q}$. Since none of the corresponding paths are self-overlapping, the reasoning of the above paragraph can be used to find the same result.

It may be impossible to decompose path $\mathcal P = \mathcal P_1 \# \dots \# \mathcal P_{m}$ such that all of its components cross with $\mathcal Q$. This happens only in the case where some edges in path $\mathcal P$ appear at more than one position. In that case, suppose for simplicity that path $\mathcal P = \mathcal P_1 \# \dots \# \mathcal P_{l-1} \# \mathcal P_{l} \# \mathcal P_{l+1} \# \dots \# \mathcal P_{o-1} \# \mathcal P_o \# \mathcal P_{o+1} \# \dots \# \mathcal P_m$ can be decomposed such that pahts $\mathcal P_{l} = \mathcal P_{o}$  are the only ones with edges in common (possibly in reversed order). The following reasoning works in the same way if there are more than a single pair of such paths. Consider the quantity

\begin{eqnarray} \label{eqn:R_overlap_ind}
\begin{aligned}
& \prod_{ i =1 }^{m} \frac{\langle \vec i \bigoplus_{j = 1}^{i} \vec{\alpha}^{\mathcal P_j} \bigoplus_{k = 1}^{m^i} \vec{\alpha}^{p_{k}^i} \vert S^+_{\mathcal P \lbrack i \rbrack} \prod_{j = 1}^{m^i} B_{p_j^i} S^+_{\mathcal P \lbrack i \rbrack}  \vert \vec i \bigoplus_{j = 1}^{i} \vec{\alpha}^{\mathcal P_j} \rangle }{ \langle \vec i \bigoplus_{j = 1}^{i-1} \vec{\alpha}^{\mathcal P_j} \bigoplus_{k = 1}^{m^i} \vec{\alpha}^{p_{k}^i} \vert S^+_{\mathcal P \lbrack i-1 \rbrack} \prod_{j = 1}^{m^i} B_{p_j^i} S^+_{\mathcal P \lbrack i-1 \rbrack}  \vert \vec i \bigoplus_{j = 1}^{i-1} \vec{\alpha}^{\mathcal P_j} \rangle  } \\
& =  \prod_{\substack{ i =1 \\ i \neq l, o }}^{m} \frac{\langle \vec i \bigoplus_{j = 1}^{i} \vec{\alpha}^{\mathcal P_j} \bigoplus_{k = 1}^{m^i} \vec{\alpha}^{p_{k}^i} \vert \prod_{j = 1}^{m^i} B_{p_j^i} \vert \vec i \bigoplus_{j = 1}^{i} \vec{\alpha}^{\mathcal P_j} \rangle }{ \langle \vec i \bigoplus_{j = 1}^{i-1} \vec{\alpha}^{\mathcal P_j} \bigoplus_{k = 1}^{m^i} \vec{\alpha}^{p_{k}^i} \vert \prod_{j = 1}^{m^i} B_{p_j^i} \vert \vec i \bigoplus_{j = 1}^{i-1} \vec{\alpha}^{\mathcal P_j} \rangle  } \\
& \times \prod_{i = l}^{o-1} \frac{ F_{\mathcal P_l} (\vec i \bigoplus_{j = 1}^{i} \alpha^{\mathcal P_j} \bigoplus_{k = 1}^{m^i} \vec{\alpha}^{p_{k}^i} \oplus \vec{\alpha}^{\mathcal P_l} )}{ F_{\mathcal P_l} (\vec i \bigoplus_{j = 1}^{i} \alpha^{\mathcal P_j} \bigoplus_{k = 1}^{m^{i+1}} \vec{\alpha}^{p_{k}^{i+1}} \oplus \vec{\alpha}^{\mathcal P_l} ) },
\end{aligned}
\end{eqnarray}
where
\begin{equation}
S^+_{\mathcal P \lbrack i \rbrack}  = \begin{cases}
S^+_{\mathcal P_l} & \textnormal{ if } l \leq i < o \\
1 & \textnormal{ otherwise.}
\end{cases}
\end{equation}
Using Lemma~\ref{lem:path_p_ind}, for any $i \in \{ l, \dots, o \}$ it is always possible to find a path decomposition and sets of plaquettes $\mathcal B_{\mathcal P_i}^{\mathcal Q, a_i}$ such that on $\textnormal{Conn} (\mathcal P_l)$, $ \bigoplus_{k = 1}^{m^{i}} \vec{\alpha}^{p_{k}^i} = \vec{\alpha}^{\mathcal Q} $. This can simply be achieved by taking individual paths of lenght $1$ in the decomposition of path $\mathcal P$. We thus find that

\begin{equation}
 \prod_{i = l}^{o-1} \frac{ F_{\mathcal P_l} (\vec i \bigoplus_{j = 1}^{i} \alpha^{\mathcal P_j} \bigoplus_{k = 1}^{m^i} \vec{\alpha}^{p_{k}^i} \oplus \vec{\alpha}^{\mathcal P_l} )}{ F_{\mathcal P_l} (\vec i \bigoplus_{j = 1}^{i} \alpha^{\mathcal P_j} \bigoplus_{k = 1}^{m^{i+1}} \vec{\alpha}^{p_{k}^{i+1}} \oplus \vec{\alpha}^{\mathcal P_l} ) } = 1,
\end{equation}
which, given Equation~\eqref{eqn:ratio_form}, leads us to the conclusion that
\begin{equation}
\mathcal R_{\mathcal P_1 \# \dots \# \mathcal P_{l-1} \# \mathcal P_l \# \mathcal P_{l+1} \# \dots \# \mathcal P_{o-1} \# \mathcal P_{o} \# \mathcal P_{o+1} \# \dots \# \mathcal P_m, \mathcal Q } = \mathcal R_{\mathcal P_1 \# \dots \# \mathcal P_{l-1} \# \mathcal P_{l+1} \# \dots \# \mathcal P_{o-1} \# \mathcal P_{o+1} \# \dots \# \mathcal P_m, \mathcal Q }.
\end{equation}
To complete the proof, it suffices to notice that we can now apply the reasoning of the first two paragraphs of this proof.


\end{proof}
\end{reptheorem}

\twocolumngrid

\subsection{The need for path concatenation}\label{Sec:app_initial_phases}
Each time Algorithm \ref{alg:F_finding} picks a configuration representative $\vec i$, an initial phase must be chosen. All of these choices yield valid string operators. One may wonder what is the physical difference between different choices. To answer this question, note that two string operators $S^+_{\mathcal P}$ and $\tilde{S}^+_{\mathcal P}$ obtained by a different choice of phases in Algorithm~\ref{alg:F_finding} are related through
\begin{equation} \label{eqn:string_op_diff_phases}
\tilde{S}^+_{\mathcal P} = S^+_{\mathcal P} \sum_{c \in \mathcal C_{\mathcal P}} e^{i \theta_c} P_c,
\end{equation}
where $ \mathcal C_{\mathcal P}$ denotes the set of all configuration classes of path $\mathcal P$,  $P_c = \sum_{\vec i \in c} \vert \vec i \rangle \langle \vec i \vert$ is the projector on the states of configuration class $c$, and $e^{i \theta_c}$ is an independent arbitrary complex phase for every class configuration.

Equivalently, given a string operator $S^{+}_{\mathcal P}$, we can obtain another string operator $\tilde{S}^{+}_{\mathcal P}$ multiplying $S^{+}_{\mathcal P}$ by an operator $S^z_{\mathcal{C}_{\textnormal{dual}}}$, where $\mathcal{C}_{\textnormal{dual}}$ is a closed loop in the dual lattice affecting only qubits in Conn($\mathcal{P}$). Since $S^z_{\mathcal{C}_{\textnormal{dual}}}$ is a loop, it may be expressed as a multiplication of vertex operators (unless it is a non-trivial loop, which may happen for closed strings). Therefore $\tilde{S}^{+}_{\mathcal P}$ still commutes with all plaquette and vertex operators and it is contained in Conn($\mathcal{P}$), satisfying properties \ref{property_1} and \ref{property_2}. It is also possible to obtain another string operator multiplying $S^{+}_{\mathcal P}$ by a linear combination of $S^z$ operators on closed loops, i.e., 
\begin{equation} \label{eq:equivalence_strings_z}
\tilde{S}^{+}_{\mathcal P} = S^{+}_{\mathcal P} \left(\mathbb{I} + c\left( \vec{\alpha}^{\mathcal{C}_1}\right) S^z_{\mathcal{C}_1}+ c\left( \vec{\alpha}^{\mathcal{C}_2}\right) S^z_{\mathcal{C}_2}+...\right),
\end{equation}
where $c(\vec{\alpha}^\mathcal{C}_i)$ are coefficients associated with the closed-string operator $S^z_{\mathcal{C}}$ and $\mathcal{C}_1$, ..., $\mathcal{C}_n$ are closed paths in the dual lattice contained in Conn($\mathcal{P}$). For any two strings generated by Algorithm \ref{alg:F_finding} differing only in the choice of initial phases, we can always find a relation of the form given by Eq. \eqref{eq:equivalence_strings_z}.

\begin{figure}
\center
\includegraphics[scale=0.25]{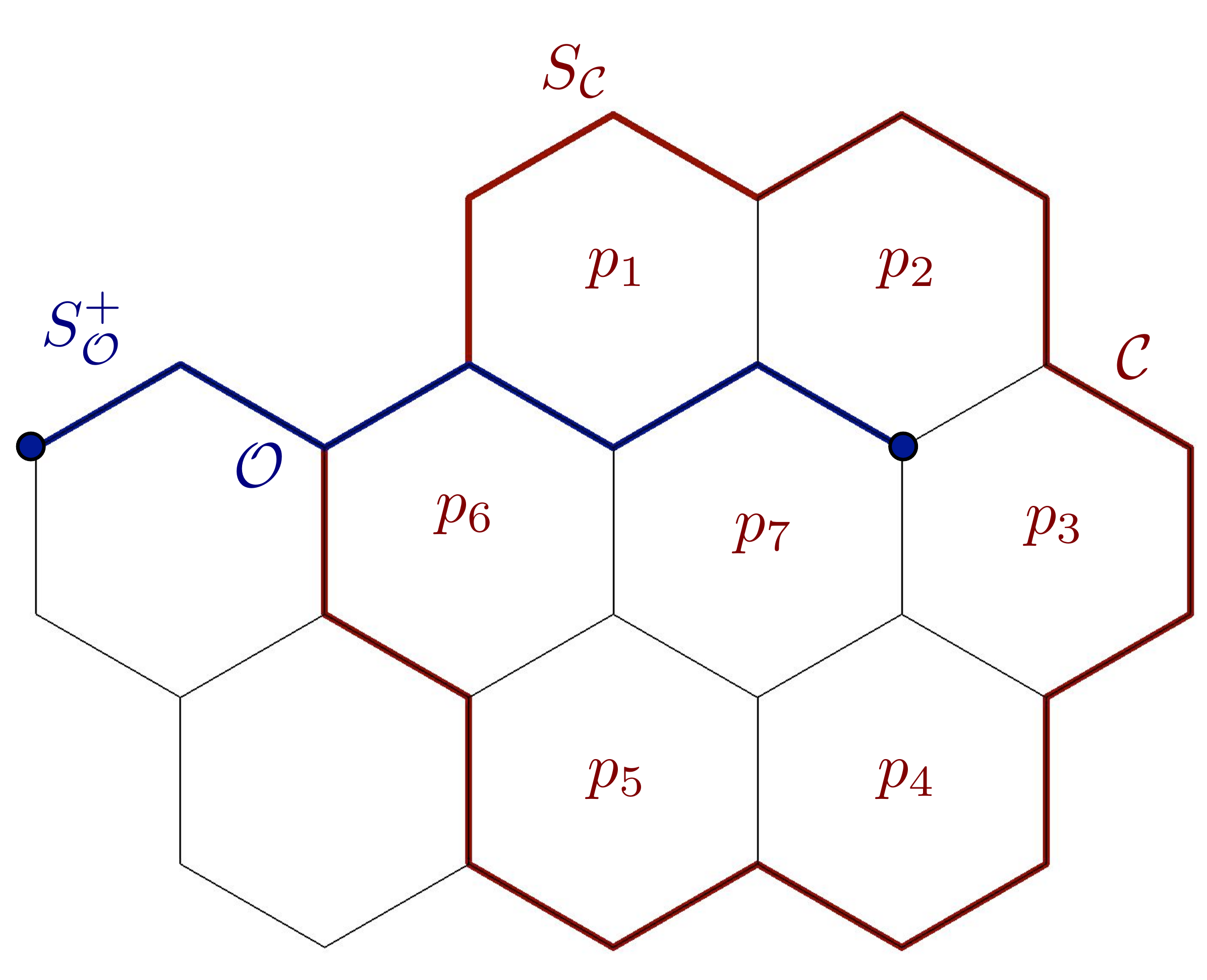}
\caption{A closed string, $S_{\mathcal{C}}$, and a positive-chirality open string, $S^+_{\mathcal{O}}$, where paths $\mathcal{O}$ and $\mathcal{C}$ cross once. $\vec \alpha^\mathcal{C}$ can be written in this particular example as  $\vec \alpha^\mathcal{C} = \vec \alpha^{p_1} \oplus \dots \oplus \vec \alpha^{p_6}$, restricted to Conn($\mathcal{C}$) . The multiplication of plaquettes inside $\mathcal C$, i.e., $B_{p_1}... B_{p_7}$ is a negative-chirality string in Conn($\mathcal{C}$).}
\label{fig:string_open_closed}
\end{figure}

\subsubsection{Closed-string operators} 
If we consider now a closed path, $\mathcal{C}$, and we use Algorithm \ref{alg:F_finding} to find a closed-string operator, we will not find in general a positive-chirality nor a negative-chirality string, but some mixing of both. Physically, this is caused by the fact that for a closed string there is no difference in the pattern of plaquette violations between positive- and negative-chirality strings, since there are no endpoints. From Eq. \eqref{eq:equivalence_strings_z} we can see that starting from a positive-chirality string, $S^{+}_{\mathcal C}$, it is possible to add $S^z$ operators forming a loop which cannot be expressed as the product of vertex operators (which we call a non trivial loop in the remainder of the section) to the linear superposition. Once this is done, the resulting operator, $\tilde{S}_{\mathcal C}$, does not have a well-defined chirality. Remember, that for an open string, we may obtain the negative-chirality string by multiplying it by an open string $S^z$, violating the plaquettes at the endpoints. For closed strings we may proceed analogously to obtain the opposite chirality by multiplying by a non-trivial closed string $S^z_{\mathcal{C'}}$. Since we are multiplying $S^{+}_{\mathcal C}$ by a linear combination of trivial and non-trivial loops of $S^z$, the chirality of $\tilde{S}_{\mathcal C}$ is no longer positive nor negative. Thus, we drop the `$+$' superscript in $\tilde{S}_{\mathcal C}$.

The mixing of chiralities becomes apparent when computing the commutator between a closed string, $S_\mathcal{C}$, and  a positive-chirality open string, $S^+_\mathcal{O}$, where paths $\mathcal{C}$ and $\mathcal{O}$ cross once (see Fig. \ref{fig:string_open_closed}):
\begin{equation}\label{eq:string_commutator}
\mathcal{R_{O,C}} = \langle \vec i \;\vert\left[ S^+_\mathcal{O}, S_\mathcal{C}\right]\vert \vec i\; \rangle= \frac{F_\mathcal{O}\left(\vec i \oplus \vec{\alpha}^\mathcal{C}\right) F_\mathcal{C}\left(\vec i \; \right)}{F_\mathcal{O}\left(\vec i \; \right) F_\mathcal{C}\left(\vec i \oplus \vec{\alpha}^\mathcal{O}\right)},
\end{equation}
where here $[\cdot\;,\cdot]$ is the commutator in the group sense, i.e., $[g,h]=g^{-1}h^{-1}gh$. In this case, notice that on $\textnormal{Conn} (\mathcal C)$, the configurations $\vec i$ and $\vec i \oplus \vec{\alpha}^{\mathcal O}$ do not belong to the same class. Using Eq.~\eqref{eqn:string_op_diff_phases}, it is thus clear that $\mathcal R_{\mathcal O, \mathcal C}$ can be modified by selecting different phases to initialize $F_{\mathcal C}$ in Algorithm~\ref{alg:F_finding}.

\end{document}